\documentclass[aps,onecolumn]{revtex4}

\usepackage{graphicx}
\usepackage{dcolumn}
\usepackage{bm}
\usepackage{amsmath}
\usepackage{amssymb}
\usepackage{amsfonts}
\usepackage{color}
\usepackage{mathtools}
\usepackage{nicefrac}
\usepackage[lofdepth,lotdepth]{subfig}

\newcommand{\period}{\hspace{0.005\linewidth} . \hspace{0.02\linewidth} }
\newcommand{\coma}{\hspace{0.02\linewidth} , \hspace{0.02\linewidth} }
\newcommand{\with}{\hspace{0.02\linewidth} \text{ with } \hspace{0.02\linewidth} }
\newcommand{\andd}{\hspace{0.02\linewidth} \text{ and } \hspace{0.02\linewidth} }
\newcommand{\for}{\hspace{0.02\linewidth} \text{ for } \hspace{0.005\linewidth} }

\newcommand{\bs}[1]{\boldsymbol{#1}}
\newcommand{\rey}[1]{\overline{#1}}
\newcommand{\integ}[1]{\big\langle #1 \big\rangle}

\newcommand{\rp}{\tau p}
\newcommand{\qi}{q^\texttt{ir}}
\newcommand{\qs}{q^\texttt{so}}

\newcommand{\Li}{\mathcal{L}^\texttt{ir}}
\newcommand{\Ls}{\mathcal{L}^\texttt{so}}
\newcommand{\Lsz}{\mathcal{L}^\texttt{so}_0}

\newcommand{\mE}{\texttt{E}}
\newcommand{\sQ}{Q^\texttt{so}}
\newcommand{\mQ}{\texttt{Q}^\texttt{so}}
\newcommand{\mT}{\texttt{T}}
\newcommand{\mF}{\texttt{E}_{ww}}
\newcommand{\mG}{\texttt{E}_{uw}}
\newcommand{\mQt}{\texttt{Q}^\texttt{tot}}
\newcommand{\mQi}{\texttt{Q}^\texttt{ir}}
\newcommand{\mR}{\texttt{E}_{\rho\rho}}

\newcommand{\rh}{\rey{\rho}}
\newcommand{\kvd}{k_\texttt{vd}}
\newcommand{\kes}{\mathcal{K}^\texttt{so}}
\newcommand{\kr}{\mathcal{K}}
\newcommand{\eps}{\varepsilon}

\newcommand{\se}{s_{e}}
\newcommand{\sq}{s_{q}}
\newcommand{\nn}{n_k}

\newcommand{\nx}{\tilde{x}}
\newcommand{\bnx}{\bs{\nx}}

\newcommand{\tf}[1]{\widehat{#1}}
\newcommand{\pp}{\mathcal{P}}
\newcommand{\proj}{P}

\newcommand{\x}{\bs{x}}
\newcommand{\K}{\bs{k}}
\newcommand{\p}{\bs{p}}
\newcommand{\q}{\bs{q}}

\newcommand{\nk}{\tilde{k}}
\newcommand{\bnk}{\bs{\nk}}

\newcommand{\ud}{\mathrm{d}}
\newcommand{\cjgt}{*}

\date{\today}

\begin{document}

\title{
Permanence of large eddies in variable-density homogeneous turbulence \\ with small Mach numbers
}

\author{O. Soulard}
\email[Corresponding author : ]{olivier.soulard@cea.fr}
\author{J. Griffond}
\author{B.-J. Gr\'ea}
\author{G. Viciconte}
\affiliation{CEA, DAM, DIF, F-91297 Arpajon, France.}

\begin{abstract}
In this work, we study the large-scale structure of homogeneous flows displaying high density contrasts and small turbulent Mach numbers.
Following Batchelor \& Proudman \cite{batchelor56}, we draw an analogy between this configuration and that of a single isolated eddy displaying density non-uniformities.
By doing so, we are able to highlight the crucial role played by the solenoidal component of the momentum in the preservation of initial conditions.
In particular, we show that the large-scale initial conditions of the spectrum of the solenoidal momentum are preserved provided its infrared exponent is smaller than $4$. This condition is reminiscent of the one usually derived in constant density flow, except that it does not apply to the velocity spectrum. The latter is actually shown to be impermanent for infrared exponents larger or equal to $2$. 
The consequences of these properties on the self-similar decay of the flow are discussed.
Finally, these predictions are verified by performing large eddy simulations of homogeneous isotropic turbulence.
\end{abstract}

\maketitle

%=======================================================================
\section{Introduction}
%=======================================================================
 Eddies much larger than the integral scale of turbulence play a central role in the  decay of constant density homogeneous turbulence \cite{loitsyanskii39,kolmogorov41b,landau54,saffman67,lesieur00,davidson04,lesieur08,llor11,mons14}.
Under certain conditions, these large eddies evolve on a time scale much longer than the one governing the decrease of kinetic energy or the growth of the integral scale. 
As a result, their initial state is preserved during the whole flow evolution: they are said to be permanent.
When this permanence is verified, the  flow is constrained by its large-scale initial condition and its decay rate can be expressed as a function of some large-scale initial parameter, such as, for instance, the initial infrared exponent.
The latter corresponds to the power law exponent of the initial kinetic energy spectrum at small wave numbers and is denoted by $\se$.
Its value is related the existence of large scale integrals of turbulence, such as Loitsyanskii's \cite{loitsyanskii39,kolmogorov41b,landau54} or Saffman's \cite{saffman67}.
Whether large eddies are permanent or not depends on the long range correlations induced by non-linear terms. 
In a constant density flow, these terms involve  quadratic products between the  components of the velocity field and their non-local propagation by the pressure field. 
When $\se \ge 2$, these non-linearities lead to a ``backscattering'' transfer of energy: large eddies receive energy from interactions involving smaller eddies, mostly those with a size on the order of the integral scale.
Several models \cite{proudman54,lesieur08,llor11} predict that this backscattering transfer has an infrared exponent of $4$ when expressed in spectral space.
Thus, for initial spectra satisfying $\se<4$, non-linear processes have a vanishingly small effect and the infrared spectrum is invariant, i.e. large eddies are permanent.
By contrast, spectra with $\se > 4$ are not invariant: they evolve towards a spectrum with an infrared exponent of $4$, equal to that of the backscattering term.
In practice, it is found that the transition between the permanent and impermanent behavior of the spectrum is not sharp and occurs over a small interval between $4$ and a lesser value close to $3.5$ \cite{lesieur08,davidson04,llor11}.

Beyond the constant density context, the properties of large eddies have also been studied for different types of variable-density flows.
In particular, when the  Mach number and the density contrast are small enough for the flow to obey the incompressible Boussinesq approximation, large eddies have been  shown to be permanent under conditions similar to the ones proposed in the constant density case. This conclusion has been reached  for several homogeneous and inhomogeneous configurations \cite{batchelor92,soulard14,soulard15,soulard18}.
Besides, several authors have also considered the large-scale properties of fully compressible turbulence, when the Mach number and the density contrast are both high.
Chandrasekhar \cite{chandrasekhar51} was one of the first to make a foray into this field of study. He showed the existence of a large-scale invariant for the density, but  did  not discuss the existence of velocity invariants.
Around the same time, Krzywoblocki \cite{krzywoblocki52} obtained the same invariant for the density and also suggested that Loitsyanskii's integral was not constant because of compressibility effects.
Those two results were generalized by Sitnikov \cite{sitnikov58}. The latter derived a series of Saffman-like integrals based on the conservation of momentum, density and total energy. He then argued that they should be invariant provided that non-linear terms decay sufficiently rapidly at large scales. More precisely, in spectral space, non-linear terms should have an infrared exponent strictly larger than $2$, as made explicit in Monin \& Yaglom \cite{monin75}.
In compressible turbulence, Monin \& Yaglom \cite{monin75} justify that this condition is respected because the sound velocity is finite and that distant points decorrelate exponentially, an argument also used by Lumley \cite{lumley66}

In between the two extreme cases of Boussinesq and fully compressible turbulence,  lies an intermediate class of variable-density flows: those having a small Mach number and a high density contrast. 
This particular regime is met in many configurations of interest, including flows driven by the Richtmyer-Meshkov and Rayleigh-Taylor instabilities \cite{zhou17a,zhou17b,boffetta17}.
Because of the small Mach number condition, the argument of Monin \& Yaglom \cite{monin75} based on the sound velocity cannot be used to guarantee a fast decay of non-linear terms at large scales. Consequently, the results of Sitnikov \cite{sitnikov58} on fully compressible turbulence lose their justification.
Because of the high density contrast condition, the Boussinesq approximation becomes invalid so that the corresponding predictions on large scales lose their basis.
Hence, as far as large scales are concerned, small Mach number high density contrast flows stand apart from Boussinesq and fully compressible flows and require a study of their own.
In this respect, a vast literature provides an in depth analysis of their statistical properties \cite{sandoval95, livescu07,livescu08,chung10,gauthier10,movahed15}.
Unfortunately, most of these works  do not address the behavior of large scales, save for brief comments in Refs. \cite{livescu07,gauthier10}. Therefore, the large-scale properties of flows having a small Mach number and a high density contrast remain mostly unexplored.

Thus, the purpose of this work is to study the large-scale structure of variable-density homogeneous  turbulence for flows having a small Mach number and a high-density contrast.
To this end, we will rely on the so-called ``variable-density'' approximation, a quasi-incompressible approximation which has been shown to provide an accurate description of this type of flow \cite{sandoval95, livescu07}.
This approximation will be recalled in Sec. \ref{sec:govern_eq}.
Starting from this approximation, we will endeavor in Sec. \ref{sec:single_eddy} to understand the role played by initial conditions and pressure forces  on large scales by looking at a simplified configuration: that of a single isolated variable-density eddy. 
Indeed, Batchelor \& Proudman \cite{batchelor56} have shown that an analogy exists between the large-scale structure of homogeneous turbulence and the properties of a single eddy, far from its core.
By adding density non-uniformities to the eddy, we will be able to highlight their role by comparison with the constant density situation studied in \cite{batchelor56}. In particular, we will discuss the role played by the solenoidal component of the momentum and its initial conditions on the  the late-time evolution of the eddy.
In Sec. \ref{sec:hom_turb}, we will  perform a spectral analysis of  variable-density homogeneous turbulence.
More precisely, we will focus on the spectrum of the solenoidal component of the momentum and propose a closure for its evolution at small wavenumbers.
The invariance conditions of this spectrum will then be studied, as well as its relation with other spectra.
In Sec. \ref{sec:consequences}, the implications of the results obtained in Sec. \ref{sec:hom_turb} will be considered. In particular, we will discuss the self-similarity of the flow and the relevance of the Boussinesq and fully compressible predictions.
In Sec. \ref{sec:validation}, we will perform large-eddy simulations (LES) of homogeneous isotropic turbulence to validate our predictions.

%=======================================================================
\section{Governing equations} \label{sec:govern_eq}
%=======================================================================

%------------------------------------------
\subsection{Variable-density approximation}
%-------------------------------------------
We consider a turbulent flow obeying the variable-density approximation  \cite{sandoval95,livescu07,livescu08,chung10}.
This approximation can be thought of as a generalization of the Boussinesq approximation to flows displaying large density fluctuations but keeping small turbulent Mach numbers. 
It belongs to the broader family of pseudo-compressible approximations and can be derived from the Navier-Stokes equations by performing an asymptotic analysis such as the one proposed for instance in \cite{soulard12c}.

A simpler way to derive the variable-density set of equations consists in considering a turbulent mixture between two incompressible fluids.
The corresponding flow is then governed by  equations for the density $\rho$, the velocity field $\bs{v}$ and the concentration of one of the fluids $c$:
\begin{subequations}\label{eq:NS_cons}
\begin{align}
\label{eq:ns_rho}
 \partial_t \rho + \partial_j(\rho v_j) &= 0
\coma
\\ \label{eq:ns_rhoc}
 \partial_t (\rho c) + \partial_j(\rho v_j c) &= \partial_j (\rho \nu_c \partial_j c)
\coma
\\
\label{eq:ns_rhou}
 \partial_t (\rho v_i) + \partial_j(\rho v_i v_j) &= - \partial_j ( p \delta_{ij} + \sigma_{ij} ) 
\coma
\end{align}
where $\nu_c =  \mu_c/\rho$ is the diffusion coefficient of the concentration field, $p$ is the pressure and $\bs{\sigma}$ is the viscosity tensor:
\begin{align}
\sigma_{ij} = - \rho \nu S_{ij} \with S_{ij} =\partial_jv_i + \partial_iv_j - \frac{2}{3} \partial_k v_k \delta_{ij}
\andd 
\nu = \mu / \rho \text{ the shear viscosity}
\period
\end{align}
\end{subequations}
Since we assumed that the two fluids being mixed are incompressible,  density variations arise from modifications in the local composition. This assumption can be expressed in terms of the following equation of mixing:
\begin{align} \label{eq:eos}
\tau = \frac{1}{\rho} = \frac{1-c}{\rho_0} + \frac{c}{\rho_1}
\end{align}
with $\rho_0$ and $\rho_1$ the constant densities of each fluid and $\tau$ the  specific volume, inverse of the density $\rho$.
Then, from Eq. \eqref{eq:ns_rho}, one has $\partial_t\rho+v_j\partial_j\rho = -\rho \partial_jv_j$ while from Eqs. \eqref{eq:ns_rhoc} and \eqref{eq:eos}, one has $\partial_t\rho + v_j\partial_j\rho = \rho \partial_j (\nu_c \partial_j \rho/\rho)$.
Thus, to ensure that the evolutions of $\rho$ and $c$ respect the equation of mixing, the following constraint on the velocity divergence must be verified:
\begin{align} \label{eq:divu}
\partial_j v_j = \partial_j a_j
\coma
\end{align}
where the velocity $a_j$ corresponds to the molecular transport of the density:
\begin{align}
a_j  = - \nu_c \partial_j \rho/\rho
\period
\end{align}
Because the two fluids being mixed are incompressible, the viscous coefficients $\nu$  and $\nu_c$ may only be functions of the local composition. Pressure and temperature variations are implicitly neglected.
Equivalently,  $\nu$  and $\nu_c$ can be expressed as functions of the density $\rho$:
$$
\nu \equiv \nu(\rho) \coma \nu_c \equiv \nu_c(\rho)
\period
$$
As a result, the diffusive velocity $\bs{a}$ takes the form of a gradient:
\begin{align}\label{eq:a_A}
a_i = \partial_i A(\rho) \with A(\rho) = A(\rho_0) - \int_{\rho_0}^\rho \frac{\nu_c(\xi)}{\xi}\ud \xi
\period
\end{align}

%--------------------------------------
\subsection{Divergence-free formulation}
%---------------------------------------
Equation \eqref{eq:divu} implies that the divergence of the difference between $\bs{v}$ and $\bs{a}$ is null. While it is by no means necessary, we find it convenient to work with a divergence-free velocity. Therefore, instead of $\bs{v}$ we will hereafter consider  the following velocity field:
$$
\bs{u} = \bs{v} - \bs{a}
\period
$$
To express the evolution of $\bs{u}$, we must first derive that of $\bs{a}$. To this end, we use the fact that $\bs{a}$ is a gradient: we take the time derivative of Eq. \eqref{eq:a_A} and obtain that:
$$
\partial_t a_i = - \partial_i( \nu_c {\partial_t \rho}/{\rho}) =  - \partial_i\big( a_k v_k - \nu_c \partial_k a_k \big)
\period
$$
Combining these different elements, Sys. \eqref{eq:NS_cons} can be rewritten by eliminating the concentration field and by using only the velocity field $\bs{u}$ and the density~$\rho$: 
\begin{subequations} \label{eq:ns_vd}
\begin{align}
\label{eq:vd_rho}
\partial_t \rho + \partial_j \big(\rho u_j \big) &= \partial_j\big( \nu_c \partial_j\rho\big)
\coma
\\
\label{eq:vd_rhou}
\partial_t(\rho u_i) + \partial_j\big( \rho u_i u_j \big) &= - \partial_j\big(p \delta_{ij} + \Sigma_{ij} \big)
\coma
\\
\label{eq:vd_divu}
\partial_j u_j &= 0
\coma
\end{align}
where $\Sigma_{ij}$  accounts for various viscous and diffusive effects
\begin{align}
\Sigma_{ij} = \sigma_{ij} 
+ \rho a_i u_j + \rho u_i a_j + \rho a_i a_j 
- \rho \Big(  u_k a_k  + a_k a_k - \nu_c \partial_k a_k  \Big) \delta_{ij}  
\period
\end{align}
\end{subequations}
Note that when expressed in terms of the velocity $\bs{u}$, one has $\sigma_{ij}=-\rho \nu(\partial_j u_i + \partial_iu_j)  + 2 \rho \nu(\partial_j a_i - \partial_ka_k \delta_{ij}/3)$.

Sys. \eqref{eq:ns_vd} defines the evolution of a flow under the variable-density approximation. 
It differs from the Boussinesq system of equations by two main aspects.
First, the viscosity tensor $\sigma_{ij}$ is replaced by a more complex expression $\Sigma_{ij}$. This is due to the fact that the actual velocity field $\bs{v}$ is not divergence-free. Instead, its divergence accounts for the modification in density due to the molecular mixing of the two species present in the flow.
When cast in terms of the divergence-free velocity $\bs{u}=\bs{v}-\bs{a}$, these molecular mixing effects find their way back into the viscous stress tensor.
Second,  and more importantly, the density in the velocity evolution equation \eqref{eq:vd_rhou} is not constant, as it is in the Boussinesq approximation. As a result, the non-linearities acting on the velocity field differ from the Boussinesq case.
 This aspect can be observed more clearly by  recasting Eq. \eqref{eq:vd_rhou}  as a non-conservative equation for the velocity field $\bs{u}$. One obtains:
\begin{subequations}
\begin{align}
\label{eq:vd_u}
\partial_t u_i + \partial_j\big( u_j u_i \big) &= - \partial_j\big(\rp \delta_{ij} + K_{ij} \big) 
+ f_i
\end{align}
where the non-symmetric viscous tensor $\bs{K}$ is defined by:
\begin{align}
K_{ij} = \tau \sigma_{ij} + a_i u_j + \Big( \nu_c \partial_k a_k -  u_k a_k - \frac{a_ka_k}{2}\Big) \delta_{ij} 
\coma
\end{align}
and where $\bs{f}$ is a force proportional to the density gradient:
\begin{align}\label{eq:force}
f_i = p\partial_i\tau - \rho \eps_{i\tau}
\period
\end{align}
with $\eps_{i\tau}$ defined by:
\begin{align}\label{eq:eps_it}
\eps_{i\tau} = (\nu_c \partial_j u_i + \nu S_{ij}) \partial_j \tau
\period
\end{align}
\end{subequations}
The first term in the expression of the force $\bs{f}$ can be understood as the pressure work exerted on an element of mass when its volume changes along a density gradient. The second term corresponds to a cross-dissipation term for the specific volume and velocity.

Thus, from Eq. \eqref{eq:vd_u}, it can be seen that, because of density variations, an additional non-linear term exists compared to the Boussinesq approximation.
This term takes the form of a force $\bs{f}$  which involves the product of the density gradient with the pressure field and a viscous/diffusive tensor.
As will be seen, this additional non-linearity modifies the properties of the pressure field and plays an important role in the behavior of large scales.
%

%-----------------------------------
\subsection{About the pressure field}
%------------------------------------

The pressure field is one of the primary factors involved in the evolution and statistical properties of large eddies. Indeed, because of its non-local nature, it is at the origin of long range correlations between distant points.
Given the incompressible constraint \eqref{eq:vd_divu}, one can derive a Poisson-like equation for the pressure by taking the divergence  of Eq. \eqref{eq:vd_u}:
\begin{align} \label{eq:poisson}
\partial^2_{jj} \left(\tau p\right)  - \partial_j ( p \partial_j \tau)
= 
-\partial^2_{ij}\big( u_i u_j +  K_{ij}\big)
-
\partial_j (\rho \eps_{j\tau} )
\period
\end{align}
When density is constant, the Poisson equation \eqref{eq:poisson} simplifies to :
\begin{align}\label{eq:poiss_cd}
\text{for } \rho = \text{Cst} \coma
\partial^2_{jj}(\rp)  
= 
-\partial^2_{ij}\big( u_i u_j \big)
\period
\end{align}
Comparing Eqs. \eqref{eq:poisson} and \eqref{eq:poiss_cd}, it can be seen that density variations affect the value of the pressure field in several ways.
In particular, the operator acting on pressure is not a Laplacian any more. Instead, it involves another term depending on the product of the pressure and the gradient of the specific volume.
This term arises because of the additional non-linearity highlighted previously in Eq. \eqref{eq:vd_rhou}.
Note that for numerical codes aiming to solve the variable-density equations, this operator modification prevents the use of efficient and standard techniques for computing the pressure field. Custom, and often costly, iterative pressure solvers must be put in place \cite{livescu07,livescu08,chung10}.

The other differences between the variable and constant density cases stem from molecular transport effects, either through the value of the velocity divergence or because of the non-linear product between $\tau$ and the gradient of the viscosity tensor in Eq. \eqref{eq:vd_u}.
In this respect, the main difference is probably the appearance of the divergence of the cross-dissipation term $\eps_{i\tau}$ as a new source in the pressure equation. The other viscous and diffusive terms only modify the definition of the stresses $K_{ij}$ but otherwise do not alter the structure of the  Poisson equation.

%=======================================================================
\section{The nature of the problem} \label{sec:single_eddy}
%=======================================================================
Before they dived into the analysis of the large-scale structure of homogeneous turbulence in their seminal article \cite{batchelor56}, Batchelor \& Proudman took a step aside and proposed a simple example in order to make ``clear the nature of the problem before [them]''.
More precisely, Batchelor \& Proudman drew an analogy between the large-scale structure of homogeneous turbulence and the properties of a single eddy, far from its core.
This analogy allowed them to highlight the central role played by  initial conditions and  by pressure forces.
In this section, we would like to take a similar detour by 
adapting the single-eddy configuration of Batchelor \& Proudman to the variable-density case. As in \cite{batchelor56}, our aim is to shed light on the remote action of the pressure field and to gain insight into the properties of large eddies in homogeneous turbulence.

\begin{figure}[!htb]
\includegraphics[width=0.3\linewidth]{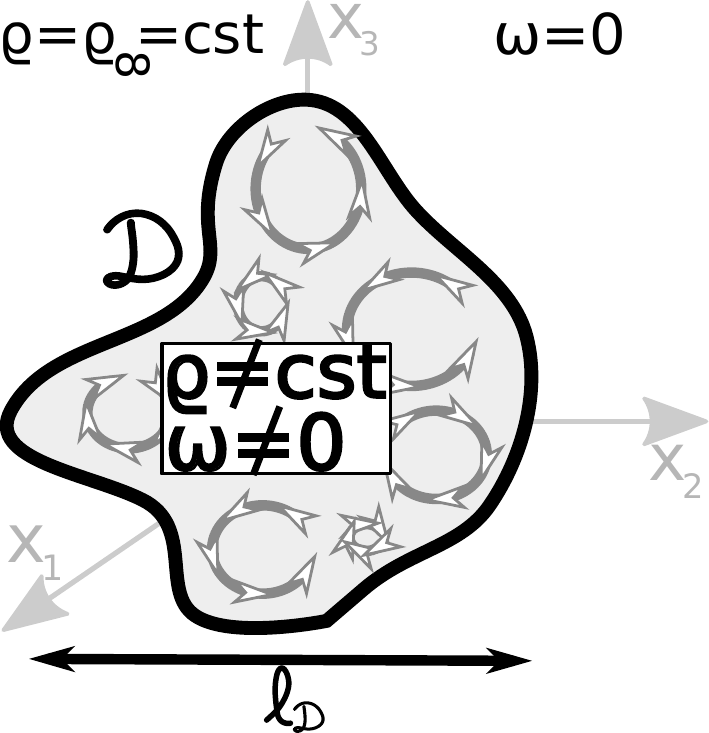}
\caption{\label{fig:vorticity}
Schematic representation of a single variable-density eddy.
}
\end{figure}

Thus, as in \cite{batchelor56}, we consider a configuration where the vorticity field $\bs{\omega}$ is null everywhere except in a finite domain  $\mathcal{D}$  located close to the origin $\x=0$. This isolated blob of vorticity is hereafter assimilated to a single eddy.
Compared to \cite{batchelor56}, we then add an extra-element:  within the domain $\mathcal{D}$ where vorticity is non-zero, we assume that the density field $\rho$ is non-uniform, while outside of $\mathcal{D}$, it is constant.
This configuration is displayed schematically in Fig. \ref{fig:vorticity}.

%---------------------------------------------
\subsection{Far-field scaling of the momentum} \label{sec:momentum}
%---------------------------------------------
The evolution of the eddy is governed by Sys. \eqref{eq:ns_vd} which expresses the conservation of the density $\rho$ and of the momentum $\rho \bs{u}$ in a divergence-free flow.
Outside the eddy, the density $\rho$ is simply equal to $\rho_\infty$ while the momentum $\rho \bs{u}$ decays to zero.
Our purpose is to express the scaling of the momentum far from the eddy core.

To this end, we start by relating the momentum to quantities localized within the boundaries of the eddy.
This can be done by performing an Helmholtz decomposition of the momentum. This operation splits the momentum into an irrotational component $\bs{\qi}$ and a solenoidal one $\bs{\qs}$:
\begin{align}
\rho \bs{u} = \bs{\qi} + \bs{\qs}
\with 
\qi_i = - \partial_i \phi \andd
\qs_i = \epsilon_{ijk} \partial_j \psi_k
\coma
\end{align}
where $\epsilon_{ijk}$ is the Levi-Civita tensor.
The scalar and vector potentials $\phi$ and $\bs{\psi}$ are given by:
$$
\phi(\x,t) = \frac{1}{4 \pi} \int \partial_j(\rho u_j)(\x',t) \frac{\ud \x'}{\left|\x - \x' \right|}
\andd
\psi_i(\x,t) = \frac{1}{4 \pi} \int \Omega_i(\x',t) \frac{\ud \x'}{\left|\x - \x' \right|}
\period
$$
where $\bs{\Omega}$ is a density-weighted ``vorticity'', different from the actual vorticity $\bs{\omega}$. The two are defined by:
$$
\Omega_i = \epsilon_{ijk} \partial_j (\rho u_k) = \rho \omega_i + \epsilon_{ijk}u_k \partial_j \rho 
\andd 
\omega_i = \epsilon_{ijk} \partial_j u_k
\period
$$
Because both the vorticity $\bs{\omega}$ and the density gradient are null outside the eddy, this is also the case for $\bs{\Omega}$.
Besides, while we have chosen to focus on the modified divergence-free velocity $\bs{u}$ instead of the actual velocity $\bs{v}$, it is worth noting that both velocity fields share the same vorticity fields: $\omega_i  = \epsilon_{ijk} \partial_j v_k = \epsilon_{ijk} \partial_j u_k$ and $\Omega_i  = \epsilon_{ijk} \partial_j(\rho v_k) = \epsilon_{ijk} \partial_j (\rho u_k)$. This is because the diffusion velocity $\bs{a}=\bs{v}-\bs{u}$ is a gradient proportional to $\bs{\partial} \rho$.

Far from the core of the eddy, the expressions for $\phi$ and $\psi$ can be Taylor-expanded. 
Injecting these expansions into the definitions of $\bs{\qi}$ and $\bs{\qs}$ as well as their sum $\rho \bs{u}$, we deduce that:
\begin{subequations} \label{eq:q_farfield}
\begin{align}
\label{eq:qi_farfield}
\text{for }  |\x| \gg \ell_\mathcal{D} \coma &
\qi_i(\x,t) = \phantom{\rho_\infty} \Li_j(t) \, M_{ji}(\bnx) \, |\x|^{-3}  + \mathcal{O}(|\x|^{-4})
\coma
\\
\label{eq:qs_farfield}
 &
\qs_i(\x,t) = \phantom{\rho_\infty} \Ls_j(t) \, M_{ji}(\bnx) \, |\x|^{-3}  + \mathcal{O}(|\x|^{-4})
\coma
\\ &\label{eq:rhou_farfield}
\rho u_i(\x,t) = \rho_\infty L_j(t) \,  M_{ji}(\bnx) \, |\x|^{-3}
+ \mathcal{O}(|\x|^{-4})
\coma
\end{align}
\end{subequations}
where $\ell_\mathcal{D}$ is the characteristic length of the domain $\mathcal{D}$, $\bnx = \x/|\x|$, $M_{ij}(\bnx) = (3\nx_i\,\nx_j - \delta_{ij} )/(4\pi) $ and where the integrals $\bs{L}$, $\bs{\Li}$ and $\bs{\Ls}$ are defined by:
\begin{align} \label{eq:integ}
\begin{split}
\Li_i(t) = - \int \big(\rho(\x,t)-\rho_\infty\big) u_i(\x,t) \ud \x
\coma
\Ls_i(t) = \frac{1}{2} \int \epsilon_{ijk} x_j \Omega_k(\x,t) \ud \x
\\
\andd
\rho_\infty {L}_i(t) = {\Li_i(t) + \Ls_i(t)} = \frac{\rho_\infty}{2} \int \epsilon_{ijk} x_j \omega_k(\x,t) \ud \x
\period
\end{split}
\end{align}
The integral $\bs{L}$ is a quantity commonly used to characterize eddies in constant density flows.
In this context, $\bs{L}$ is called the linear impulse of the eddy \cite{davidson04} and coincides with its linear momentum provided the velocity field decays sufficiently rapidly. 
 However, for a variable-density eddy, the connection between $\bs{L}$ and the momentum is lost. It is $\bs{\Ls}$ which ensures this role, while $\bs{L}$ is instead linked to the average velocity of the eddy.
More precisely, if the momentum of the eddy $\rho \bs{u}$ converges faster than $|\x|^{-3}$, the integral $\bs{\Ls}$ can be shown to be equal to the linear momentum $\int \rho \bs{u} \ud \x$ while $\bs{L}$ becomes equal to $\int \bs{u} \ud \x$.
Despite its inherent inadequacy, we will nonetheless keep the denomination ``linear impulse'' to refer to $\bs{L}$. The integrals $\bs{\Ls}$ and $\bs{\Li}$ will be referred to as ``linear solenoidal impulse'' and ``linear irrotational impulse'' to stress their origins.

Note that the linear solenoidal impulse $\bs{\Ls}$ is linked to the self-induced  translational momentum of the eddy. It gives an indication of the displacement of the eddy, when seen as a region carrying vorticity. 
As for the linear irrotational impulse $\bs{\Li}$, it can be interpreted as the integral of the flux of density relative to its uniform constant value outside the eddy. It gives an indication of the displacement of the eddy, when seen as a region carrying density non-uniformities.

Equation \eqref{eq:rhou_farfield} highlights that the far-field momentum scaling depends on whether the eddy has a linear impulse or not:
$$
\text{if } \bs{L} = 0 \coma \rho \bs{u} \propto |\x|^{-4}
\andd
\text{if } \bs{L} \ne 0 \coma \rho \bs{u} \propto |\x|^{-3} 
\period 
$$
Given that density is constant for $|\x| \gg \ell_\mathcal{D}$, this conclusion also applies to the scaling of the velocity far-field.
In this respect, the result derived for the momentum/velocity scaling is strictly equivalent to the one obtained for a constant density eddy \cite{davidson04,batchelor56}. However, the behavior of $\bs{L}$ -- and consequently the far-field momentum/velocity scaling -- differs strongly whether in a constant or a variable-density case. This aspect is linked to the pressure field and will be discussed in Sec \ref{sec:press_field}.

Another significant point is that the irrotational and solenoidal components of the momentum have independent scalings, based respectively on whether $\bs{\Li}$ and $\bs{\Ls}$ are null or not.
Therefore, the overall scaling and prefactor of the far-field momentum can actually stand for a superposition of scalings and prefactors having distinct properties. This aspect will be discussed in Sec \ref{sec:invariant}.

To conclude this subsection, we would like to stress again that the Helmholtz decomposition we are considering is based on the momentum $\rho\bs{u}$ and not on the velocity field.
To highlight further this aspect, we list in table \ref{tab:notations} some of the quantities introduced so far along with their solenoidal and irrotational components. Anticipating on Secs. \ref{sec:hom_sea} and \ref{sec:hom_turb}, we also add to this table the spectra of $\bs{u}$ and $\rho \bs{u}$. Their precise definitions are given in Sec. \ref{sec:hom_turb}.
\begin{table}[!hbt]
\begin{tabular}{l c cc cc}
\hline\hline
  &  & & $\underbrace{\text{Solenoidal part}}$ & & $\underbrace{\text{Irrotational part}}$    \\
Velocity  & $\bs{v}$  &=& $\bs{u}$ &+& $\bs{a}$\\
Momentum  & $\rho \bs{v}$  &=& $\bs{\qs}$ &+& $\bs{\qi} + \rho \bs{a}$\\
$\bs{u}$-based momentum &  $\rho \bs{u}$ &=&  $\bs{\qs}$  &+& $\bs{\qi}$ \\
Linear impulse & $\rho_\infty \bs{L}$  &=& $\bs{\Ls}$   &+& $\bs{\Li}$ \\
Spectrum of $\bs{u}$  &  $\mE$ &=&  $\mE$  &+&  0 \\
Spectrum of $\rho \bs{u}$ &  $\mQt$   &=& $\mQ$   &+&  $\mQi$ \\
\hline
\hline
\end{tabular}
\caption{\label{tab:notations}  Solenoidal and irrotational components of the velocity, momentum, linear impulse and turbulent spectra.  }
\end{table}

%--------------------------
\subsection{Pressure field} \label{sec:press_field}
%--------------------------
The properties of the pressure field are set by the Poisson equation \eqref{eq:poisson}.
As already mentioned, this equation is different from the one obtained in the constant density case (Eq. \eqref{eq:poiss_cd}).
Indeed, it is not a Laplacian which acts on the pressure because $\tau$ is not constant.
This difference is such that an expression for the pressure as a function of the velocity field alone cannot be found in general.
Still, an implicit solution can be expressed by inverting the Laplacian acting on $\tau p$. This leads to:
\begin{equation} \label{eq:p_integ}
4 \pi  \,\tau p(\x) = \int\Big( \partial^2_{ij}\big( u_i u_j +  K_{ij}\big)  - \partial_j f_j \Big)   (\x')
 \frac{\ud \x'}{\left|\x - \x' \right|}
\period
\end{equation}
By expanding $\left|\x - \x' \right|$, we obtain that, far from the eddy core:
\begin{align} \label{eq:blob_press}
 \text{for } |\x| \gg \ell_\mathcal{D} \coma 
\tau p(\x) =& 
\phantom{+\,\,}
|\x|^{-2} \frac{\nx_i}{4\pi} \; \int f_i \ud\x
  \\ \nonumber &
 +
|\x|^{-3} M_{ij}(\bnx) 
\left( \int u_iu_j \ud \x + \int K_{ij} - \frac{1}{2} \big( x_i f_j  + x_j f_i \big) \; \ud \x \right)
  \\ \nonumber &
+\mathcal{O}(|\x|^{-4})
\period
\end{align}
The integral involving $f_i = p \partial_i \tau - \eps_{i\tau}$ converges because we assumed that $\tau$ is constant outside the domain $\mathcal{D}$, so that $\partial_i \tau = 0$ and $\eps_{i\tau}=0$ for $\x \notin \mathcal{D}$. Those involving the velocity field converge because $\bs{u}$ decays at least as fast as $|\bs{x}|^{-3}$.
As previously, we use the constant density case as a reference. 
For $\rho = \text{Cst}$, Eq. \eqref{eq:blob_press} simplifies to: 
$$
 \text{for } \rho = \text{Cst} \andd |\x| \gg \ell_\mathcal{D} \coma 
\tau p(\x) =  |\x|^{-3} M_{ij}(\bnx) \int u_iu_j \ud \x
+\mathcal{O}(|\x|^{-4})
\period
$$
Comparing this simplified expression to its full version \eqref{eq:blob_press}, it can be seen that the presence of density variations modifies the scaling of the pressure field: its decay is slower ($\sim |\x|^{-2}$) for a variable-density single eddy than for a constant density one ($\sim |\x|^{-3}$).
This difference has strong implications for the momentum.
Indeed, if we inject the momentum and pressure scalings (Eqs. \eqref{eq:rhou_farfield} and \eqref{eq:blob_press}) into the momentum equation \eqref{eq:vd_rhou}, we obtain the following orders of magnitude:
\begin{align} \label{eq:rhou_scaling}
 \text{for } |\x| \gg \ell_\mathcal{D} \coma 
\underbrace{\partial_t \rho u_i}_{ \rho_\infty |\x|^{-3}  \partial_t {L}_j M_{ji}  + \mathcal{O}(|\x|^{-4})}  = \underbrace{-\partial_i p}_{\begin{minipage}{0.23\linewidth}\begin{center} \begin{itemize} \footnotesize \item  Var. dens. : $\mathcal{O}(|\x|^{-3})$ \item Const. dens. : $\mathcal{O}(|\x|^{-4})$ \end{itemize} \end{center} \end{minipage}}  
- \underbrace{\partial_{j} \Sigma_{ij}}_{\mathcal{O}(|\x|^{-5})}
- \underbrace{\partial_j(\rho u_j u_i)}_{\mathcal{O}(|\x|^{-7})}  
\end{align}
Hence, in the constant density case, the  pressure field decays too rapidly to modify the term of the velocity field proportional to $\bs{L}$.
As a result, the linear  impulse $\bs{L}$ is an invariant: 
$$
 \text{for } \rho = \text{Cst} \coma
\partial_t \bs{L} = 0
\period
$$
Given Eq. \eqref{eq:rhou_farfield}, this means that if initially $\bs{L} \ne 0$ then $\bs{u}(\x) \propto |\x|^{-3} $ at all times and is stationary for $|\x| \gg \ell_\mathcal{D}$.
Besides, if initially $\bs{L} = 0$ then $\bs{u}(\x) \propto |\x|^{-4} $ at all times and is not necessarily stationary for $|\x| \gg \ell_\mathcal{D}$. 
The pressure field is too weak to modify the initial scaling of the velocity for a constant density eddy.

By contrast, for a variable-density eddy, the decay of the pressure field is slower and its gradient can now affect the evolution of the leading order of the velocity field, the one proportional to $\bs{L}$.
A full identification of the leading terms of the pressure gradient and momentum variation in Eq. \eqref{eq:rhou_scaling} leads to:
\begin{align} \label{eq:lin_imp}
 \text{for } \rho \ne \text{Cst} \coma
\partial_t \bs{L} = \int \bs{f} \ud \x
\period
\end{align}
Therefore, Eq.  \eqref{eq:lin_imp} shows that the integral $\bs{L}$ is generally not an invariant of a variable-density single eddy.
If we except special cases like barotropic inviscid flows, the source term in the evolution of $\bs{L}$  generally does not vanish: $\partial_t \bs{L} \ne 0$.
In particular, when $\bs{L}$ is initially null, it does not necessarily stay so and conversely when $\bs{L} \ne 0$.
Hence, the momentum far-field scaling is not set by the initial value of $\bs{L}$ and is not necessarily preserved in time. Besides, even when $\bs{L}\ne0$, the velocity and momentum are not necessarily stationary for $|x| \gg \ell_\mathcal{D}$.

Thus, because of density non-uniformities localized within the eddy, the pressure field casts its influence further and displays a stronger gradient at distant points. In turn, this enhanced gradient can modify the velocity far-field and its scaling. This modification coincides with a variation of the value of the linear impulse $\bs{L}$, which is not an invariant of the eddy.
This behavior of $\bs{L}$ and of the momentum is in stark contrast with the one encountered in constant density flows. Nonetheless, a continuous transition between the two configurations exists, as detailed in App. \ref{app:eddy_remarks}.

%-------------------------------------------------------------------
\subsection{Helmholtz components of the momentum}
\label{sec:invariant}
%-------------------------------------------------------------------
At the end of Sec. \ref{sec:momentum}, we stressed that the  far-field scaling of the momentum resulted from the superposition of two independent scalings: one coming from the irrotational component of its Helmholtz decomposition and the other one from its solenoidal component.
The far-field properties of these two components depend on the value of the two integrals $\bs{\Li}$ and $\bs{\Ls}$.
The evolutions of these integrals can be deduced from Eqs. \eqref{eq:ns_vd}.   By noting that  $\frac{1}{2} \epsilon_{ijk} x_j \Omega_k = \rho u_i - \frac{1}{2} \partial_j(x_j \rho u_i - x_k\rho u_k \delta_{ij} )$, we obtain that:
\begin{align} \label{eq:evol_integ}
\partial_t \bs{\Li} = -\rho_\infty \int \bs{f} \ud \x
\andd
\partial_t \bs{\Ls} = 0
\period
\end{align}

The evolution of $\bs{\Li}$ is proportional to that of $\bs{L}$ (Eq. \eqref{eq:lin_imp}). Consequently, we can apply to $\bs{\Li}$ and to the scaling of $\bs{\qi}$ the same observations as those detailed for $\bs{L}$ and  $\rho \bs{u}$ in Sec. \ref{sec:press_field}. In particular, the value of $\bs{\Li}$ and the far-field scaling of $\bs{\qi}$ are not fixed by their initial state and are not necessarily preserved in time. 
But the most important point shown by Eq. \eqref{eq:evol_integ} is that the linear solenoidal impulse is an invariant. Denoting by $\bs{\Lsz}$ its initial value, one has:
$$
\bs{\Ls}(t) = \bs{\Lsz}
\period
$$
As a result, the far-field behavior of the solenoidal momentum $\bs{\qs}$ is set by initial conditions:
\begin{itemize}
\item if $\bs{\Lsz} \ne 0$, the solenoidal momentum is invariant at distant points and is proportional to $ |\x|^{-3}$, 
\item if $\bs{\Lsz} = 0$,  the solenoidal momentum is proportional to $ |\x|^{-4} $ at distant points and at all times but the corresponding prefactor is not necessarily invariant.
\end{itemize}
These conclusions are similar to the ones drawn for $\bs{L}$ and $\rho \bs{u}$ when density is constant. And indeed,  in that case, the momentum has only one component: the solenoidal one. Therefore, one has:
$$
\text{for } \rho = \text{Cst} \coma 
\rho \bs{u} = \bs{\qs} \andd \rho_\infty \bs{L} = \bs{\Ls}
\period
$$
Thus, there is no difference between the constant and variable-density cases as far as the solenoidal momentum $\bs{\qs}$ and its related integral $\bs{\Ls}$ are concerned.
The mentioned discrepancy for the total momentum $\rho \bs{u}$ and the linear impulse $\bs{L}$ stem entirely from the existence of an irrotational component of the momentum, which is linked to density non-uniformities, and which has a time-dependent scaling.

Upon this particular point, the  evolutions of $\bs{\Li}$ and $\bs{L}$ are not necessarily known with precision.
Still, their long-time behavior can be deduced in the following way.
Starting from the definition of $\bs{\Li}$ given in Eq. \eqref{eq:integ} and applying the Schwartz inequality, one finds that:
$$
|\bs{\Li}| \le \rho_\infty \sqrt{ \int \rho(\tau -\tau_\infty)^2 \ud \x  \int  \rho u_i u_i \ud\x }
\coma
$$
with $\tau_\infty = 1/\rho_\infty$. The first integral in the right-hand side of the inequality corresponds to the variance of the specific volume within the blob and the second to the total kinetic energy. They satisfy:
$$
\partial_t \int \rho(\tau -\tau_\infty)^2 \ud \x = - 2 \int \rho \nu_c \partial_j \tau \partial_j \tau \ud \x
\andd
\partial_t \int \rho u_i u_i \ud\x =  -2 \int \rho (-K_{ij} \partial_j u_i) \ud\x
\period
$$
Thus, the variance of the specific volume is a decreasing function of time so that it eventually tends to $0$. As for the kinetic energy, its evolution is only driven by molecular processes and is also expected to decrease at late times, even though the positivity of the dissipation $-K_{ij} \partial_j u_i$ cannot be guaranteed at all times.
In any case, the total kinetic energy remains bounded as there is no production mechanism in the flow.
As a result, $\bs{\Li}$ tends to $0$ at late times:
$$
\text{For } t \to \infty \coma \bs{\Li} \to 0
\period
$$
Given the relation $\bs{L} = (\bs{\Li}+\bs{\Ls})/\rho_\infty$ and the invariance of $\bs{\Ls}$, we also deduce that:
$$
\text{For } t \to \infty \coma \bs{L} \to \bs{\Lsz}
\period
$$
These relations allow to determine the late time scalings of the irrotational momentum and of the velocity field.
Indeed, they show that, at late times and distant points ($t \to \infty$ and $|\x| \gg \ell_\mathcal{D}$):
\begin{itemize}
\item the irrotational momentum $\bs{\qi}$ scales as $|\x|^{-4}$ even if initially it scaled as $|\x|^{-3}$, 
\item  if $\bs{\Lsz} \ne 0$, the velocity  scales as $ |\x|^{-3}$ even if initially it scaled as $|\x|^{-4}$, 
\item if $\bs{\Lsz} = 0$,  the velocity  scales as $ |\x|^{-4}$ even if initially it scaled as $|\x|^{-3}$. 
\end{itemize}
Thus, the initial value of $\bs{\Ls}$ not only sets the far-field properties of $\bs{\qs}$ at all times, it also sets the far-field properties of the momentum $\rho \bs{u}$ at late times.

To conclude this section, we would like to mention the existence of other invariants for a variable-density eddy, even though we will not use them in the remaining of this work.
The invariance of $\Ls$ corresponds to the conservation of the linear momentum. Other invariants can be built on the conservation of mass and of the angular momentum.
Thus, one can show that:        
\begin{align}\label{eq:other_inv}
\partial_t \mathcal{M} = 0
\andd
\partial_t \bs{\mathcal{H}} =0
\with \mathcal{M} = \int (\rho-\rho_\infty) \ud \x 
\andd 
\mathcal{H}_i = \frac{1}{3} \int x_i x_j \Omega_j - x_j x_j \Omega_i \ud \x
\period
\end{align}

%-------------------------------------------------
\subsection{Homogeneous sea of independent eddies} \label{sec:hom_sea}
%-------------------------------------------------

\begin{figure}[!htb]
  \begin{minipage}{0.48\linewidth}
    \begin{center}
      \includegraphics[width=0.5\linewidth]{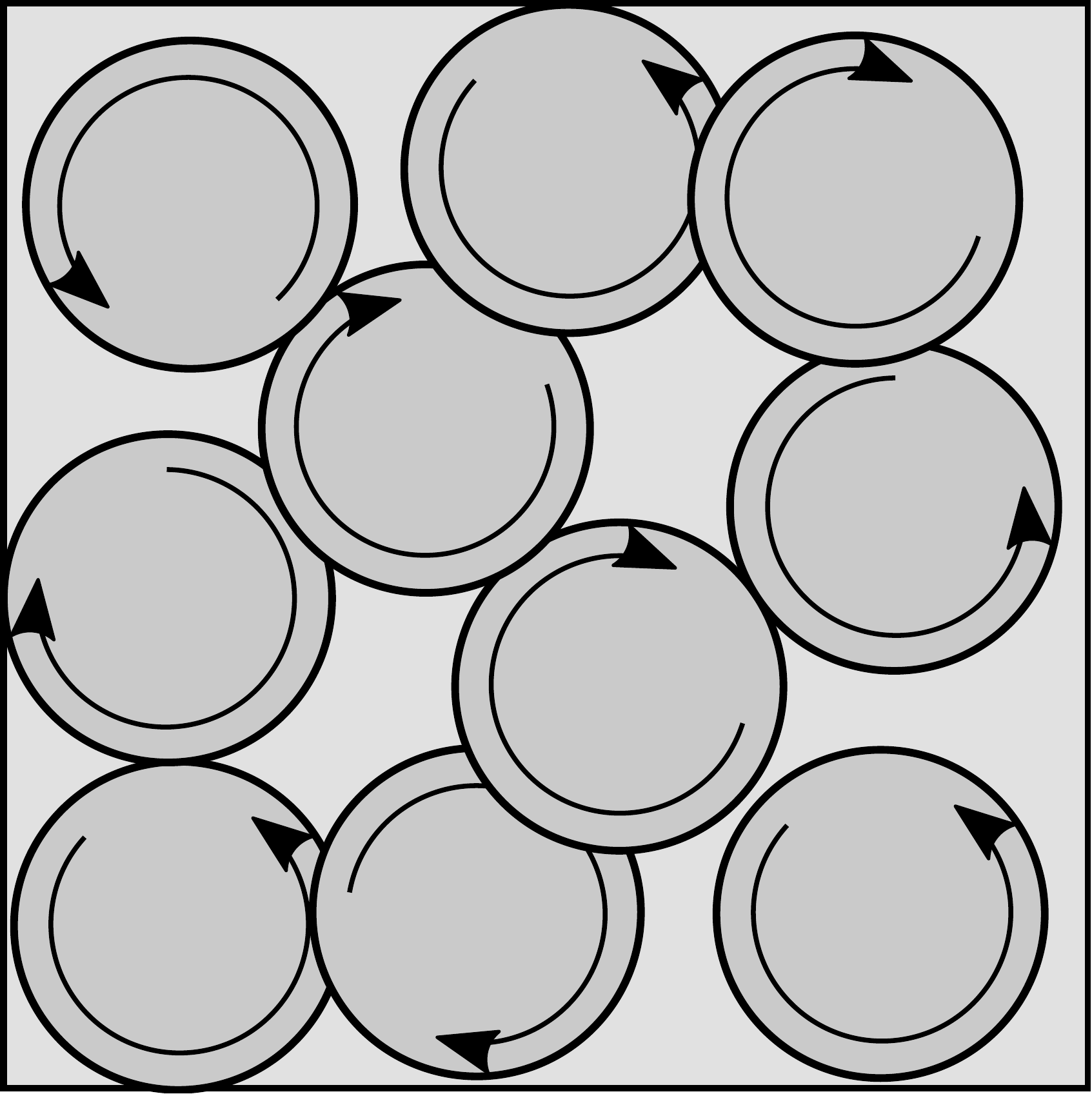}

      If $\bs{L}=0$ \;,\; $\mE(k) \propto k^4$  for $k\to 0$

      (Batchelor spectrum).
    \end{center}

  \end{minipage}
  \begin{minipage}{0.48\linewidth}
    \begin{center}
      \includegraphics[width=0.5\linewidth]{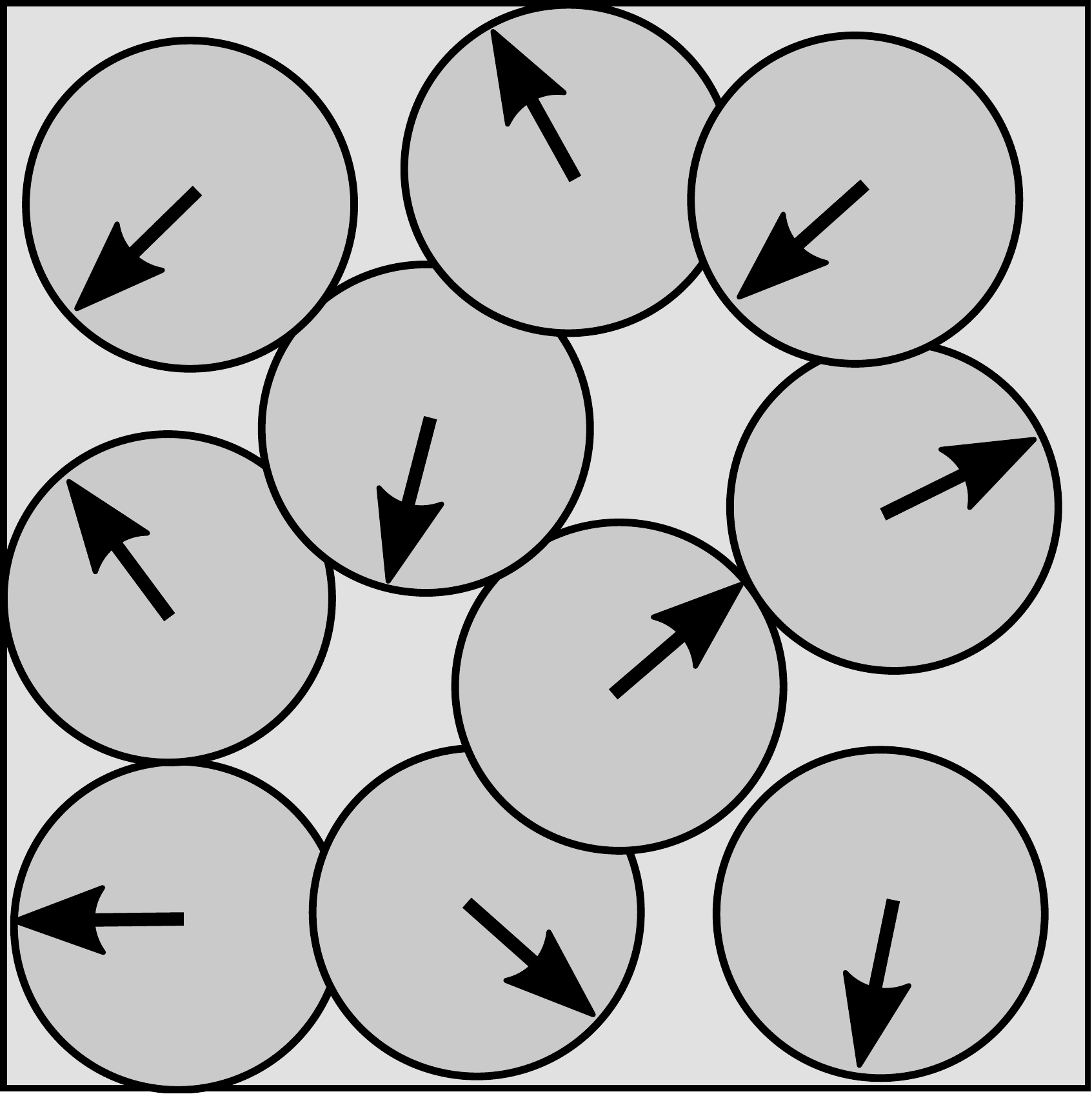}

      If $\bs{L}\ne0$ \;,\; $\mE(k) \propto k^2$ for $k\to 0$

      (Saffman spectrum).
    \end{center}

  \end{minipage}

\caption{\label{fig:saff_batch} 
Schematic representation of a superposition of independent eddies with and without linear impulse.
}
\end{figure}

The results derived so far can be transposed to homogeneous turbulence by considering a superposition of independent eddies instead of a single isolated one \cite{saffman67,davidson04,llor11}.
The large-scale properties of the velocity spectrum $\mE$ emerging from this homogeneous sea of eddies have been studied in \cite{saffman67,davidson04,llor11}. It has been shown in these references that the infrared scaling of $\mE$ depends on whether the eddies have a linear impulse $\bs{L}$ or not. More precisely, it has been shown that, for small wavenumbers $k$, 
if $\bs{L}=0$, $\mE(k) \propto k^4$ and if $\bs{L} \ne 0$, $\mE(k) \propto k^2$.
The $k^4$ and $k^2$ spectra are respectively called  Batchelor and  Saffman spectra. The difference between the two spectra is illustrated in Fig. \ref{fig:saff_batch}.

For variable-density eddies, we showed that the linear impulse $\bs{L}$ is time dependent. Therefore, for a homogeneous collection of independent variable-density eddies, the velocity spectrum $\mE$ is not permanent. It may even transition from a Batchelor $k^4$ spectrum to a Saffman $k^2$ spectrum and reciprocally. 
This stands in contradiction with the depiction of the permanence of large eddies exposed in the introduction.

But looking at the issue of the permanence of large eddies in terms of the velocity spectrum $\mE$ may not be the most pertinent choice to make.
Indeed, we have shown in Sec. \ref{sec:invariant} that a variable-density single eddy did possess an invariant,  $\bs{\Ls}$, called linear solenoidal impulse and associated with the solenoidal component of the momentum $\bs{\qs}$.
The idea is then to focus on the properties of the spectrum $\mQ$ of $\bs{\qs}$ instead of the spectrum $\mE$ of $\bs{u}$.
For the homogeneous sea of eddies considered here, we can  apply to $\mQ$ the same techniques as those used to study $\mE$ in Refs. \cite{saffman67,davidson04,llor11}.
We obtain that the infrared scaling of $\mQ$ depends on whether the eddies have a linear solenoidal impulse $\bs{\Ls}$ or not. For small wave numbers, we have:
$$
\text{if } \bs{\Ls}=0 \coma \mQ(k) \propto k^4
\andd
 \text{if } \bs{\Ls} \ne 0 \coma \mQ(k) \propto k^2
\period
$$
As opposed to $\bs{L}$, $\bs{\Ls}$ is an invariant quantity. As a result, a Batchelor or a Saffman $\mQ$ spectrum  remains so and a Saffman $\mQ$ spectrum is in addition permanent at small wave numbers. 
This description of the large scale behavior of $\mQ$ appears compatible with the formulation of the permanence of large eddies given in the introduction.
In the next section, we aim to analyze further the properties of $\mE$ and $\mQ$ at small wave numbers and verify if their properties agree with the study of a homogeneous sea of eddies.

%=======================================================================
\section{Large-scale properties of the solenoidal momentum spectrum} \label{sec:hom_turb}
%=======================================================================

The study of a single variable-density eddy has put forward several elements which are relevant to our understanding of variable-density homogeneous turbulence.
First,  density non-uniformities have an influence on the momentum and velocity far-fields which is mediated by the pressure field. Because of this influence, the momentum and velocity scalings are not permanent.
By contrast, the solenoidal component of the momentum has a permanent far-field scaling, associated with the large scale invariant $\bs{\Ls}$.
Second, when a homogeneous superposition of eddies is considered,  these results translate into distinct behaviors for the turbulent spectra: the spectrum $\mE$ of the velocity $\bs{u}$ is found to vary at large scales, while the spectrum $\mQ$ of the solenoidal momentum $\bs{\qs}$ is found to keep its initial infrared scaling at all times and to be permanent at large scales if $\bs{\Ls}\ne 0$.

Combined together, these elements suggest that the standard formulation of the permanence of large eddies given in the introduction applies to variable-density density turbulence for the spectrum $\mQ$ of $\bs{\qs}$ but not for the spectrum $\mE$ of $\bs{u}$.
The purpose of this section is to confirm these expectations.

%-------------------------------------------------
\subsection{Evolution of the fluctuating field}
%-------------------------------------------------
To begin with, we decompose the flow into a mean and a fluctuating part.
For any quantity $X$, we denote by $\rey{X}$  its ensemble mean and by $X'=X-\rey{X}$ its fluctuation. 
Averaging equations \eqref{eq:vd_rhou}, \eqref{eq:vd_rho} and \eqref{eq:vd_u} yields:
\begin{align}\label{eq:mean}
\partial_t \rey{\rho} = 0 \coma \partial_t \rey{\rho \bs{u}} = 0
\andd
\partial_t \rey{\bs{u}} = \rey{\bs{f}}
\period
\end{align}
The mean density and momentum are constant. Without loss of generality, we can choose:
$$
\rey{\rho \bs{u}} = 0
\period
$$
As for the mean velocity $\rey{\bs{u}}$, it is not constant but only depends on time because of the homogeneity of the flow.
Hence, its effect on the fluctuating field is only to advect it globally. This effect can be accounted for by using the following change of coordinates:
$$
\bs{x}^* = \bs{x} - \int_0^t \bs{\rey{u}}(s) \ud s
\period
$$
Note that $\bs{\rey{u}}=0$ if the flow is isotropic. However, the  assumption of isotropy is not required and will not be used in the forthcoming analysis.
Applying this change of coordinates and dropping the star exponential notation for simplicity, we can then write that the fluctuating velocity and density fields evolve according to:
\begin{subequations}  \label{eq:vd_fluc}
\begin{align} \label{eq:fluc_rho}
\partial_t \rho' + \partial_j\big(\rho' u'_j  \big) &= \partial_j \big(\nu_c \partial_j \rho'\big) 
\coma
\\
\label{eq:fluc_rhou}
\partial_t (\rho u_i)' + \partial_j \big( (\rho u_i)' u'_j\big) &= -\partial_j \big( p' \delta_{ij} + \Sigma'_{ij} \big)
\coma
\\
\label{eq:fluc_divu}
\partial_j u'_j &= 0
\period
\end{align}
\end{subequations}

%-------------------------------------------------
\subsection{Evolution of $\bs{\qs}$ in spectral space}
\label{sec:qs_spec}
%-------------------------------------------------
The next step consists in going to spectral space. 
For any quantity $X(\x,t)$, we denote by $\tf{X}(\K,t) ~~~=~~~$ $\frac{1}{(2\pi)^3} \int X(\x,t) e^{-\imath \K\cdot\x}\ud \x$ its Fourier transform at a given wave vector $\K$ and at time $t$. Applying the Fourier transform to Sys. \eqref{eq:vd_fluc} allows to derive the evolution equations of $\tf{\rho'}$ and $\tf{(\rho u_i)'}$.
However, in this section, we will only focus on the spectral properties of the solenoidal component of the momentum $\bs{\qs}$ and will only write the  evolution equation of its Fourier transform. 
This equation is derived directly from the evolution of $\tf{(\rho u_i)'}$ since both quantities are related by:
$$
\tf{\qs_i}(\K,t) = P_{ij}(\bnk) \tf{(\rho u_j)'}(\K,t)
\coma
$$
with $\proj_{ij}$ the projector on incompressible fields:
$$
\proj_{ij}(\bnk) = \delta_{ij} - {\nk}_i {\nk}_j
\with \bnk=\K / k
\period
$$
Thus, applying the Fourier transform to Eq. \eqref{eq:fluc_rhou} and multiplying the result by $P_{ij}$, we obtain that, for $\K \ne 0$:
\begin{align} \label{eq:tf_qs}
\partial_t \tf{\qs_i} = -\imath k \mathcal{P}_{ijk}(\bnk) \Big( \tf{(\rho u_j)' u'_k}
+ \tf{\Sigma'_{jk}}
\Big)
\coma
\end{align}
with the non-symmetric tensor $\mathcal{P}_{ijk}(\bnk)$ defined by:
$$
\mathcal{P}_{ijk}(\bnk) = \proj_{ij}(\bnk) \nk_k
\period
$$
In the right-hand side of Eq. \eqref{eq:tf_qs}, one recognizes the Fourier transforms of the different terms of Eq. \eqref{eq:fluc_rhou}. The first term on the right-hand side of Eq. \eqref{eq:tf_qs} corresponds to the non-linear product of the velocity and momentum appearing in Eq. \eqref{eq:fluc_rhou}.  
The second one  accounts for the viscous and diffusive effects carried by $\bs{\Sigma'}$ in Eq. \eqref{eq:fluc_rhou}.  
But the most noticeable feature of Eq. \eqref{eq:tf_qs} is the absence of a pressure term. The pressure gradient which appears in  Eq. \eqref{eq:fluc_rhou} disappears after selecting the solenoidal component of $(\rho u_i)'$.
In the single eddy configuration studied in Sec. \ref{sec:single_eddy}, we stressed that it was the pressure field which was responsible for the long range correlations which made impermanent the far-field scaling of the velocity field.
Its absence in Eq. \eqref{eq:tf_qs}  suggests that these long-range pressure-generated correlations will only indirectly affect the evolution of  $\tf{\bs{\qs}}$.

%-----------------------------------------------------
\subsection{Spectrum $\mQ$ of the solenoidal momentum $\bs{\qs}$}
%-----------------------------------------------------
The modulus spectrum of the solenoidal momentum $\bs{\qs}$ is defined by:
$$
\mQ(k,t) = k^2 \oint \sQ(\K,t) \ud \bnk
\coma
$$
where $\oint \cdot \ud\bnk$ is the integral over the surface of the sphere unity and where $\sQ$ is the spectral density of $\bs{\qs}$:
$$
\sQ(\K,t) \delta\left(\K-\K'\right)
= \frac{1}{2} \rey{\tf{\qs_i}(\K,t) {\tf{\qs_i}^{\cjgt}(\K',t)}} 
\period
$$
Using the identity $\tf{\qs_i}(\K,t) = \tf{\qs_i}^{(0)}(\K) + \int_0^t \partial_t \tf{\qs_i}(\K,s) \ud s$, with $\tf{\qs_i}^{(0)}$ the value of $\tf{\qs_i}$ at $t=0$, the evolution equation of $\sQ(\K,t)$ can be formally written as:
\begin{align} \label{eq:Q}
 \partial_t \sQ(\K,t) =  \mathcal{T}^{(0)}(\K,t) & + \int_0^t \mathcal{T}(\K,t,s)  \ud s\coma
\\
\nonumber
\with   
\mathcal{T}^{(0)}(\K,t)\delta\left(\K-\K'\right)
= \Re\left( \rey{\tf{\qs_i}^{(0)}\hspace{-0.5ex}(\K) \partial_t\tf{\qs_i}^\cjgt(\K',t)   } \right)
& \andd 
\mathcal{T}(\K,t,s) \delta\left(\K-\K'\right)
= \Re\left( \rey{\partial_t\tf{\qs_i}(\K,t)\partial_t\tf{\qs_i}^\cjgt(\K',s) } \right)
\period
\end{align}
The notation $\Re$ refers to the real part of a given quantity.
The second component of the transfer term can be expressed further by substituting $\partial_t\tf{\qs_i}$ with its value given by Eq. \eqref{eq:tf_qs}. One finds that $\mathcal{T}(\K,t,s) \delta\left(\K-\K'\right)$ is the real part of:
\begin{align} \label{eq:Tnl}
\rey{\partial_t\tf{\qs_i}(\K,t)\partial_t\tf{\qs_i}^\cjgt(\K',s)} 
=&
k^2 \; P_{ik}(\bnk) \nk_j \nk_l \; 
\rey{
\Big(\tf{(\rho u_i)' u'_j}(\K,t)+ \tf{\Sigma'_{ij}}(\K,t) \Big) 
\Big(\tf{(\rho u_k)' u'_l}^\cjgt(\K',s)+ \tf{\Sigma'_{kl}}^\cjgt(\K',s) \Big) 
} 
\period
\end{align}
Finally, the evolution of the spectrum $\mQ$ is given by:
\begin{align} \label{eq:mQ}
\partial_t \mQ(k,t) =   \mT(k,t) \coma
\with 
\mT(k,t)
= k^2 \oint \mathcal{T}^{(0)}(\K,t) \ud\bnk
+
k^2 \int_0^t \oint \mathcal{T}(\K,t,s)\ud \bnk \; \ud s
\period
\end{align}

%-------------------------------------------------------------
\subsection{Modeling non-linear interactions at large scales} \label{sec:model}
%-------------------------------------------------------------
Our focus is on the evolution of the spectrum $\mQ$ at small wave numbers.
 More precisely, we denote by $k_e(t)$ the peak wave number of $\mQ$ at time $t$. The expression ``small wave number'' or ``large scales'' will hereafter refer to scales satisfying the condition:
$$
\text{large scale range } \equiv \;\;\; k \ll k_e(t)
\period
$$
Note that, as time increases, the integral scale of turbulence increases, i.e. $k_e(t)$  decreases. Thus, a wave number belonging to the large scale range at time $t$ also belongs to the large scale range at time $s<t$:
$$
\for s<t \coma k \ll k_e(t) < k_e(s) 
\period
$$
The large-scale evolution of $\mQ$ is driven by the non-linear transfer term $\mT$, or equivalently by $\mathcal{T}^{(0)}$ and $\mathcal{T}$.
These quantities are not known in terms of second-order spectral correlations. Therefore, their properties must be modeled. The assumptions which we use to achieve this objective are detailed below.

To begin with, we assume that the initial value $\tf{\bs{\qs}}^{(0)}$ is uncorrelated with the time derivative of $\tf{\bs{\qs}}$ at time $t$. Thus, we set:
$$
\text{for } k \ll k_e(t) \coma \mathcal{T}^{(0)}(\K,t) = 0 
\period
$$
Next, we assume that viscous and diffusive effects are negligible at large scales.
Hence, in the expression of $\mathcal{T}$, we  neglect all contributions coming from the viscous/diffusive tensor $\bs{\tf{\Sigma}}$.
As a result, using  \eqref{eq:Tnl}, we can write that:
$$
\text{for } k \ll k_e(t) \coma 
\mathcal{T}(\K,t,s) \delta\left(\K-\K'\right)
= 
k^2 \; P_{ik}(\bnk) \nk_j \nk_l \; 
 \Re\Big(
\rey{
\tf{(\rho u_i)' u'_j}(\K,t)\tf{(\rho u_k)' u'_l}^\cjgt(\K',s) 
}
\Big) 
\period
$$
Thus, the transfer term $\mathcal{T}$ depends on a fourth-order two-time correlation involving convolution products between the fluctuations of $\bs{u}$ and $\rho \bs{u}$.
In order to model this correlation, we assume that the spectra and co-spectra of the fluctuating velocity and momentum $\bs{u'}$ and $(\rho \bs{u})'$ peak at a  wave number close to $k_e(t)$.
Then, we assume that the largest contribution of a correlation involving  $\bs{u'}$ and $(\rho \bs{u})'$ comes from a range of wave-numbers close to or larger than $k_e(t)$, while smaller wave-numbers only provide a marginal contribution. 
This energy  containing range is denoted by:
$$
\text{Energy containing range} \equiv k \gtrsim k_e(t)
\period
$$
Note that this assumption cannot be verified if the infrared exponent $\se$ of the turbulent spectra is equal or smaller than $2$. This limit value corresponds to a spectrum being constant at large scales and is thus not compatible with the existence of a range containing most of the energy. The subsequent analysis is thus limited to $\se>2$.
Applying these assumptions to the fourth-order correlation appearing in the definition of $\mathcal{T}$, we obtain that:
\begin{align} \label{eq:model_Tnl}
\text{for } k \ll k_e(t) \coma 
\rey{\tf{(\rho u_i)' u'_j}(\K,t)\tf{(\rho u_k)' u'_l}^\cjgt(\K',s) }
 &=
 \iint S_{ijkl}(\p, \K\!-\!\p,\q,\K'-\q;t,s) \ud \p \ud \q \; \delta(\K-\K')
\\ \nonumber
& \approx 
\iint_{p \gtrsim k_e(t) , q \gtrsim k_e(s)} \hspace{-12ex} S_{ijkl}(\p, \K\!-\!\p,\q,\K'-\q;t,s) \ud \p \ud \q \; \delta(\K-\K')
\\ \nonumber
& \approx 
\iint_{p \gtrsim k_e(t) , q \gtrsim k_e(s)} \hspace{-12ex} S_{ijkl}(\p, \!-\!\p,\q,-\q;t,s) \ud \p \ud \q \; \delta(\K-\K')
\coma
\end{align}
with $S_{ijkl}(\bs{a}, \bs{b},\bs{c},\bs{d}; t,s) \delta(\bs{a}+\bs{b} \!-\!\bs{c}\!-\!\bs{d})\!=\!\rey{\tf{(\rho u_i)'}(\bs{a},t)\tf{u'_j}(\bs{b},t)\tf{(\rho u'_k)}^\cjgt(\bs{c},s)\tf{u'_l}^\cjgt(\bs{d},s)}$.
The first equality is the definition of the convolution product. The first approximation is a direct expression of our main assumption and the second one is a Taylor expansion in the limit $k\ll k_e$.
This overall procedure is nothing more than the distant interaction hypothesis usually used to simplify spectral models like the eddy-damped quasi-normal model (EDQNM) \cite{lesieur08,sagaut06}.

~\\ \indent
The end result here is that  $\rey{\tf{(\rho u_i)' u'_j}(\K,t)\tf{(\rho u_k)' u'_l}^\cjgt(\K',s) }$ only depends on $t$ and $s$ but not on the wave vector $\K$.
Thus, the non-linear transfer term $\mathcal{T}$ can be simplified into the following expression:
\begin{align} \label{eq:nonlin}
\text{for } k\ll k_e(t) \coma 
\mathcal{T}(\K,t,s) = k^2 \; \mathcal{T}^\texttt{mod}(\bnk,t,s)
\coma
\end{align}
\begin{align*}
\with &
\mathcal{T}^\texttt{mod}(\bnk,t,s) \delta\left(\K-\K'\right)
 =P_{ik}(\bnk) \nk_j \nk_l \; 
 \Re\Big(
\left.
\rey{
\tf{(\rho u_i)' u'_j}(\K,t)\tf{(\rho u_k)' u'_l}^\cjgt(\K',s) 
}
\right|_{p,q\gtrsim k_e}
\Big) 
\period
\end{align*}
The notation $\left. \rey{\;\cdot\;}\right|_{p,q\gtrsim k_e}$ refers to the restriction of the fourth-order correlations to the energetic range detailed in Eq. \eqref{eq:model_Tnl}. As explained above, this restriction is independent from the wave number $\K$, which explains why  $\mathcal{T}^\texttt{mod}$ only depends on $\bnk$ and time.

Combining our different assumptions into Eq. \eqref{eq:mQ}, we eventually obtain the following modeled evolution for $\mQ$ at large scales:
\begin{align} \label{eq:dQdt}
\text{for } k\ll k_e(t) \coma 
\partial_t \mQ(k,t) =  k^4 \; \mT^\texttt{mod}(t) 
\with 
{\mT}^\texttt{mod}(t) =  \int_0^t \oint \mathcal{T}^\texttt{mod}(\bnk,t,s)\ud \bnk\; \ud s 
\period
\end{align}
Thus, following our assumptions,  non-linear interactions display a $k^4$ scaling at small wave numbers.
This classical scaling is the one predicted by several models \cite{proudman54,lesieur08,llor11} for the non-linear transfer term  of the velocity spectrum  in constant density incompressible turbulence.
The difference here is that this scaling is established for the transfer term of the solenoidal momentum spectrum $\mQ$ and not for the velocity spectrum $\mE$.
A complete analysis of the velocity transfer term is proposed in App. \ref{app:vel_spec} and significant differences between the constant and variable-density cases are exhibited.

%-----------------------------------------------------
\subsection{Invariance of $\mQ$ at small wave numbers} \label{sec:perm_mQ}
%-----------------------------------------------------
Starting from the modeled equation \eqref{eq:dQdt}, we are now ready to discuss the permanence of large-eddies in variable-density homogeneous turbulence.
At initial time, we suppose that the spectrum $\mQ(\K)$  obeys a power law:
$$
\text{for } k\ll k_e(t=0) \coma\mQ(k,t=0) = C_q k^{\sq}
\coma
$$
with $C_q$ a constant and $\sq$ the initial infrared exponent of the spectrum $\mQ$.
Integrating Eq. \eqref{eq:dQdt} yields:
\begin{align} \label{eq:mQ_ls}
\text{for } k\ll k_e(t) \coma
\mQ(k,t) = C_q k^{\sq}+ k^4 \int_0^t{\mT}^\texttt{mod}(s) \ud s 
\period
\end{align}
Comparing the infrared exponents of the different terms of Eq. \eqref{eq:mQ_ls}, we see that, in the limit $k\to 0$, initial conditions become predominant over non-linear terms only if $\sq<4$. For $\sq>4$,  the non-linear component  becomes prevalent and for $\sq=4$ both terms contribute equally to $\mQ$.
Therefore,  in the limit $k\to0$, we can conclude the following:
\begin{itemize}
\item if $\sq < 4$, $\mQ$ is invariant at small wavenumbers,
\item if $\sq > 4$, $\mQ$ is not invariant: it transitions to a spectrum with an infrared exponent  $s=4$, 
\item if $\sq=4$, the infrared slope remains unchanged and equal to $4$ but large scales do not necessarily remain constant.
\end{itemize}
In constant density flows, the latter type of mixed behavior is observed not only for $\se=4$ but also for an interval of values of $\se$ close to $4$. As a result,  the permanence of large eddies is observed to occur strictly for $\se < 3.5$, instead of $4$ \cite{lesieur00,meldi12,eyink00,comte66}. 
For variable-density flows, such a gray area may also be expected and the permanence of large eddies might only be strictly verified for $\sq \le 4-\eta$, with $\eta$ a parameter on the order of $0.5$.
Still, the current theory does not predict this gray interval and only simulations may answer the question of its existence and extent.

To sum up, the permanence of large eddies in variable-density turbulence follows a description similar to the one exposed in the introduction for constant density turbulence, provided one looks at the spectrum $\mQ$ of the solenoidal momentum.
This conclusion is the main result of this work. 
It also agrees with what could be deduced from the consideration of homogeneous collections of independent eddies, as described in Sec. \ref{sec:hom_sea}.

%=======================================================================
\section{Consequences of the permanence of $\mQ$} \label{sec:consequences}
%=======================================================================
In Sec. \ref{sec:hom_turb}, we showed that the usual formulation of the principle of permanence of large eddies applies to the spectrum $\mQ$ of the solenoidal component of the momentum.
Several consequences of this prediction are explored in this section.
First, we assess how this prediction fares with the results established in the context of Boussinesq and fully compressible turbulence. More precisely, we discuss how it impacts their validity.
Second, we examine how the permanence of $\mQ$ affects the self-similar decay of the flow.
Finally, the large-scale properties of the density spectrum are compared with those of $\mQ$.

%--------------------------------------
\subsection{Comparison with the Boussinesq and fully compressible cases}
\label{sec:bouss_comp}
%--------------------------------------
In the introduction, we highlighted that large scales analyses of variable-density turbulence have already been performed in the context of Boussinesq and fully compressible flows.
We explain here why these analyses cannot be extrapolated to the context studied here, i.e for small Mach number high density contrast flows.

Let us recall that, for Boussinesq flows, results similar to the constant density case have been obtained whereby the velocity spectrum $\mE$ is invariant (or evolves linearly) if its initial infrared exponent is smaller than $4$ \cite{batchelor92,soulard14,soulard15,soulard18}.
As for fully compressible flows, Sitnikov \cite{sitnikov58} argued that the integral
$$
\Lambda^\texttt{tot.} = \int \rey{ \rho u_i(\x) \;\, \rho u_i(\x+\bs{r}) } \ud \bs{r}
$$
is invariant.
In spectral terms, this means that the spectrum $\mQt$ of the total momentum (solenoidal + irrotational) is invariant at small wavenumbers when its infrared exponent is equal to $2$.

Neither of these results applies to small Mach number high-density contrast flows.
Indeed, while we do not know the precise evolution of $\mE$ or $\mQt$, we still know that density fluctuations decrease with time. From Eq. \eqref{eq:fluc_rho}, we can indeed deduce that:
$$
\partial_t \rey{\rho'^2} = - 2 \rey{\nu_c \partial_j \rho' \partial_j\rho'}
\period
$$
Therefore, at late times, $\rey{\rho'^2} \to 0$ and the difference between $\mE$, $\mQt$ and $\mQ$ vanishes:
\begin{align} \label{eq:late_mE}
\for t \to \infty \coma \mE \to \mQ \andd \mQt \to \mQ
\period
\end{align}
We just showed that $\mQ$ is invariant at large scales for $\sq <4$ and goes to a $k^4$ spectrum otherwise.
Therefore, the late time properties of $\mE$ and $\mQt$ are entirely set by the initial value of $\mQ$ and its initial infrared exponent.
This conclusion mirrors the one obtained for a single eddy in Sec. \ref{sec:invariant}.

Now, $\bs{\qi}$ and $\bs{\qs}$ are two independent variable and their spectra can be  initialized independently.
As a result, the initial scalings of $\mQt$ and $\mQ$ are also independent from one another. Hence, $\mQt$ may initially display a $k^2$ spectrum and $\mQ$ a $k^{\sq}$ spectrum, with $\sq \ne 2$. In that case, $\mQt$ will evolve towards a $k^{\sq}$ or a $k^4$ spectrum at late times depending on whether $\sq<4$ or not.
This shows that the total momentum spectrum $\mQt$ is not necessarily constant at large scales, even for an initial infrared exponent of $2$. In agreement, Sitnikov's momentum integral is not an invariant of variable-density homogeneous turbulence with small Mach numbers.

As for the velocity spectrum $\mE$, it is related to $\mQ$ in a way detailed in App. \ref{app:evol_mE}.
Similarly to $\mQt$, the initial scaling of $\mE$ and $\mQ$ can differ.
As a result, $\mE$ is not permanent even if initially it scales with an infrared exponent smaller than $4$.

For both $\mE$ and $\mQt$, the impermanence of the large-scale spectrum can be traced back to pressure correlations. The latter induce a non-linear transfer term scaling as $k^2$. This aspect is detailed in App. \ref{app:vel_spec} for the velocity spectrum. Note that a non-linear term scaling as $k^2$ corresponds to the limit put forward by Monin \& Yaglom to guarantee the convergence of large-scale integrals. Their convergence condition is consequently not respected in the flows considered in this work.

As a last remark, we would like to stress that if $\Lambda^\texttt{tot.}$ is generally not an invariant, the following integral is:
$$
\Lambda^\texttt{so} = \int \rey{ \qs_i(\x) \, \qs_i(\x+\bs{r}) } \ud \bs{r}
\period
$$ 
This integral is the extension of Saffman's integral to variable-density turbulence. 
An equivalent of Loitsyanskii's integral can also be found. It can be expressed as:
$$
I^\texttt{so} = - \int r^2 \rey{ \qs_i(\x) \, \qs_i(\x+\bs{r}) } \ud \bs{r}
\period
$$
The convergence and invariance of the latter integral is subjected to the same kind of discussion as its constant density counterpart \cite{llor11,davidson04}. Besides, it is also related to the conservation of the angular solenoidal momentum on a sphere of volume $V$, $\int_V \bs{r} \wedge \bs{\qs} \ud \bs{r}$, when $V \to \infty$ \cite{llor11,davidson04}. The latter integral is also equal to $\int_V \bs{r} \wedge (\rho \bs{u}) \ud \bs{r}$ because the irrotational contribution of the momentum disappears by symmetry when integrated on the surface of a sphere of finite radius. 
This integral has been proposed as an invariant by Lumley \cite{lumley66}, even though non-linear terms can actually modify its value, as is the case in constant density turbulence \cite{llor11,davidson04}.

%------------------------------------------
\subsection{Self-similar decay of the flow} \label{sec:self_sim}
%------------------------------------------
Let us define by  $\kes$ the energy carried by the solenoidal component of the momentum:
$$
\kes(t) = \frac{\rey{{\qs_i}'{\qs_i}'}(t)}{2 \rey{\rho}^2} = \frac{1}{\rey{\rho}^2}\int_0^\infty \mQ(k,t) \ud k
\period
$$
When the flow reaches a self-similar state, $\kes$ obeys a power law:
$$
\kes(t) \propto t^{-\nn}
\period
$$
The value of $\nn$ can be determined provided an additional assumption is verified: the spectrum $\mQ$ must have a self-similar shape for both large and energetic scales, i.e. for $k \lesssim k_e(t)$.
With this hypothesis, we can indeed write that:
$$
\kes(t) \propto  \frac{1}{\rey{\rho}^2}\int_0^{\epsilon k_e(t)} \mQ(k,t) \ud k
\coma
$$
with $\epsilon$ a small parameter.
Now,  when $\sq < 4$, we showed that $\mQ$ is invariant at small wavenumbers. Injecting this result in the previous expression leads to:
$$
\text{for } \sq < 4 \coma 
\kes(t) \propto  k_e^{\sq +1}(t)
\period
$$
Finally, because  the flow is self-similar, we can deduce from dimensional analysis that:
$$
k_e(t) \propto (t \sqrt{\kes(t)})^{-1}
\period
$$
The following expression of $\nn$ ensues:
\begin{align} \label{eq:nk}
\text{for } \sq < 4 \coma 
\nn= \frac{2 (\sq+1)}{\sq + 3}
\period
\end{align}
This formula is identical to the one derived in the constant density case, save for one very important point: it is the infrared exponent $\sq$ of the solenoidal momentum spectrum $\mQ$ which appears in it and not the exponent $\se$ of the velocity spectrum.

As in the constant density case, one  expects that Eq. \eqref{eq:nk} will be accurate for $\sq \le 4- \eta$, with $\eta \approx 0.5$ delimiting the interval discussed in Sec. \ref{sec:perm_mQ}.
For $ 4-\eta < \sq \le 4 $, corrections similar to the ones introduced in the constant density context \cite{lesieur00,meldi12,comte66,eyink00} will then be required. 
One of the simplest corrections consists in modifying formula  \eqref{eq:nk} as follows \cite{eyink00,lesieur00}:
\begin{align} \label{eq:nk_modif}
\nn= \frac{2 (\min(\sq,4 -\eta) + 1)}{\min(\sq,4 - \eta) + 3}
\period
\end{align}
For $\sq<4-\eta$, this formula is identical to  Eq. \eqref{eq:nk}, while for $\sq\ge 4-\eta$, the exponent $\nn$ is found to saturate at a smaller value. With the prescription $\eta=0.55$ one finds a maximum value of $\nn=1.38$ instead of $1.41$ with Eq. \eqref{eq:nk}.
Note that for $\sq>4$, the infrared exponent evolves towards $4$ so that the preceding formula still applies.

As a last remark, we stress that when the flow is self-similar, the decay exponent of $\kes$ is the same as the exponent of the kinetic energy $\kr$:
$$
\kr(t) \propto \kes(t) \propto t^{-\nn}
\period
$$

%-----------------------------------
\subsection{About the density field}
%-----------------------------------
So far, we voluntarily left the density field out of the discussion. The aim was to stress the importance of the solenoidal momentum $\bs{\qs}$ and its spectrum $\mQ$. 
Now that we have clarified the behavior of these quantities, we would like to make a brief comment on $\rho'$ and its spectrum $\mR$.

The same procedure as the one used to study $\bs{\qs}$ can be used with $\rho'$. Applying the Fourier transform to Eq. \eqref{eq:fluc_rho}, we obtain the evolution equation of $\tf{\rho'}$. From there. we deduce the governing equation of the density spectrum $\mR$. To model this equation at large scales, we neglect viscous terms as well as the correlation between the initial value of $\tf{\rho'}$ and its time derivative.
Finally, we make a distant approximation hypothesis such as the one detailed in Sec. \ref{sec:model} to close the fourth-order correlations involving $\rho'$ and $\bs{u}'$.
As a result, we obtain the following modeled evolution for $\mR$:
$$
\partial_t \mR = k^4 \mathcal{T}^\texttt{mod}_{\rho\rho} (t)
\period
$$
Thus, initial density spectra with an infrared exponent $s_\rho < 4$ are  invariant in the limit $k\to0$ while spectra with $s_\rho >4$ are not. They evolve towards a spectrum with an infrared exponent  of $4$. 
In this sense, the large-scale properties of $\mR$ obey a classical description akin to those of $\mQ$.

Note that this description is consistent with the prediction of Chandrasekhar \cite{chandrasekhar51}, Krzywoblocki \cite{krzywoblocki52} and Sitnikov \cite{sitnikov58} that the integral 
$$
\Lambda_{\rho\rho} = \int \rey{\rho(\x) \, \rho(\x+\bs{r})} \ud \bs{r}
$$
is an invariant of variable-density homogeneous turbulence.
Note also that this invariant is the counterpart of the single eddy integral defined for the density in Eq. \eqref{eq:other_inv}.

%-----------------------------------
\subsection{About the anisotropy of large scales}
%-----------------------------------
The analysis of Sec. \ref{sec:hom_turb} does not assume the isotropy of the turbulent field. In particular, the arguments used to study the permanence of the spectrum $\mQ$ can be shown to be valid for each component of the spectral correlation $\rey{\tf{\qs_i}(\K,t) {\tf{\qs_j}^{\cjgt}(\K',t)}}$.
As a result, the anisotropy of this spectral correlation is expected to be preserved at small wavenumbers whenever $s_q<4$. In turn, this will lead to a partial return to isotropy of one-point velocity statistics, in a way similar to the one described in Ref. \cite{mons14b} for constant density flows.
Note that these aspects will not be dealt with in the validation process. Only isotropic flows will be considered in the next section.

\newpage
%=======================================================================
\section{Validation} \label{sec:validation}
%=======================================================================

%-----------------------------------------------------------------------
\subsection{Description of the simulations}
%-----------------------------------------------------------------------
In order to validate the results derived in the previous sections, we perform several implicit large eddy simulations (ILES) of homogeneous isotropic turbulence (HIT). 
The reason for performing  ILES and not DNS is that we are only interested in large scales and do not need to capture the behavior of small scales in detail. Compared to a DNS,  ILES allows to extend the simulated range of large scales  and thus to improve their observation. 

The simulations are performed with the code \textsc{Triclade}, a massively parallel code intended to solve turbulent mixing of perfect gases in a variable-density context \cite{shanmuganathan13}.  A shock capturing scheme provides just enough numerical viscosity and diffusivity to ensure stability. More precisely, for this work, the monotonic upstream centered scheme for conservation laws (MUSCL) finite-volume Godunov method referred to as M$5$ in \cite{kim05} is used. 
It is accurate to fifth order in space and is combined with a low-storage strong stability preserving Runge-Kutta scheme of third-order time accuracy. A one dimensional monotonicity preserving (MP) limiter is used in reconstructing the primitive variables. 
A  Harten-Lax-van Leer-Contact (HLLC) numerical flux is used at each cell interface. To obtain good performances in the low Mach number limit, the low Mach number correction of Thornber et al. \cite{thornber08} is used.

~\\ \indent
All simulations are set in  a box of size $2\pi$, discretized by $1024^3$ cells.
The isotropic initial state is prescribed in two different ways corresponding to the two series of simulations described below.

\subsubsection{First series of simulations}
%.........................................
For these simulations, we initialize the values of the solenoidal momentum $\bs{\qs}$ and of the density field $\rho$.
 Their mean values is set to:
$
\rey{\bs{\qs}} = 0 \andd
\rey{\rho} = 1
\period
$
Then, in Fourier space, the phases of the fluctuating part of these quantities are randomly set while their modulus are chosen so that their  spectra verify:
\begin{align}\label{eq:mQ0}
\mQ = \frac{\rey{\qs_i\qs_i}|^{(0)}}{2} \frac{2^{\frac{\sq+3}{2}}}{k_0 \Gamma(\frac{\sq+1}{2})}\Big( \frac{k}{k_0}\Big)^{\sq} e^{-2 (k/k_0)^2}
\andd
\mR =  \frac{\rey{\rho'^2}|^{(0)}}{2} \frac{2^{\frac{\sq+3}{2}}}{k_0 \Gamma(\frac{\sq+1}{2})}\Big( \frac{k}{k_0}\Big)^{\sq} e^{-2 (k/k_0)^2}
\coma
\end{align}
with $\Gamma$ the Gamma function.
Both fluctuating fields are chosen uncorrelated.
Finally, the irrotational component of the momentum is set according to its definition
$
\qi_i = -\partial_i \phi
$
where $\phi$ is solution to the following equation:
\begin{align}\label{eq:poisson_phi}
\partial_j ( \tau \partial_j \phi) = \qs_j \partial_j \tau
\period
\end{align}
This equation stems from the incompressibility constraint of the fluctuating velocity field and is solved using a modified Poisson solver.
Knowing $\bs{\qi}$ and $\bs{\qs}$, we can initialize $\rho \bs{u}= \bs{\qi} + \bs{\qs}$, which is the actual quantity computed by the code \textsc{Triclade}.

The different parameters appearing in these formulas are chosen in order to verify at initial time:
$$
M_t = \frac{\sqrt{\rey{u'_iu'_i}}}{a_\text{sound}}=0.2
\coma
k_\text{peak} = \frac{\sqrt{\sq}}{2} k_0 = 25
\coma
\frac{\sqrt{\rey{\rho'^2}}}{\rey{\rho}} = 0.4
\coma
\sq \in \{1,2,3,4, 10 \}
\period
$$

\subsubsection{Second series of simulations}
%............................................
In addition to this first series of simulations, we also perform two more simulations for which the velocity spectrum is initialized instead of $\mQ$. Hence, for these last two simulations, we initialize $\bs{u}$ and $\rho$ so that:
\begin{align}\label{eq:mE0}
\mE = \frac{\rey{u'_iu'_i}|^{(0)}}{2} \frac{2^{\frac{\se+3}{2}}}{k_0 \Gamma(\frac{\se+1}{2})}\Big( \frac{k}{k_0}\Big)^{\se} e^{-2 (k/k_0)^2}
\andd
\mR =  \frac{\rey{\rho'^2}|^{(0)}}{2} \frac{2^{\frac{\se+3}{2}}}{k_0 \Gamma(\frac{\se+1}{2})}\Big( \frac{k}{k_0}\Big)^{\se} e^{-2 (k/k_0)^2}
\period
\end{align}
The velocity and density are chosen uncorrelated.  We also set $\rh = 1$ and $\rey{\rho \bs{u}}=0$, as in the first series of simulations.
However, instead of varying the infrared exponent $\se$ as in the previous simulations, we set $\se=6$ and vary the value of the density contrast. More precisely, the two simulations are defined by the following parameters:
$$
M_t = \frac{\sqrt{\rey{u'_iu'_i}}}{a_\text{sound}}=0.2
\coma
k_\text{peak} = \frac{\sqrt{\sq}}{2} k_0 = 25
\coma
\se = 6 
\coma
\frac{\sqrt{\rey{\rho'^2}}}{\rey{\rho}} \in \{0.02 , 0.75 \}
\period
$$

%%%%%% Qso %%%%%%%%%
\begin{figure}[!htb] 
\begin{center}
\hfill
\subfloat[~$\sq=1$]{\label{fig:s1_Qso}
\includegraphics[width=0.44\linewidth]{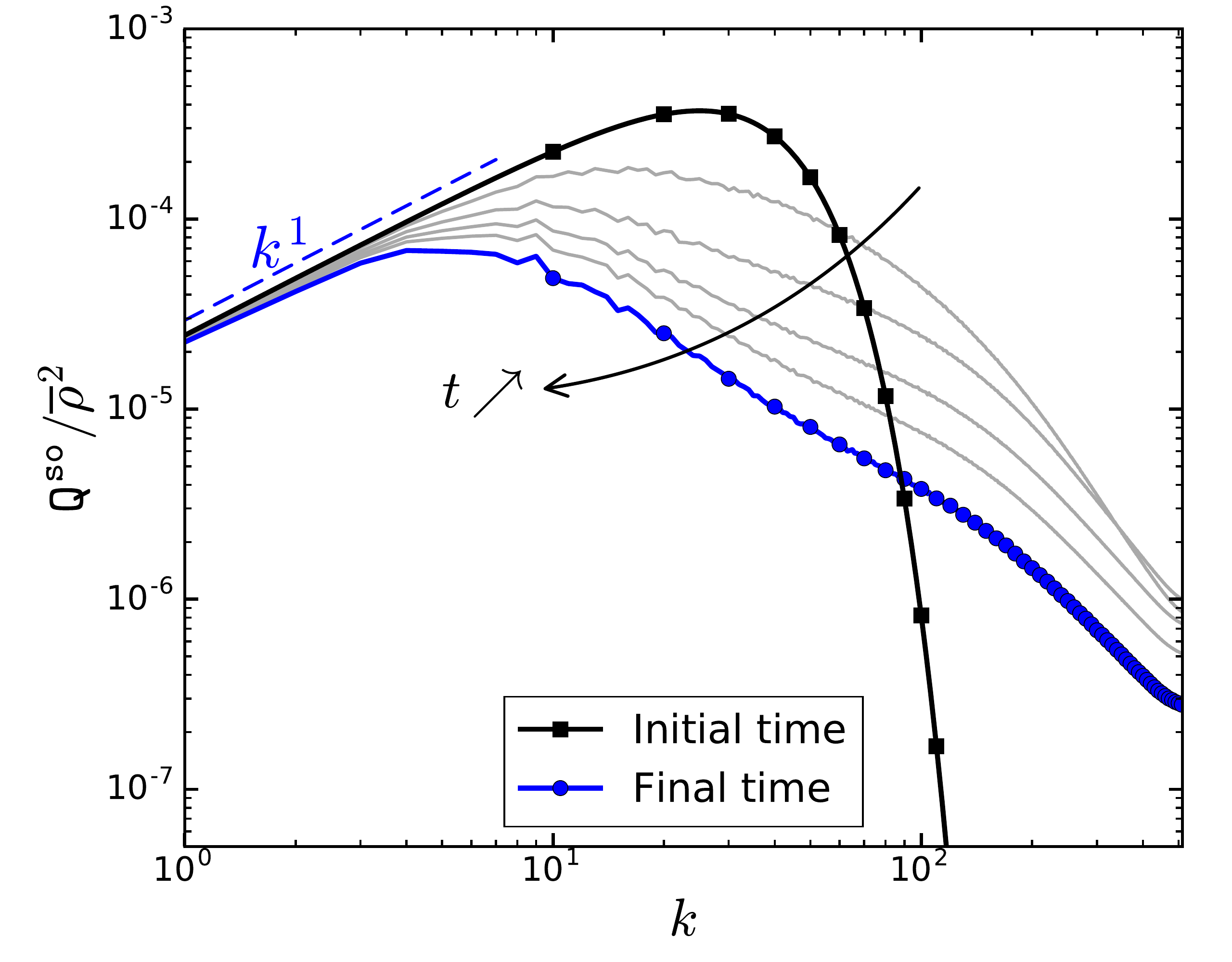} 
}
\hfill
\subfloat[~$\sq=2$]{\label{fig:s2_Qso}
\includegraphics[width=0.44\linewidth]{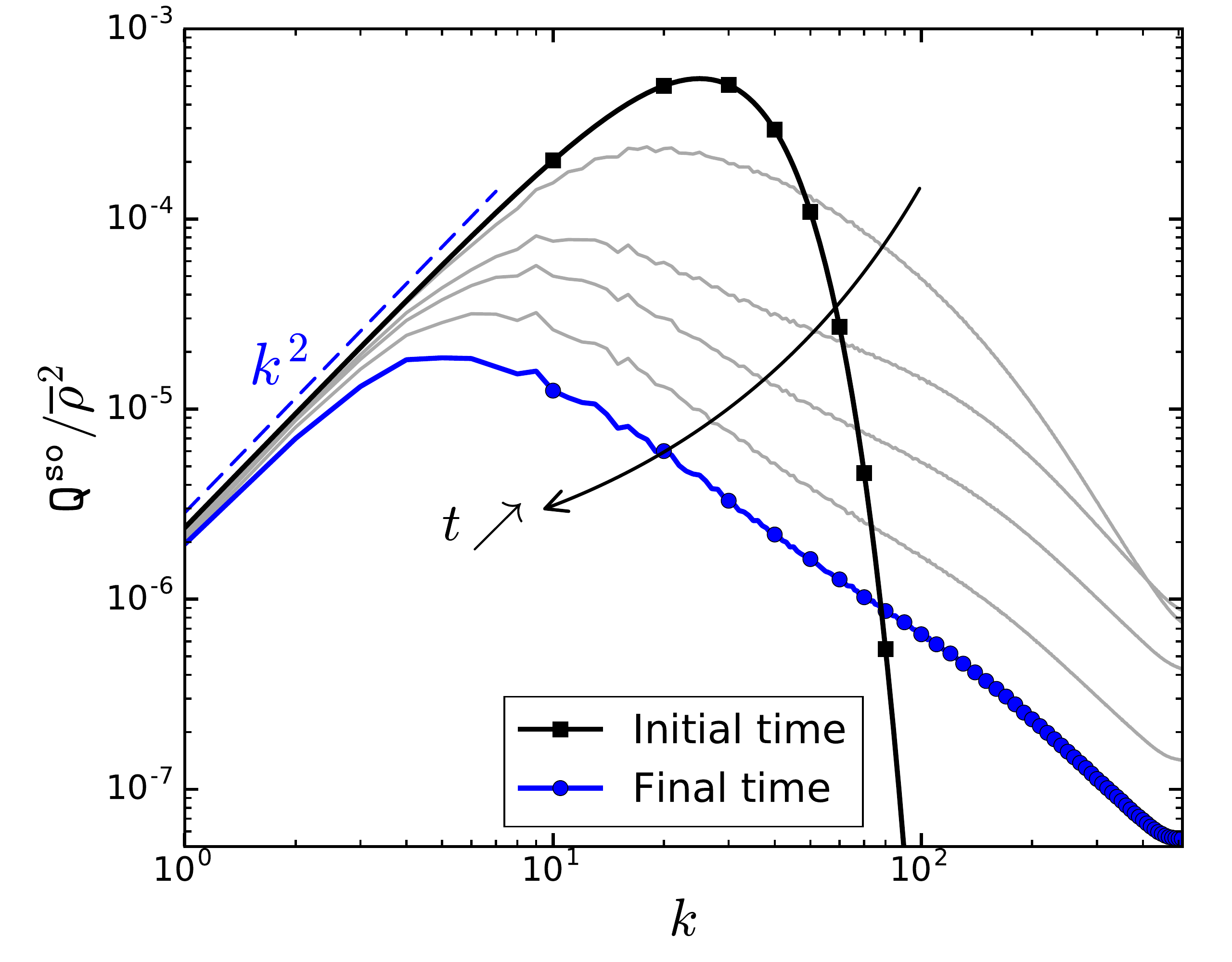}
}
\hfill~
\\
\hfill
\subfloat[~$\sq=3$]{\label{fig:s3_Qso}
\includegraphics[width=0.44\linewidth]{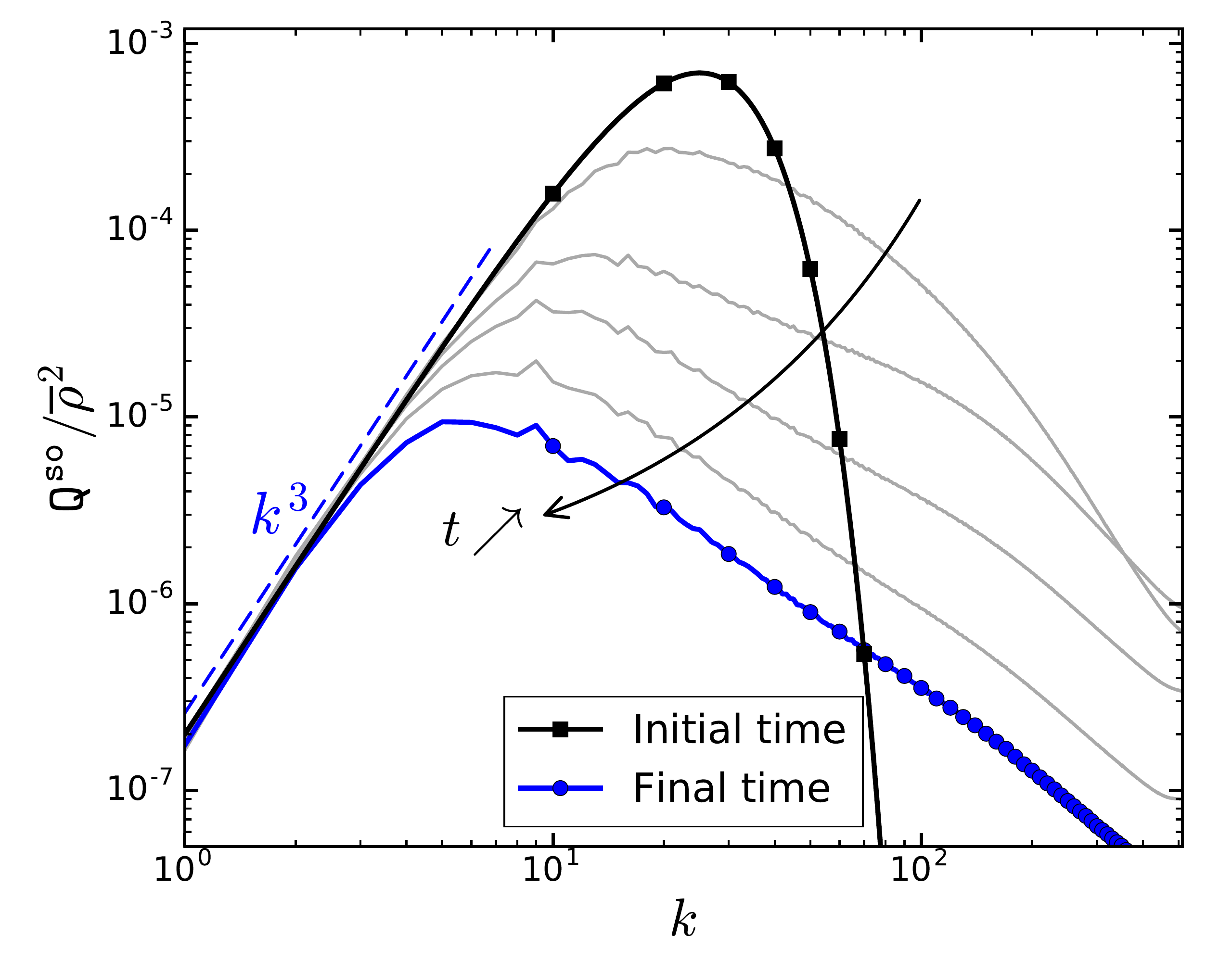} 
}
\hfill
\subfloat[~$\sq=4$]{\label{fig:s4_Qso}
\includegraphics[width=0.44\linewidth]{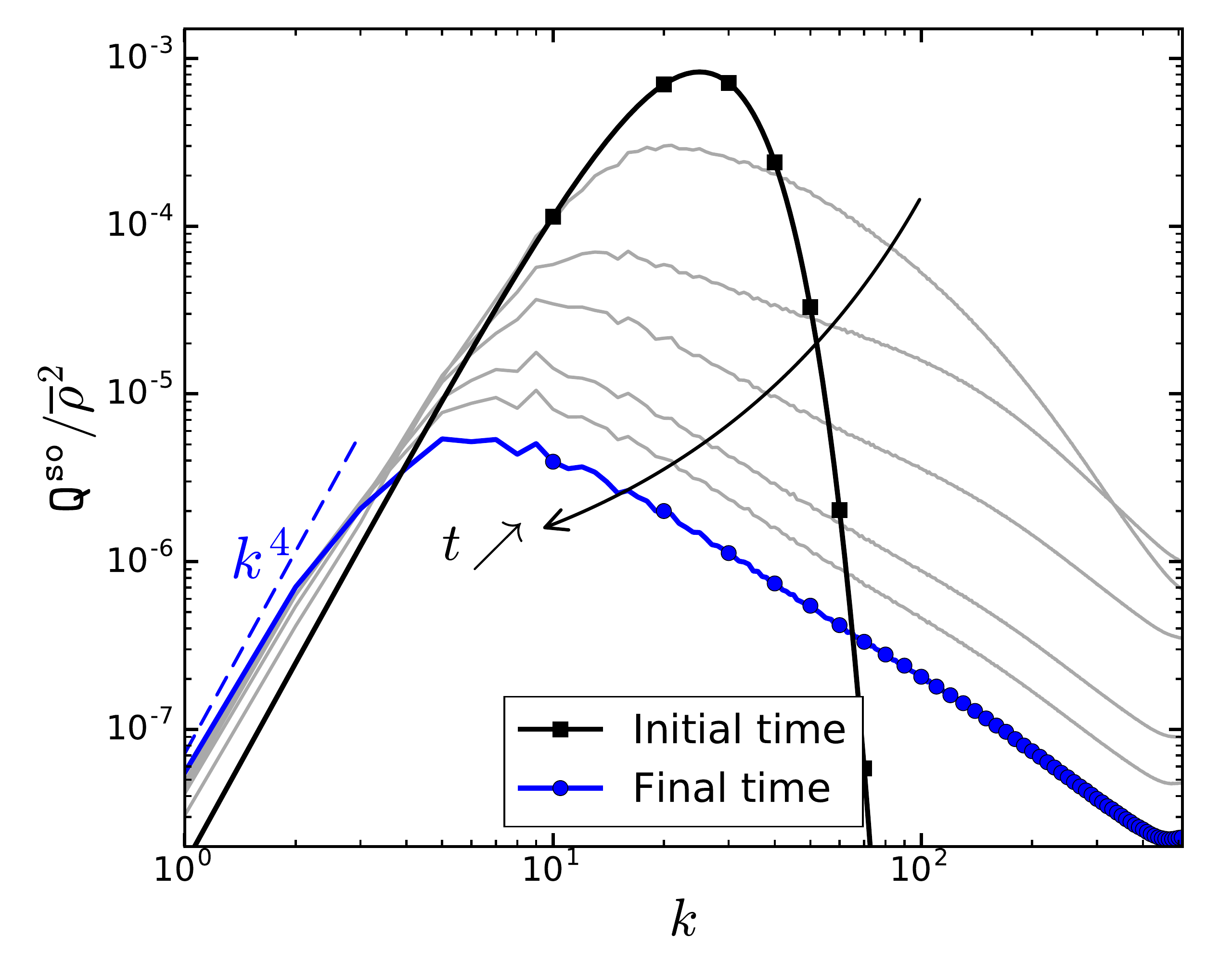}
}
\hfill~
\\
\subfloat[~$\sq=10$]{\label{fig:s10_Qso}
\includegraphics[width=0.44\linewidth]{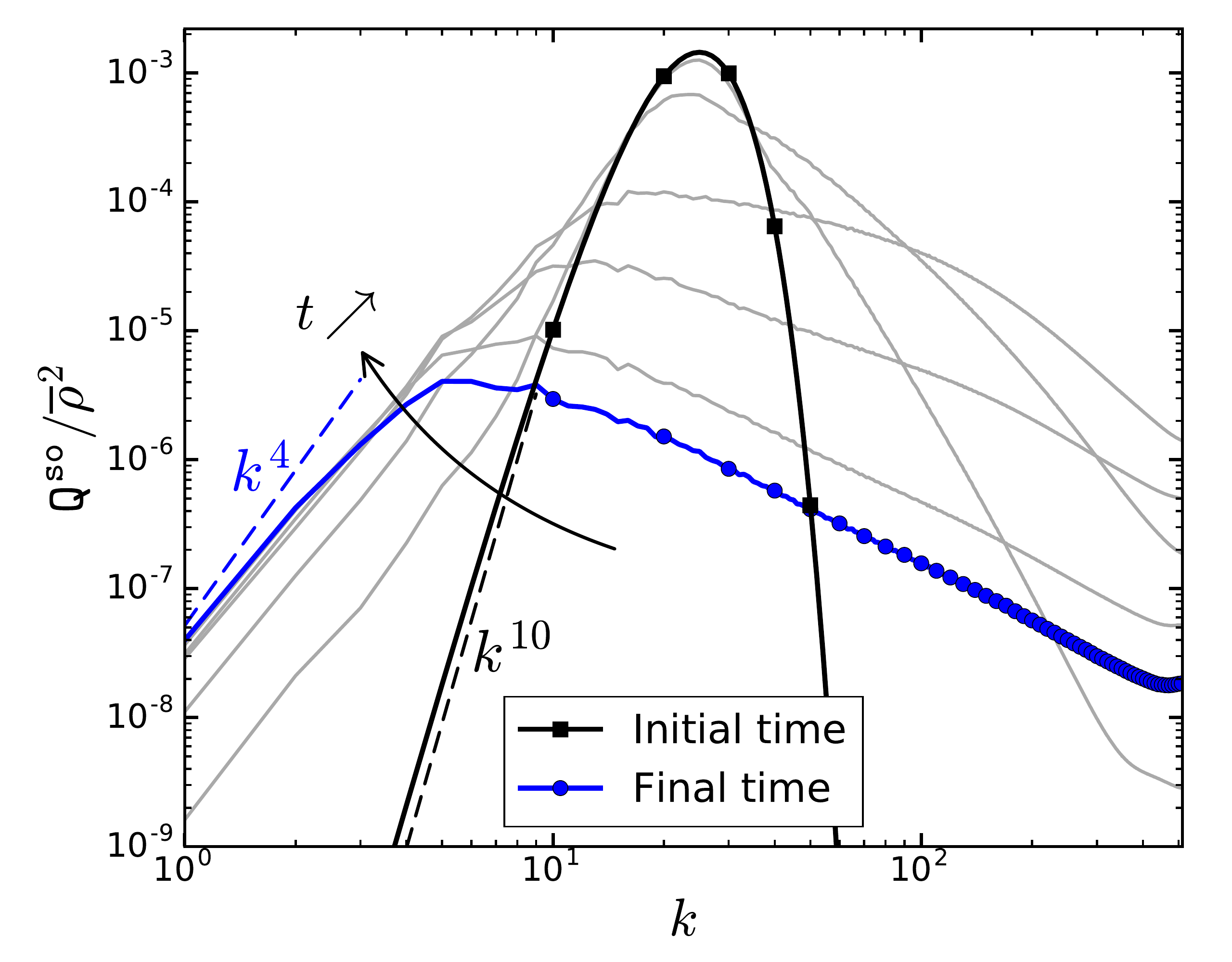} 
}
\end{center}
\caption{\label{fig:Qso} 
Evolution of the spectrum $\mQ$ of the solenoidal momentum with ${\mQ}^{(0)}$ given by Eq. \eqref{eq:mQ0} and for different values of $\sq$.
}
\end{figure}

%%%%%% Eso %%%%%%%%%
\begin{figure}[!htb] 
\begin{center}
\hfill
\subfloat[~$\sq=1$]{\label{fig:s1_Eso}
\includegraphics[width=0.44\linewidth]{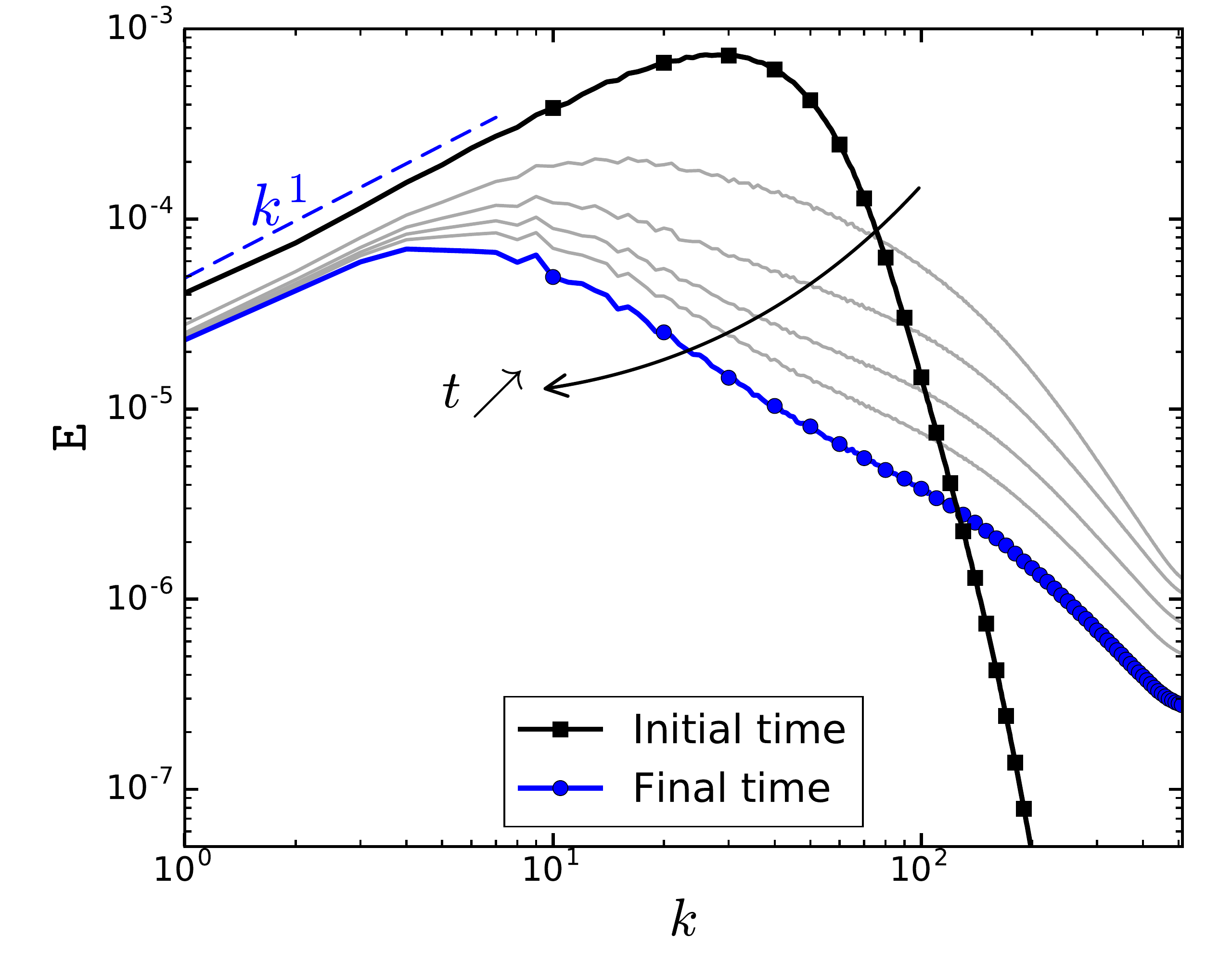} 
}
\hfill
\subfloat[~$\sq=2$]{\label{fig:s2_Eso}
\includegraphics[width=0.44\linewidth]{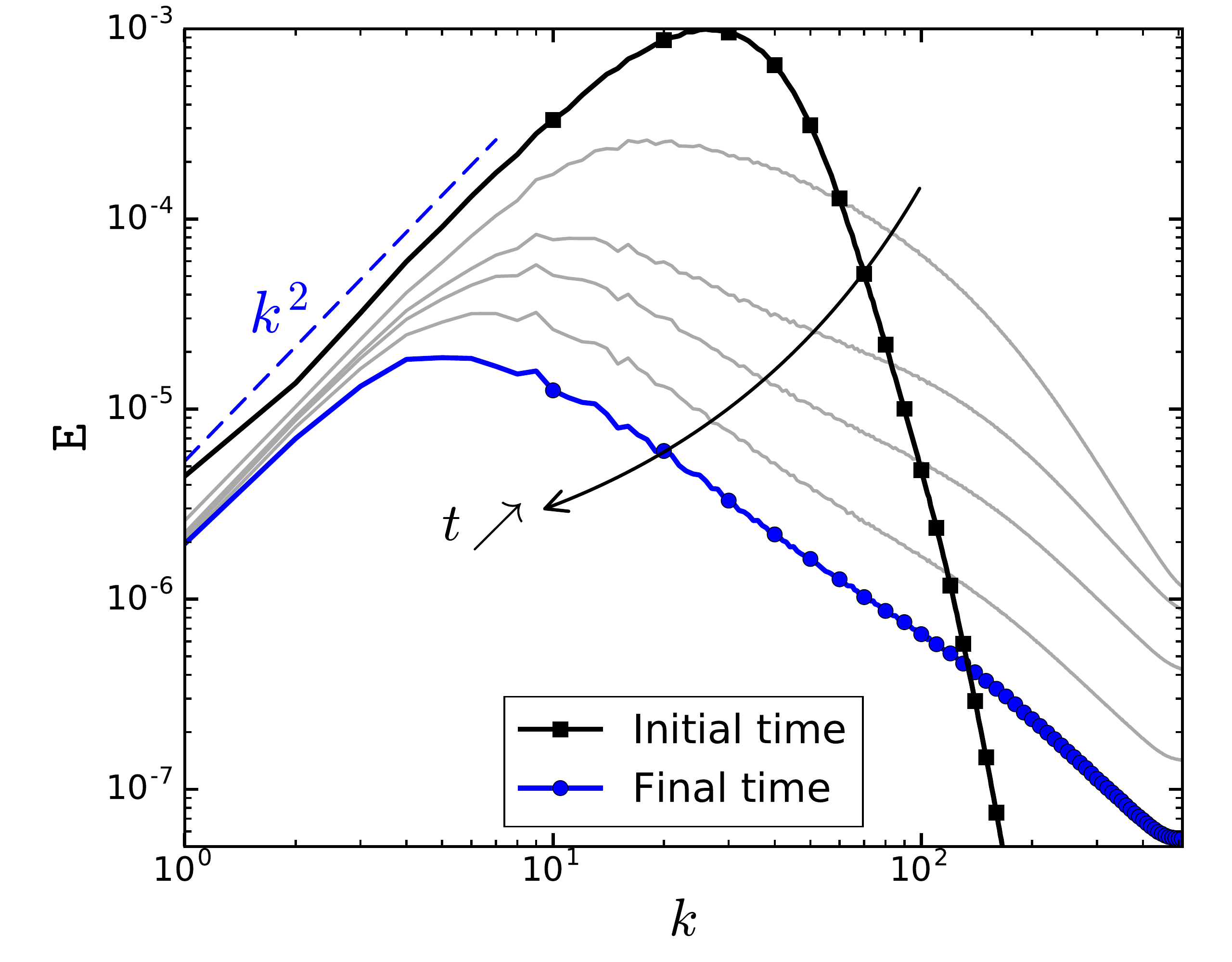}
}
\hfill~
\\
\hfill
\subfloat[~$\sq=3$]{\label{fig:s3_Eso}
\includegraphics[width=0.44\linewidth]{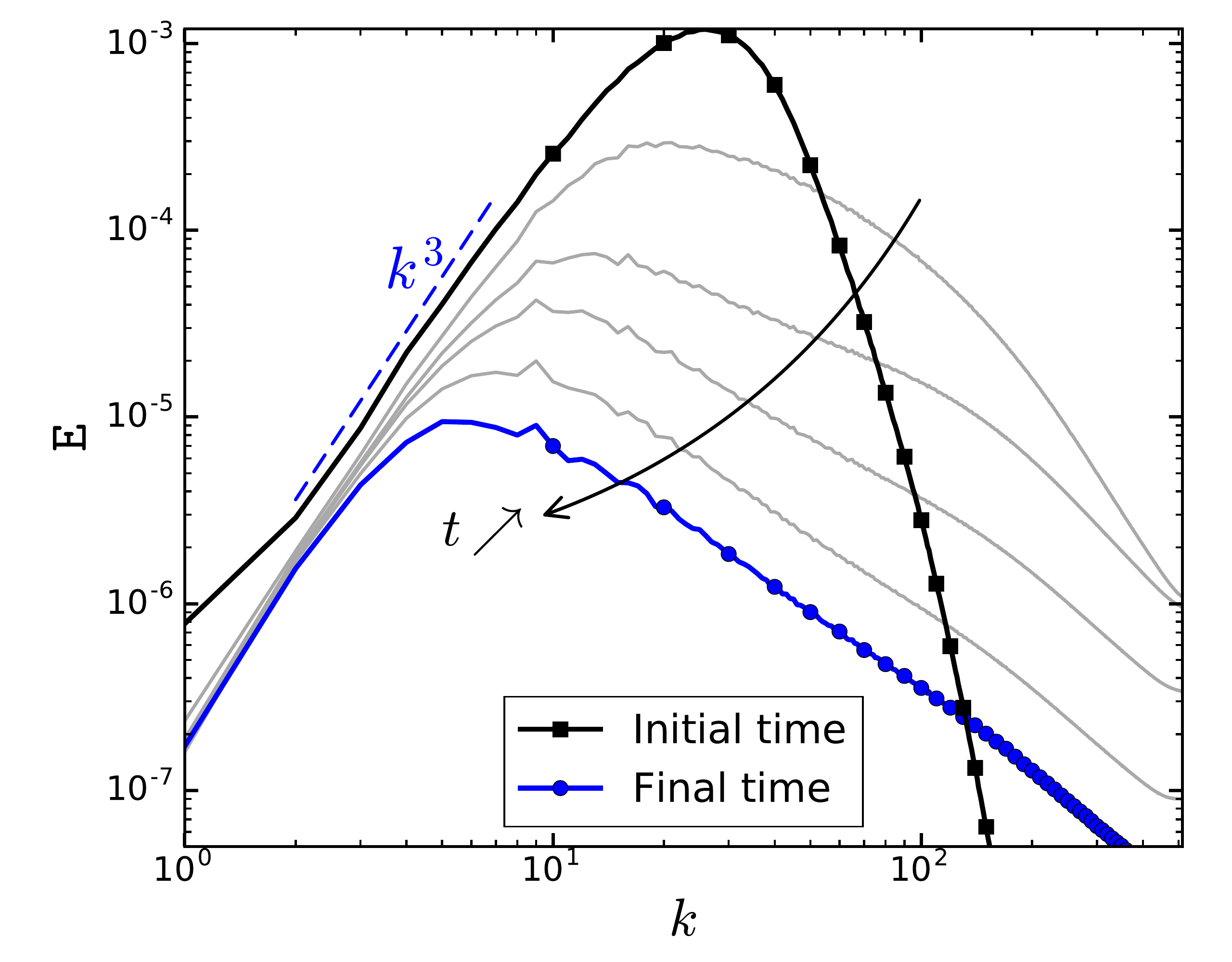} 
}
\hfill
\subfloat[~$\sq=4$]{\label{fig:s4_Eso}
\includegraphics[width=0.44\linewidth]{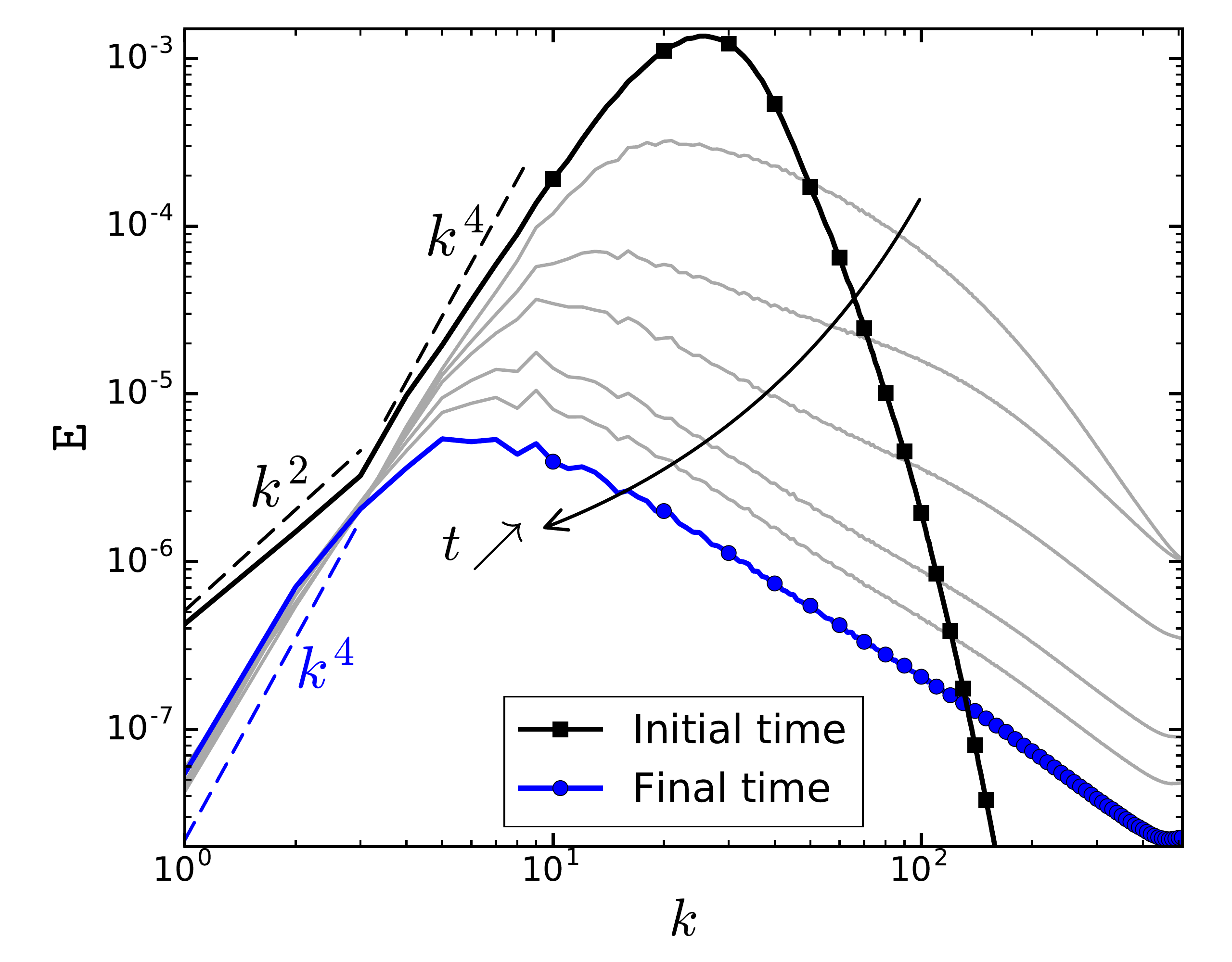}
}
\hfill~
\\
\subfloat[~$\sq=10$]{\label{fig:s10_Eso}
\includegraphics[width=0.44\linewidth]{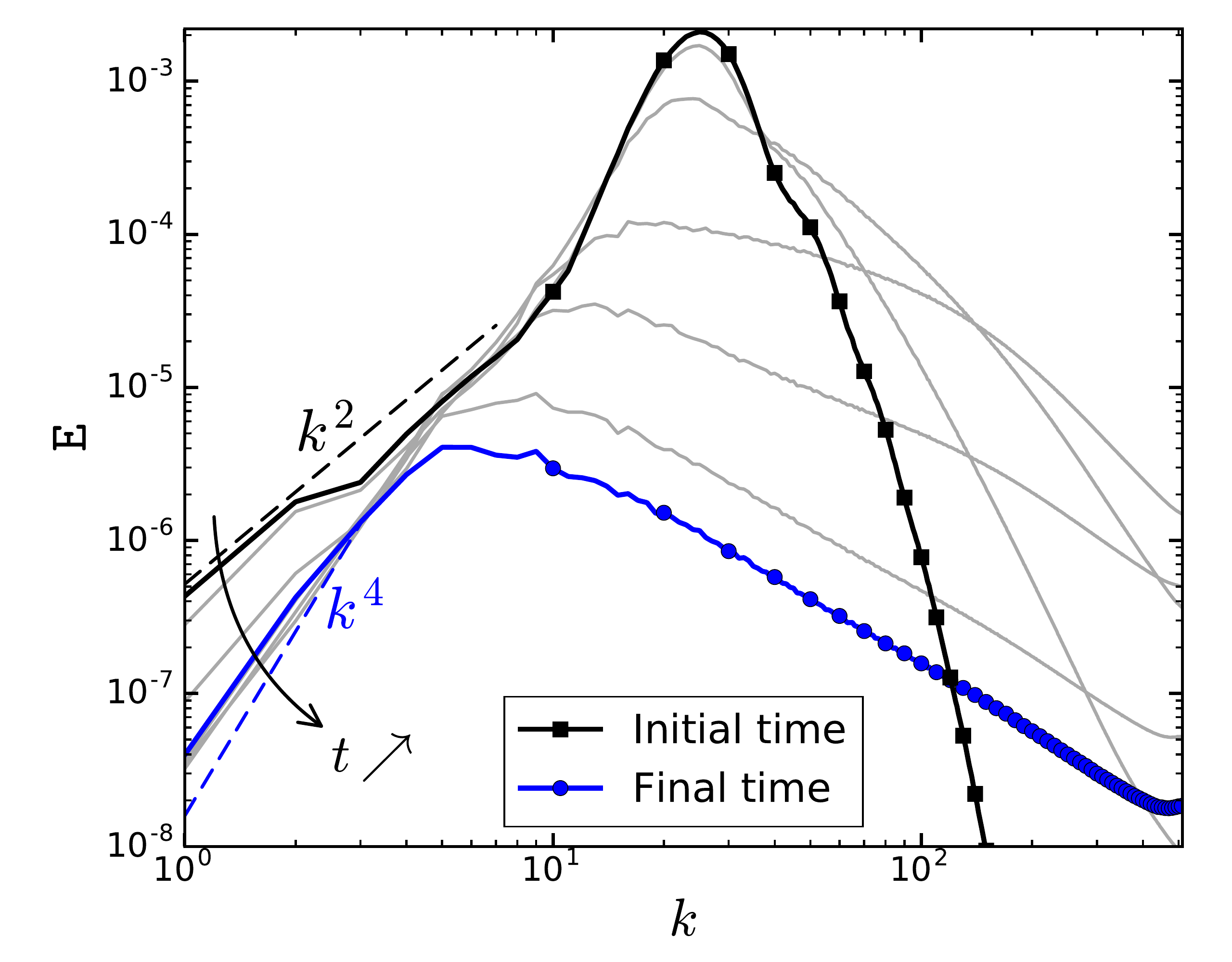} 
}
\end{center}
\caption{\label{fig:Eso} 
Evolution of the velocity spectrum $\mE$ with ${\mQ}^{(0)}$ given by Eq. \eqref{eq:mQ0} and for different values of $\sq$.
}
\end{figure}

%%%%%% Qir %%%%%%%%%
\begin{figure}[!htb] 
\begin{center}
\hfill
\subfloat[~$\sq=1$]{\label{fig:s1_Qir}
\includegraphics[width=0.44\linewidth]{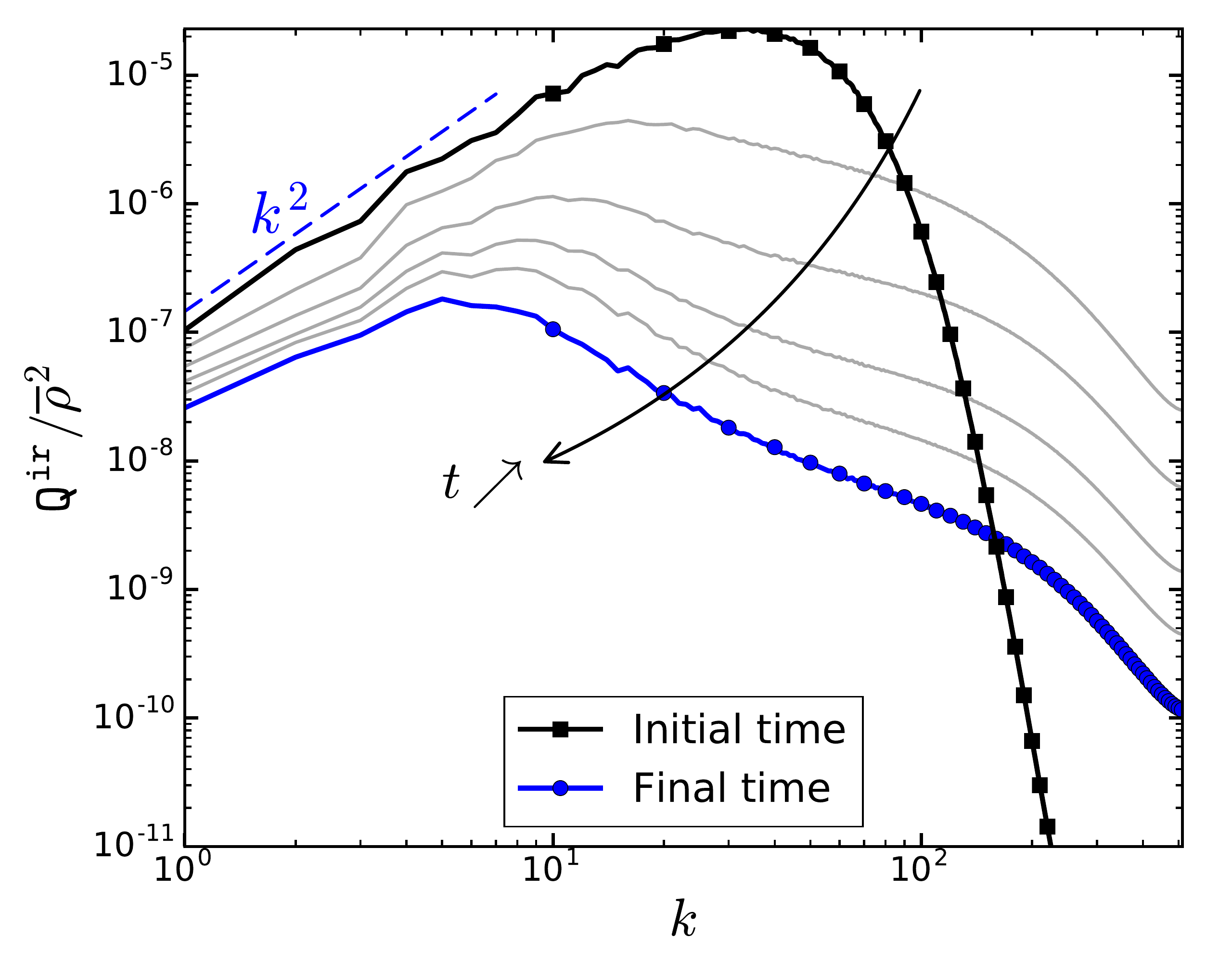} 
}
\hfill
\subfloat[~$\sq=2$]{\label{fig:s2_Qir}
\includegraphics[width=0.44\linewidth]{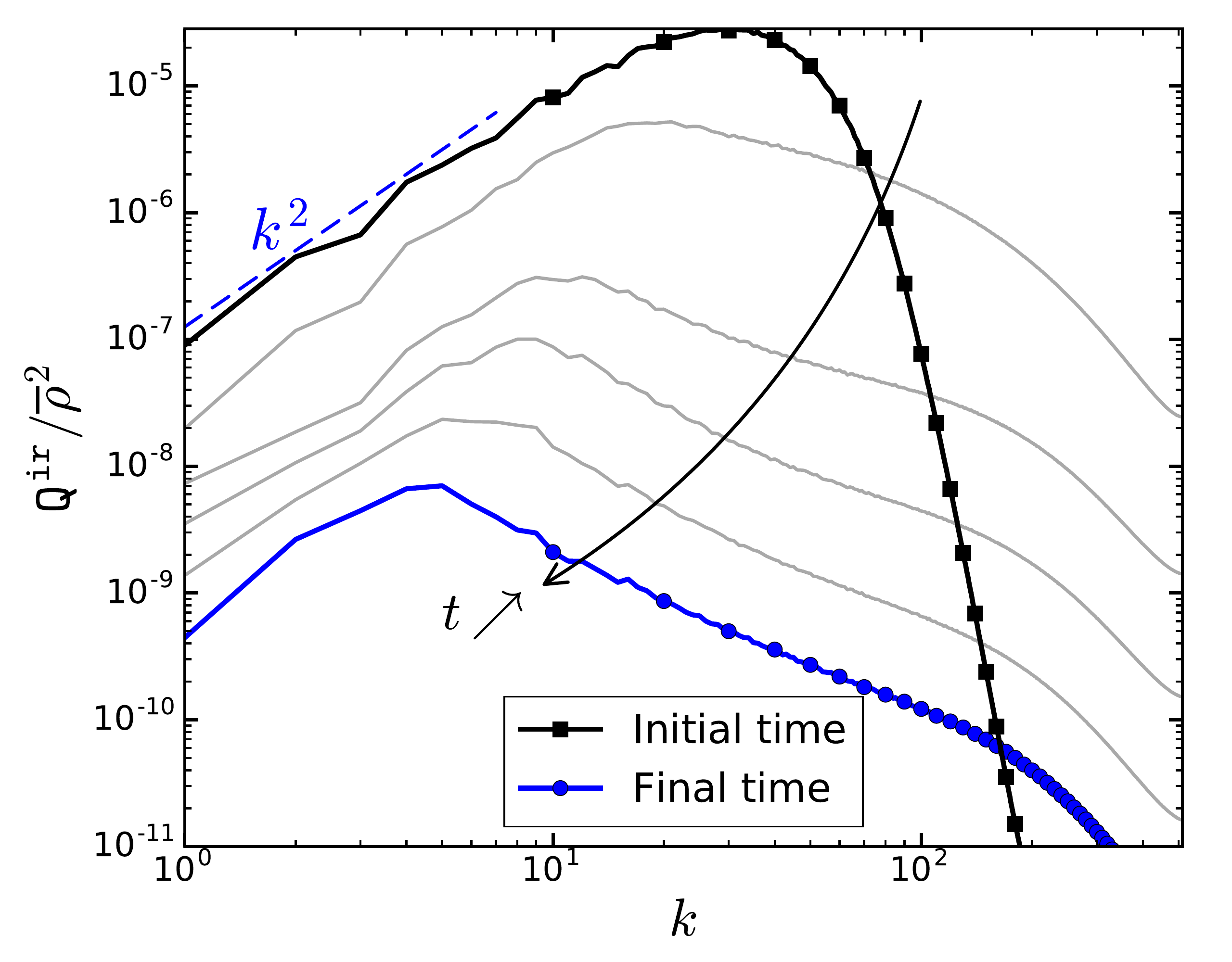}
}
\hfill~
\\
\hfill
\subfloat[~$\sq=3$]{\label{fig:s3_Qir}
\includegraphics[width=0.44\linewidth]{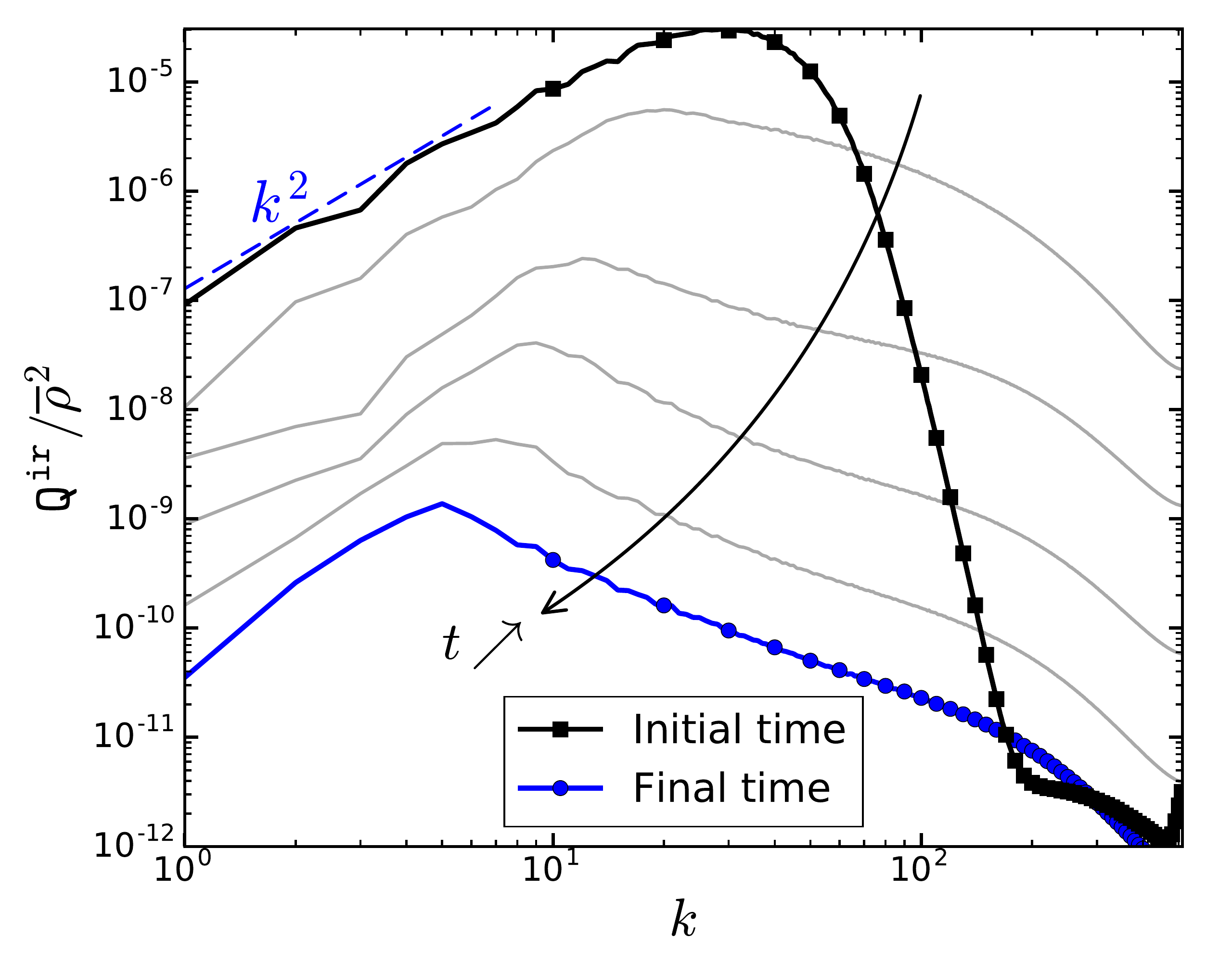} 
}
\hfill
\subfloat[~$\sq=4$]{\label{fig:s4_Qir}
\includegraphics[width=0.44\linewidth]{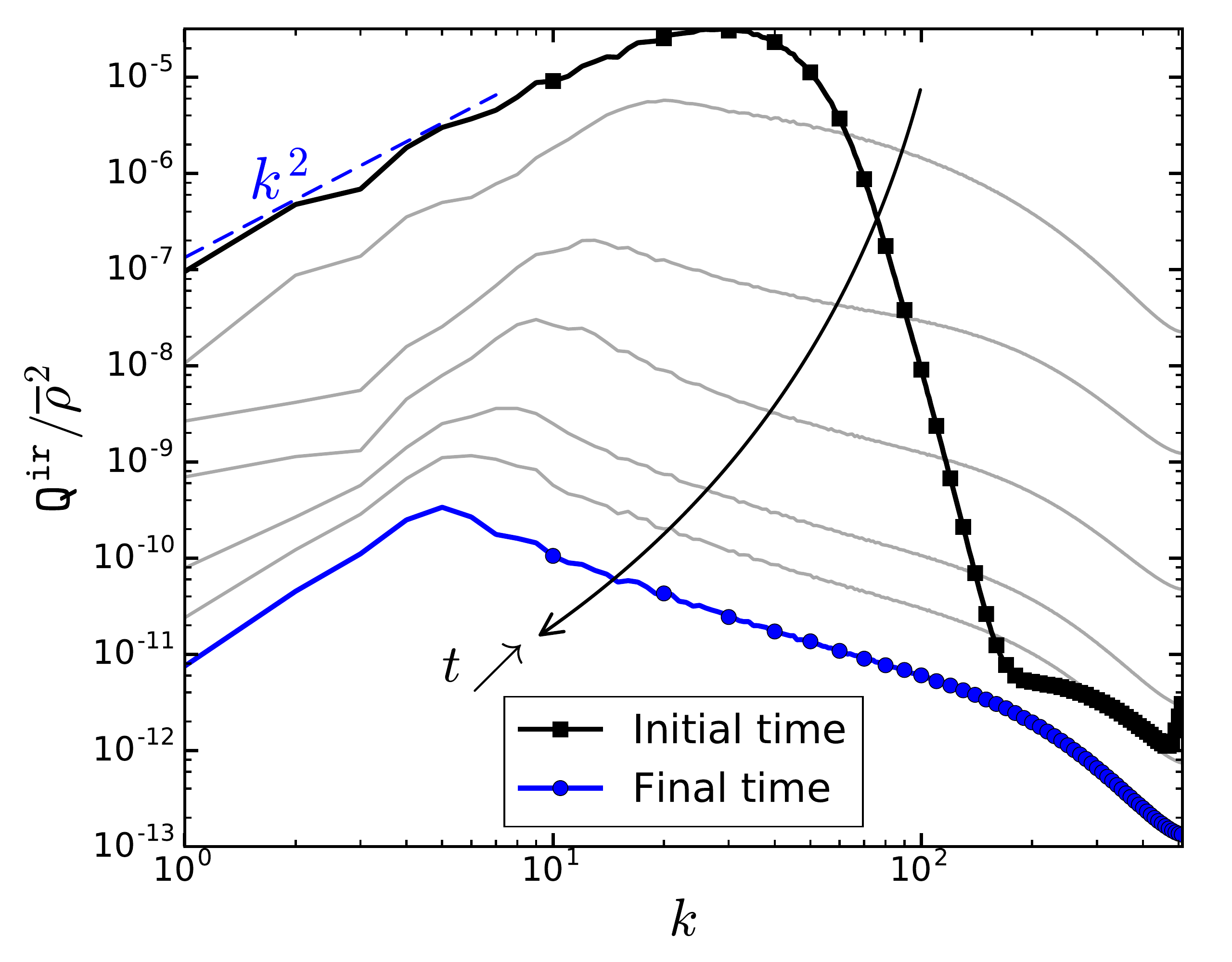}
}
\hfill~
\\
\subfloat[~$\sq=10$]{\label{fig:s10_Qir}
\includegraphics[width=0.44\linewidth]{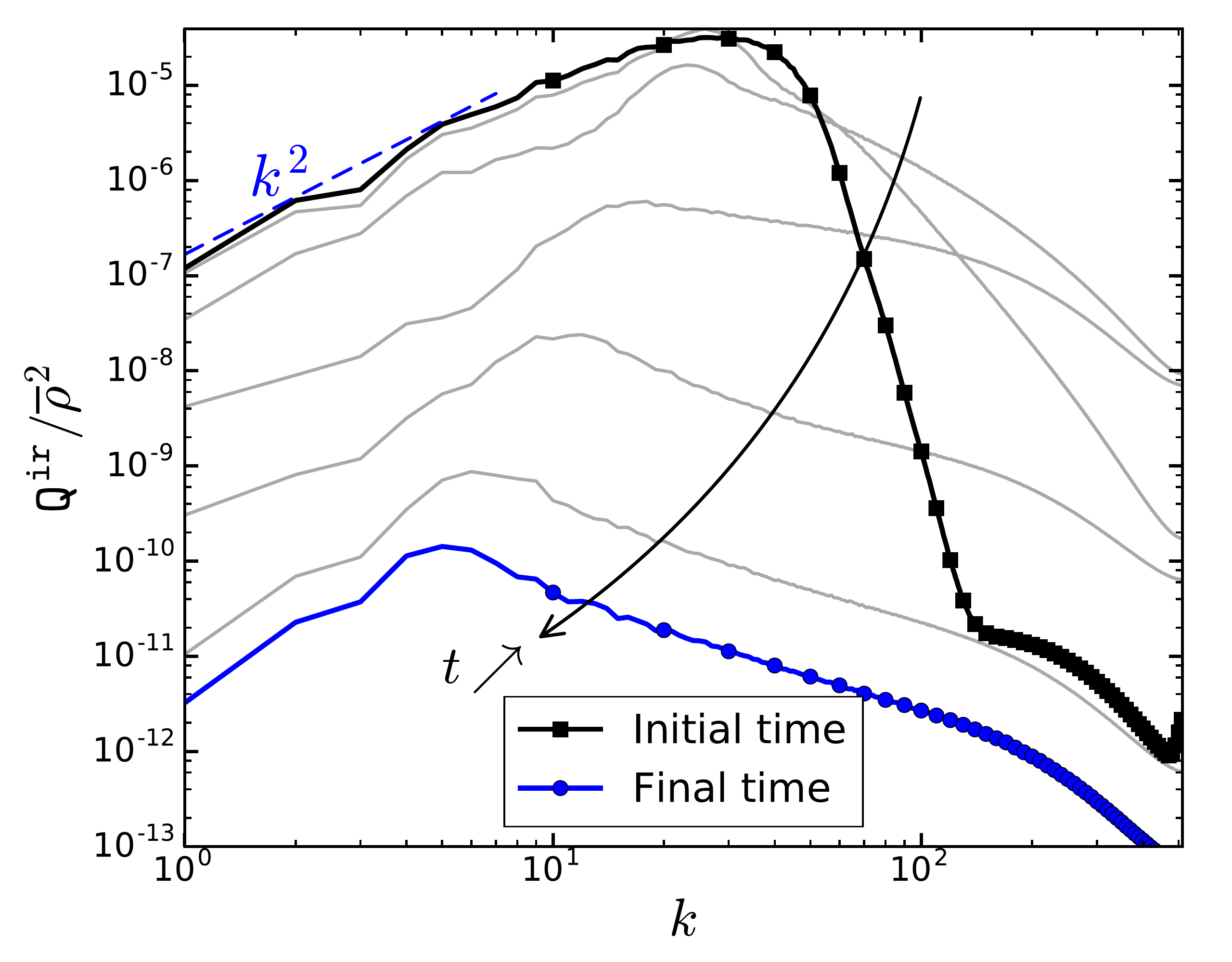} 
}
\end{center}
\caption{\label{fig:Qir} 
Evolution of the spectrum $\mQi$ of the irrotational momentum with ${\mQ}^{(0)}$ given by Eq. \eqref{eq:mQ0} and for different values of~$\sq$.
}
\end{figure}

\newpage
%----------------------------------------
\subsection{Evolution of the spectra when $\mQ$ is imposed at initial time}
%----------------------------------------
The main prediction of this work is expressed in Sec. \ref{sec:perm_mQ}. It states that the spectrum $\mQ$ of the solenoidal component of the momentum $\bs{\qs}$ is invariant at small wavenumbers provided $\sq<4$. Otherwise, if $\sq>4$, it transitions to a $k^4$ spectrum  and if $\sq=4$, its infrared amplitude varies but not its scaling.
To verify these predictions, we consider the first series of simulations for which $\mQ$ is imposed at initial time according to Eq. \eqref{eq:mQ0} and $\sq$ is varied  from $1$ to $10$.
Figure \ref{fig:Qso} shows the spectrum $\mQ$ obtained at different times from these simulations.
It can be seen that for $\sq \le 3$, $\mQ$ remains approximately constant at large scales.
For $\sq = 4$, the infrared exponent remains constant but the prefactor of the infrared power law displays a visible evolution in time. 
Finally, for $\sq=10$, the infrared slope transitions from its initial value to approximately $4$.
All of these observations are coherent with the predictions obtained in Sec.  \ref{sec:perm_mQ} and recalled above. 
They corroborate that a standard description of the permanence of large eddies applies to the spectrum $\mQ$ of the solenoidal momentum.

~\\\indent
Another important result described in Sec. \ref{sec:bouss_comp} is that the large-scale properties of the velocity spectrum $\mE$ differ strongly whether in a Boussinesq or variable-density flow. 
In the Boussinesq limit, $\mE$ becomes equal to $\mQ/\rey{\rho}^2$ so that its initial conditions at small wavenumbers are preserved under the same conditions as $\mQ$, i.e. when its infrared exponent $\se$ is smaller than $4$.
However, when the density contrast increases and the flow ceases to be in the Boussinesq limit, $\mE$ and $\mQ$ differ. And while the permanent behavior of $\mQ$ remains unaffected, that of $\mE$ is modified and mostly lost. 
This is what is shown in Fig. \ref{fig:Eso}. For all values of $\sq$, one can observe that the initial and final states of $\mE$ at small wavenumbers are distinct.
Of particular interest is the case $\sq=10$.  As explained in App. \ref{app:evol_mE}, the relation between $\mE$ and $\mQ$ involves non-linear convolution products that spur $k^2$ scalings at small wavenumbers. When the scaling of $\mQ$ is imposed, a $k^2$ range can be shown to exist for wavenumbers smaller that a limit wavenumber proportional to $k_e (   {\rey{\tau'^2}}/{\rey{\tau}^2} )^{1/(\sq-2)}$. For a given value of density contrast, the higher $\sq$ is, the higher the limit wavenumber is.
Thus, for $\sq=10$, $\mE$ displays at initial time a clear $k^2$ range. The latter can only be guessed for $\sq=4$ and is absent for $\sq<3$.
The $\sq=10$ case is noteworthy because had the permanence of large-eddies applied to $\mE$, its initial $k^2$ scaling and prefactor would have been preserved in time.
But since it is not the case, we observe that $\mE$ evolves at small wavenumbers from its initial $k^2$ scaling to a $k^4$ scaling.
As explained in Sec. \ref{sec:bouss_comp}, the final state of $\mE$ corresponds to the one of $\mQ$.

~\\\indent
Finally, Fig. \ref{fig:Qir} displays the spectrum $\mQi$ of the irrotational component of the momentum at different times and for the different values of $\sq$.
The main observation is that $\mQi$ decays at all scales and for all simulations, as expected from Sec. \ref{sec:bouss_comp}. There is no permanence of large-eddies for this spectrum. This is important because it shows that the sum $\mQ+\mQi$ which is equal to the spectrum $\mQt$ of the full momentum $\rho \bs{u}$ is not permanent at small wavenumbers. 
This property was predicted in Sec. \ref{sec:bouss_comp} and means that Sitnikov's integral \cite{sitnikov58} is not an invariant.

Note that $\mQi$ displays an initial infrared slope of $2$ for all simulations. This  property is imposed by the initialization procedure whereby $\mQi$ is obtained by solving the Poisson equation \eqref{eq:poisson_phi}.

%----------------------------------------
\subsection{Evolution of the spectra when $\mE$ is imposed at initial time}
%----------------------------------------

%%%%%% Qso %%%%%%%%%
\begin{figure}[!htb] 
\begin{center}
  \subfloat[~$\nicefrac{\sqrt{\rey{\rho'^2}}}{\rey{\rho}} = 0.02$]{
    \label{fig:Qso_a}
    \includegraphics[width=0.48\linewidth]{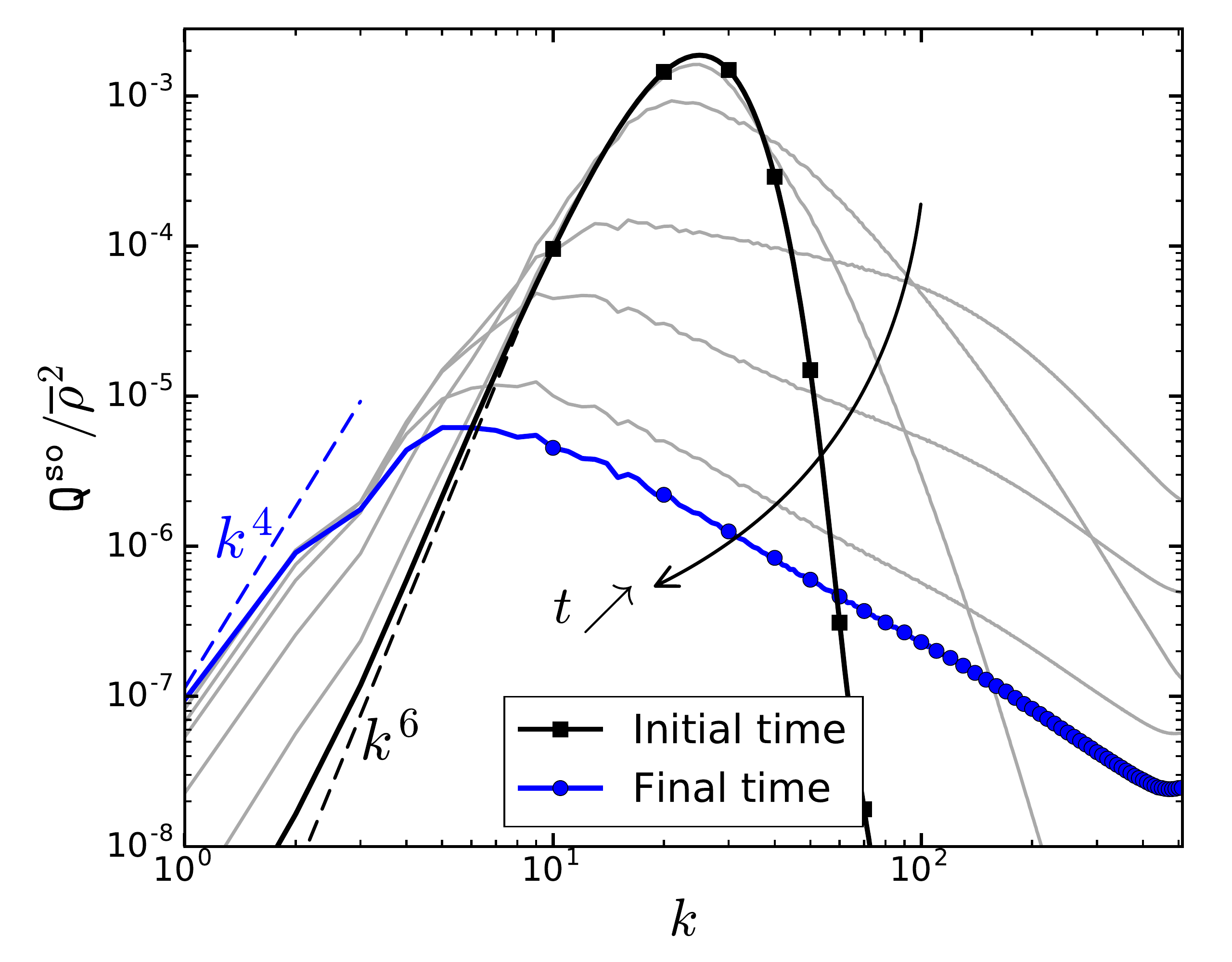}
  }
  \subfloat[~$\nicefrac{\sqrt{\rey{\rho'^2}}}{\rey{\rho}} = 0.75$]{
    \label{fig:Qso_b}
    \includegraphics[width=0.48\linewidth]{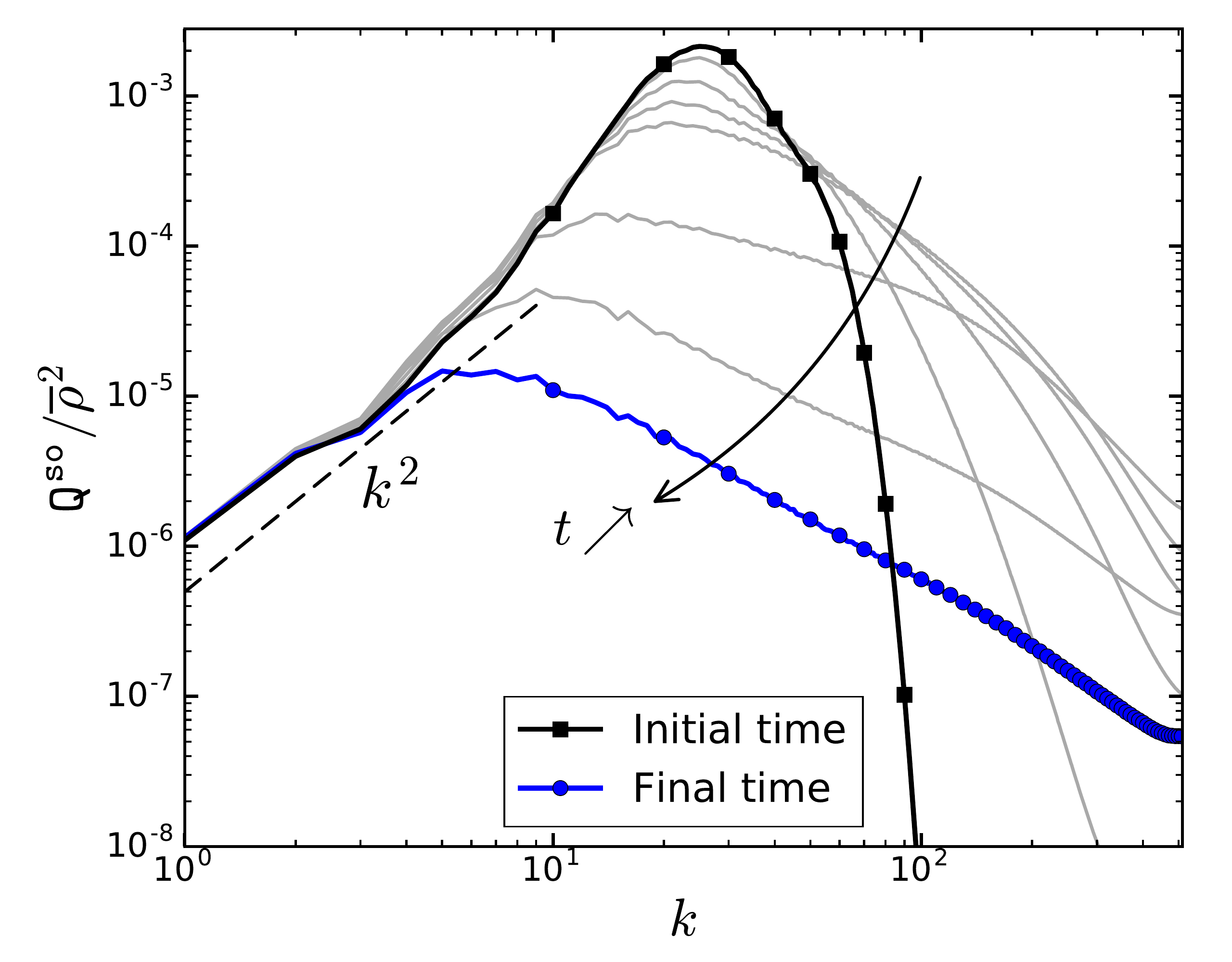}
  }
\end{center}
\caption{\label{fig:Qso_2}
Evolution of the spectrum $\mQ$ of the solenoidal momentum with $\mE^{(0)}$ given by Eq. \eqref{eq:mE0}, for $\se=6$ and for different density contrasts.
}
\end{figure}

%%%%%% Eso %%%%%%%%%
\begin{figure}[!htb] 
\begin{center}
  \subfloat[~$\nicefrac{\sqrt{\rey{\rho'^2}}}{\rey{\rho}} = 0.02$]{
    \label{fig:Eso_a}
    \includegraphics[width=0.48\linewidth]{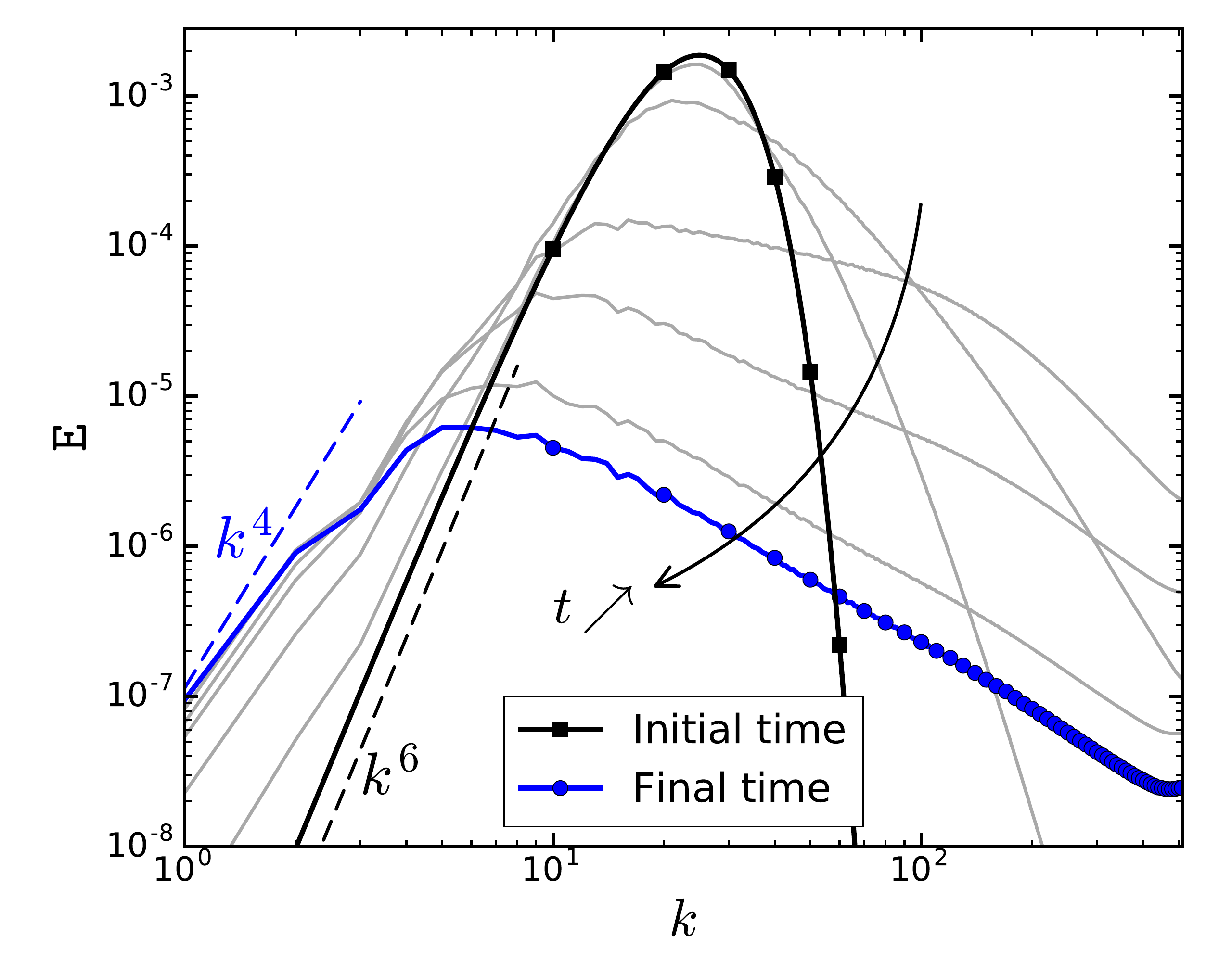}
  }
  \subfloat[~$\nicefrac{\sqrt{\rey{\rho'^2}}}{\rey{\rho}} = 0.75$]{
        \label{fig:Eso_b}
  \includegraphics[width=0.48\linewidth]{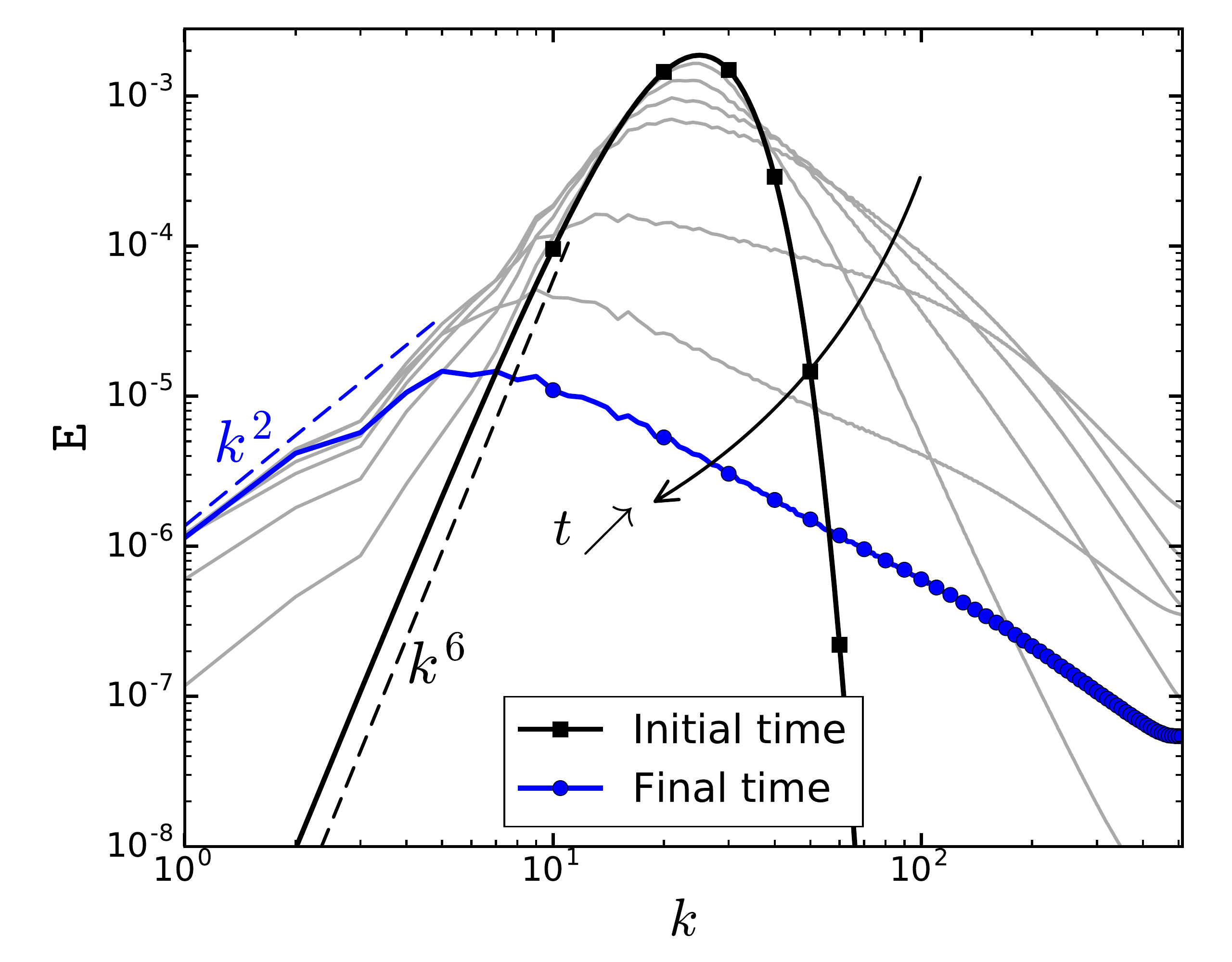}
  }
\end{center}
\caption{\label{fig:Eso_2}
Evolution of the velocity spectrum $\mE$ with $\mE^{(0)}$ given by Eq. \eqref{eq:mE0}, for $\se=6$ and for different density contrasts.
}
\end{figure}

We now turn our attention to the second series of simulations we performed, i.e. the ones where it is $\mE$ and not $\mQ$ which is initialized with a fixed infrared slope.
As explained in App. \ref{app:evol_mE}, when $\mE$ obeys a power law at initial time, $\mQ$ has possibly two large-scale ranges separated by a wavenumber $\kvd$. For $k\ll\kvd$, $\mQ$ varies like $k^2$ and for $\kvd \ll k \ll k_e$, $\mQ \approx \rey{\rho}^2 \mE \propto k^{\se}$.
The limit wavenumber $\kvd$ varies proportionally to $(\nicefrac{\rey{\tau'^2}}{\rey{\tau}^2})^{1/(\se-2)}$. Hence, for small density contrasts, one will have $\mQ \approx \rey{\rho}^2 \mE$ over the whole observable large scale range. However, for high-density contrasts, one will observe $\mQ \propto k^2$.
This difference in initial conditions can be seen by comparing the thick black curves of  Figs. \ref{fig:Qso_2} and \ref{fig:Eso_2}. While the velocity spectrum $\mE$ has the same $k^{6}$ initial condition in the high and small density contrast simulations (Figs. \ref{fig:Eso_a} and \ref{fig:Eso_b}), $\mQ$ displays very different initial scalings (Figs. \ref{fig:Qso_a} and \ref{fig:Qso_b}). In agreement with the previous explanation, one observes a $k^{6}$ initial scaling for the small density contrast simulation and a $k^2$ scaling for the high-density contrast one.

In the small density contrast simulation (Figs. \ref{fig:Qso_a} and \ref{fig:Eso_a}), we observe that both $\mE$ and $\mQ/\rey{\rho}^2$ remain approximately equal at all times. Starting from their $k^{6}$ initial condition, they both evolve towards a $k^4$ spectrum. This corresponds to the classical behavior predicted for constant density flows.
When the density contrast is high, we observe in Fig. \ref{fig:Qso_b} that $\mQ$ is invariant at small wavenumbers. Its initial $k^2$ scaling and the corresponding prefactors are preserved in time.  This confirms the predictions obtained in Sec. \ref{sec:perm_mQ} about the permanence of $\mQ$.
As for $\mE$, we see in Fig. \ref{fig:Eso_b} an evolution which is very distinct from the one that would be observed in a constant density flow. Starting from its $k^{6}$ initial condition, $\mE$ does not transition towards a $k^4$ spectrum but towards a $k^2$ spectrum.
This evolution is coherent with the conclusions reached in Sec. \ref{sec:hom_turb} and which predict that $\mE$ tends to $\mQ$ at large times.

It is interesting to compare Fig. \ref{fig:Eso_b} with Fig. \ref{fig:s10_Eso}.
They both show somewhat opposite evolutions of $\mE$ at small wave numbers: in the first case, $\mE$ increases from a $k^{6}$ to a $k^2$ spectrum while in the second it decreases from a $k^2$ to a $k^4$ spectrum. Both behaviors are fully explained by the present theory and are related to the permanence of $\mQ$.

%----------------------------------------
\subsection{Non-linear transfer terms}
%----------------------------------------

The conditions under which $\mQ$ is permanent are derived in Sec. \ref{sec:model} by showing that the non-linear transfer term  $\mT$ associated to $\mQ$ scales as $k^4$ for $\sq \le 2$.
The validity of this scaling can be indirectly inferred from Fig. \ref{fig:Qso} and in particular from Fig. \ref{fig:s10_Qso} which shows the transition of $\mQ$ from a scaling steeper than $k^4$ to a scaling close to $k^4$.

\begin{figure}[!htb] 
\begin{center}
  \subfloat[~Non-linear transfer term $\mT$ associated with $\mQ$]{
    \label{fig:NL_Qso}
    \includegraphics[width=0.48\linewidth]{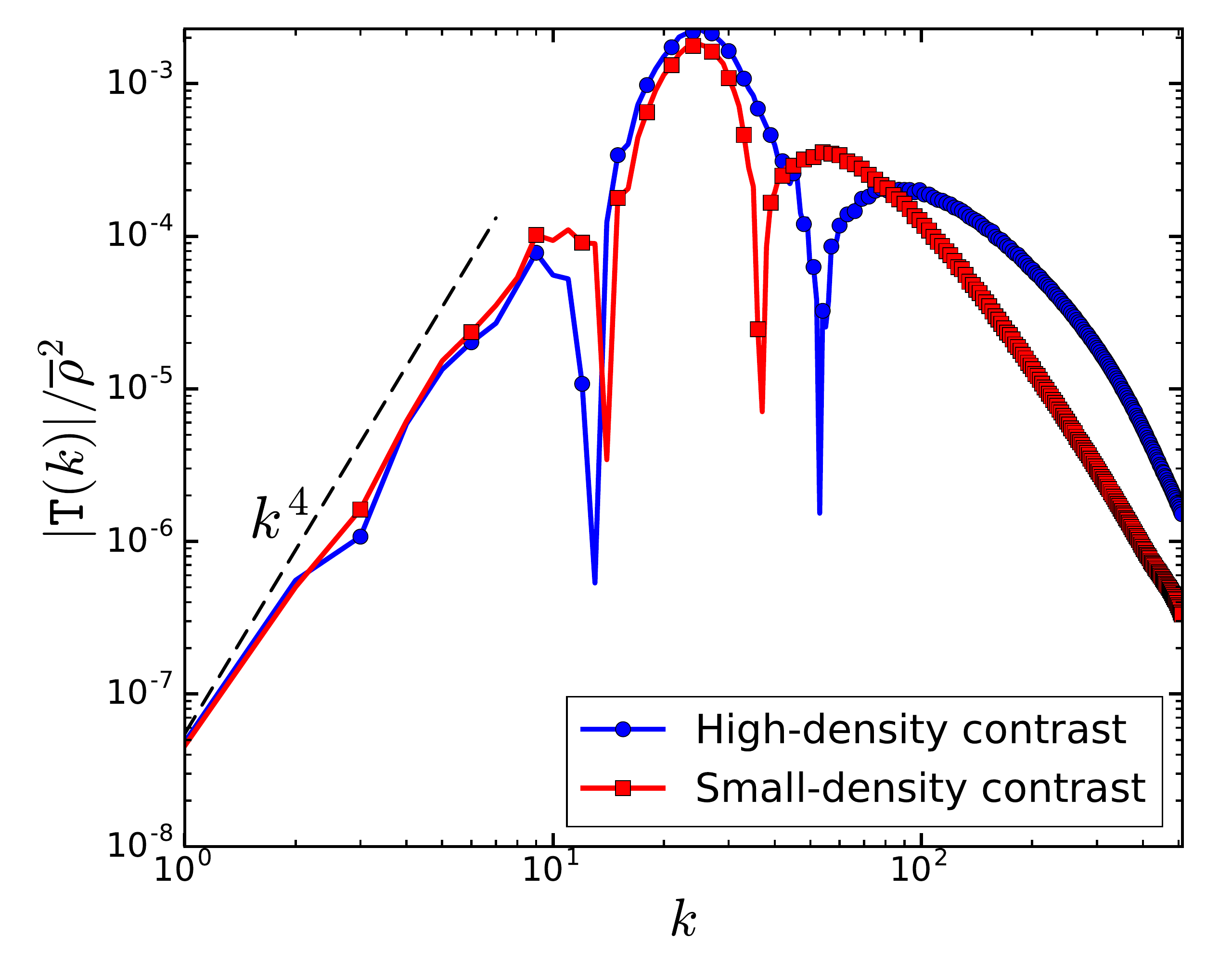}
  }
  \subfloat[~Non-linear transfer term $\mT_\mE$ associated with $\mE$]{
        \label{fig:NL_Eso}
  \includegraphics[width=0.48\linewidth]{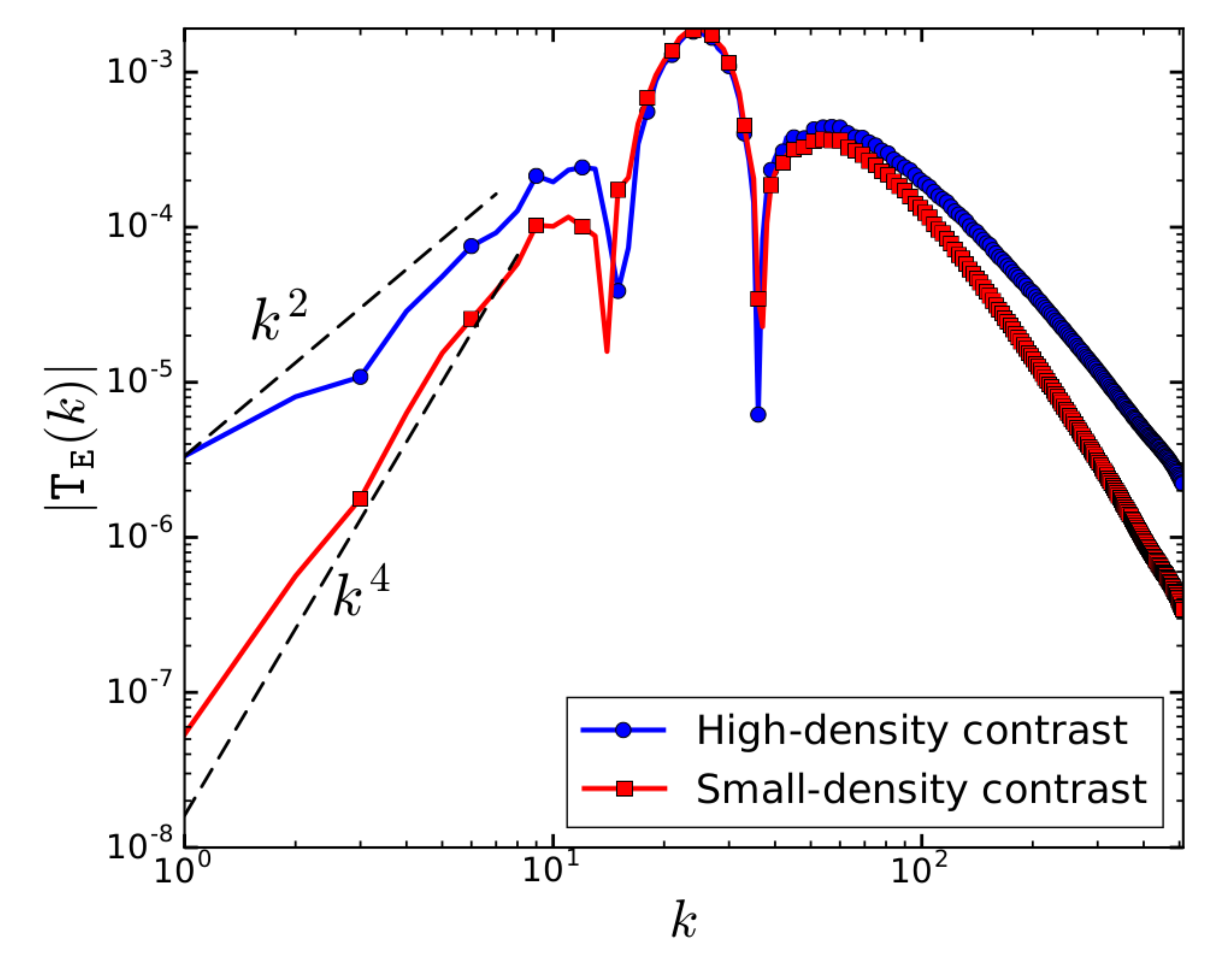}
  }
\end{center}
\caption{\label{fig:NL_QE}
Absolute value of the non-linear transfer terms $\mT$ of $\mQ$ (a) and $\mT_\mE$ of $\mE$ (b). The result is displayed at time $t=2 \tau_0$ for the series of  two simulations for which $\mE$ is imposed at initial times.
}
\end{figure}

A direct verification of this crucial assumption is useful for comforting the whole derivation and is proposed in Fig. \ref{fig:NL_QE}.
Thus, in Fig. \ref{fig:NL_Qso}, the non-linear transfer term $\mT$ defined by Eq. \eqref{eq:mQ} is shown for the last two simulations performed by initializing the velocity spectrum and having two different initial density contrasts.
In order for the large scale range to be large enough, the result is displayed at an early time $t=2 \tau_0$, where $\tau_0$ is the initial turn-over time defined by :
$$
\tau_0 = \frac{\sqrt{\int \mE \ud k}}{{\int k \mE \ud k}}
\period
$$
It can be seen that whether  the density contrast is small or high, the transfer term $\mT$ displays a $k^4$ scaling.
Besides, it should be remembered that $\mQ$ scales initially as $k^6$ in the small-density case and as $k^2$ in the high-density case considered in this figure.  Hence, the scaling of $\mT$ is independent from this initial property. This can also be verified on the first series of simulations for which $\mQ$ is imposed at initial time. While not displayed here, for $\sq \ge 2$, a $k^4$ scaling is observed for $\mT$.

This property of the non-linear transfer term $\mT$ of $\mQ$ should be contrasted with that of the non-linear transfer term $\mT_\mE$ associated with the velocity spectrum $\mE$. This second transfer term is defined by Eq. \eqref{eq:mt_me} and is analyzed more precisely in App. \ref{app:vel_spec}.
Its value, at  time $t=2 \tau_0$ and for the last two simulations performed by initializing the velocity spectrum, is shown in Fig. \ref{fig:NL_Eso}.
In the small-density contrast case, it can be seen that $\mT_\mE$ displays a $k^4$ scaling, as expected in that case.
However, in the high-density contrast case, $\mT_\mE$ displays a $k^2$ scaling.
As explained in App. \ref{app:vel_spec}, the latter scaling is associated with the Fourier correlation between $\bs{u'}$ and $p' \partial_i \tau'$.  Note that it is this very term that is also responsible for the modification of the far-field velocity scaling of a variable density single eddy (see Sec. \ref{sec:press_field}).

%----------------------------------------
\subsection{Self-similar decay exponent}
\begin{figure}[!htb] 
\begin{center}
  \subfloat[~Evolution of $\nn^\text{simu}(t)$]{
    \label{fig:nk_evol}
    \includegraphics[width=0.48\linewidth]{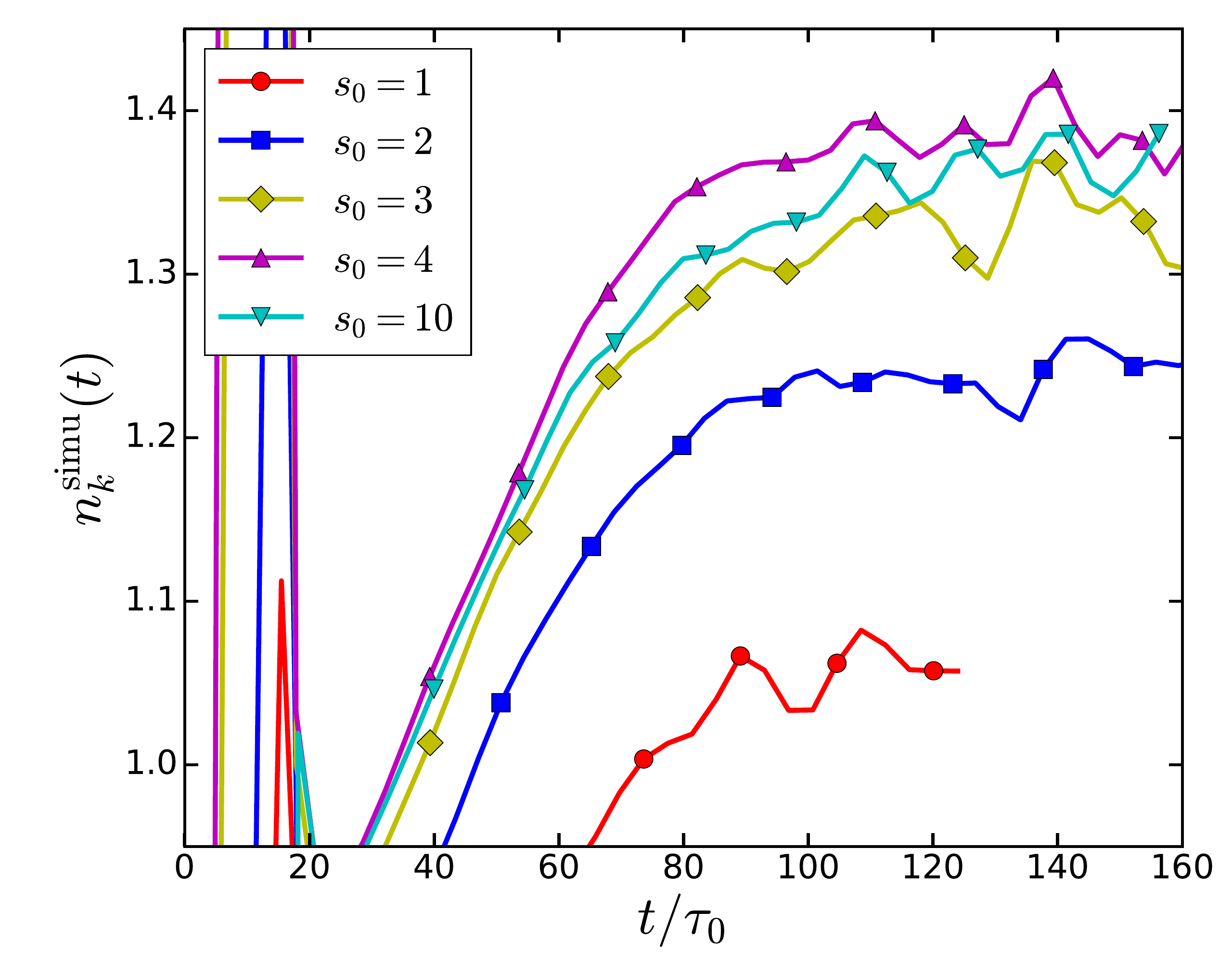}
  }
  \subfloat[~Comparison between $\integ{\nn^\text{simu}}$ and formula \eqref{eq:nk_modif}]{
        \label{fig:nk_sq}
  \includegraphics[width=0.48\linewidth]{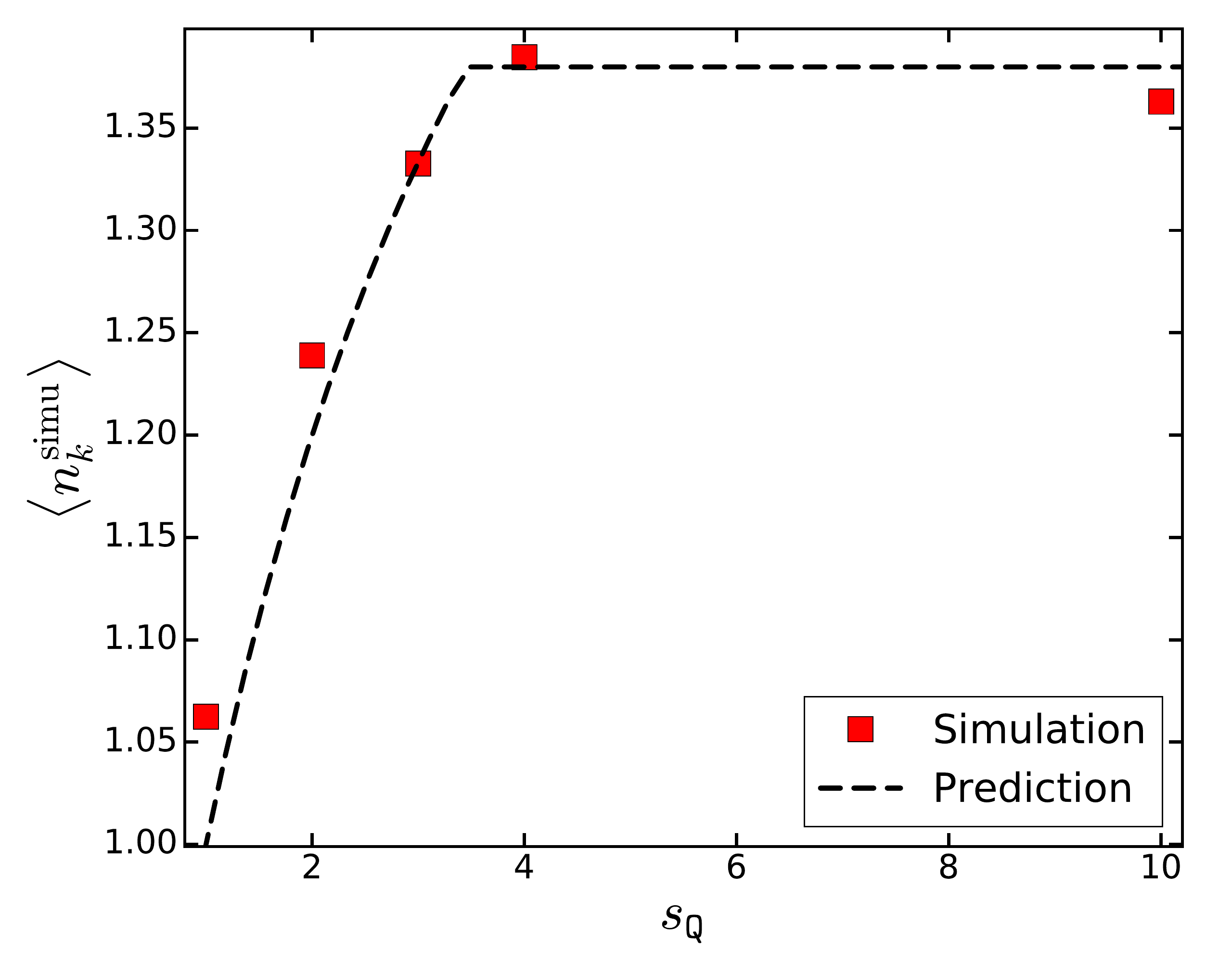}
  }
\end{center}
\caption{\label{fig:nk}
Instantaneous and averaged measures of the decay exponent of the kinetic energy $\kr$.
}
\end{figure}

%----------------------------------------
In Sec. \ref{sec:self_sim}, we explained how the decay exponent $\nn$ of the kinetic energy can be expressed  as a function of the infrared exponent $\sq$ of $\mQ$.  This function is given in Eq. \eqref{eq:nk_modif}.
To verify this prediction, we estimate a decay exponent from our simulations by using the time derivatives of the kinetic energy $\kr$:
$$
  \nn^\text{simu}(t) = \frac{[\partial_t\kr]^2}{(\kr \partial_{tt}^2\kr - [\partial_t\kr]^2)}
\period
  $$
When $\kr$ obeys a power law, $\nn^\text{simu}(t)$ is constant and equal to the corresponding power-law exponent, independently from any time-shifts.
The evolution of $\nn^\text{simu}(t)$ is shown in Fig. \ref{fig:nk_evol} for the first series of simulations for which $\sq$ is imposed. It can be seen that a plateau with oscillations is reached for all simulations for $t>100 \tau_0$.
The averaged value of the observed plateaus is denoted by $\integ{\nn^\text{simu}}$ and is computed as:
$$
\integ{\nn^\text{simu}} = \frac{ \int_{t_1}^{t_2} \nn^\text{simu}(s) \ud s }{t_2-t_1}
\with t_1 = 100 \tau_0 \andd t_2 = 160 \tau_0
\period
$$
This value is compared against its prediction given by Eq. \eqref{eq:nk_modif} in Fig. \ref{fig:nk_sq}.
A satisfactory agreement is observed: predictions and simulations differ by no more than $6.5 \%$.

%=======================================================================
\section{Conclusions}
%=======================================================================
In this work, we studied the large-scale structure of variable-density  homogeneous turbulence with small Mach numbers and high density contrasts.
As a first step, we adapted the analogy between single eddies and homogeneous turbulence  proposed in \cite{batchelor56} to the variable-density context.  This preliminary study allowed to point out the preeminent role played by the solenoidal component $\bs{\qs}$ of the momentum. This component is indeed  associated with an invariant scaling far from the eddy core. By contrast, the velocity far-field scaling is not invariant because of the pressure field which casts a larger shadow in variable-density flows than in constant density flows.

When transposing these results to homogeneous turbulence, we focused on the spectrum $\mQ$ of the solenoidal component of the momentum $\bs{\qs}$. We showed that $\mQ$ is invariant at large scales under the same conditions as those encountered in constant density flows: when its initial infrared exponent $\sq$ is smaller than $4$, $\mQ$ is invariant at large scales. Otherwise, it is not and evolves towards a $k^4$ spectrum.
To obtain this result, we derived the evolution equation for $\mQ$ and applied a distant interaction hypothesis to simplify its non-linear transfer term. Under these conditions, the latter is found to scale as $k^4$.
By contrast, the velocity spectrum $\mE$ is generally not invariant when $\se \ge 2$. Indeed, in the evolution of $\mE$, correlations between the pressure and density fields give rise to a  non-linear transfer term having a $k^2$ scaling.
These predictions were verified by performing large-eddy simulations (LES) of homogeneous isotropic turbulence. The simulations allowed to check the permanence of $\mQ$ when $\sq <4$ and the impermanence of $\mE$ when $\se>2$.

%=======================================================================
% BIBLIOGRAPHY
%=======================================================================

\appendix

%=======================================================================
\section{Additional remarks about the single eddy configuration}  \label{app:eddy_remarks}
%=======================================================================

In Sec. \ref{sec:press_field}, we noted that the distinct behaviors of the velocity field  between constant and variable-density eddies arise from the scaling of the pressure field, as given by Eq. \eqref{eq:blob_press}.
This equation also allows to understand how the transition between the variable and constant density cases is made. 
Indeed, the leading order of the pressure field in the constant density case appears as a next order term in the asymptotic expression Eq. \eqref{eq:blob_press}.
Therefore, the comparison  of the prefactors of the first two terms in Eq. \eqref{eq:blob_press} allows to define the region where the variable-density scaling of the pressure field becomes predominant compared to its constant-density scaling.
This comparison  leads to the definition of a length scale $\ell_P$:
\begin{align} \label{eq:lp}
\ell_P = \frac{\int u_ju_j \ud \x}  {\sqrt{\partial_t L_i \;\partial_t L_i} }
\period
\end{align}
Let us recall that $\bs{L}$ divided by the eddy volume corresponds to the spatially-averaged translational velocity of the eddy.
Therefore, the length $\ell_P$ roughly compares  the kinetic energy of the eddy to its translational acceleration.
It represents the distance that an accelerated eddy must travel to see its translational kinetic energy reach a value comparable to its initial kinetic energy.
Another interpretation is that the ratio $\ell_P$ over $\ell_\mathcal{D}$ compares the orders of magnitude of the centrifugal and translational accelerations of the eddy.

\begin{figure}[!htb]
\includegraphics[height=0.4\linewidth,angle=-90]{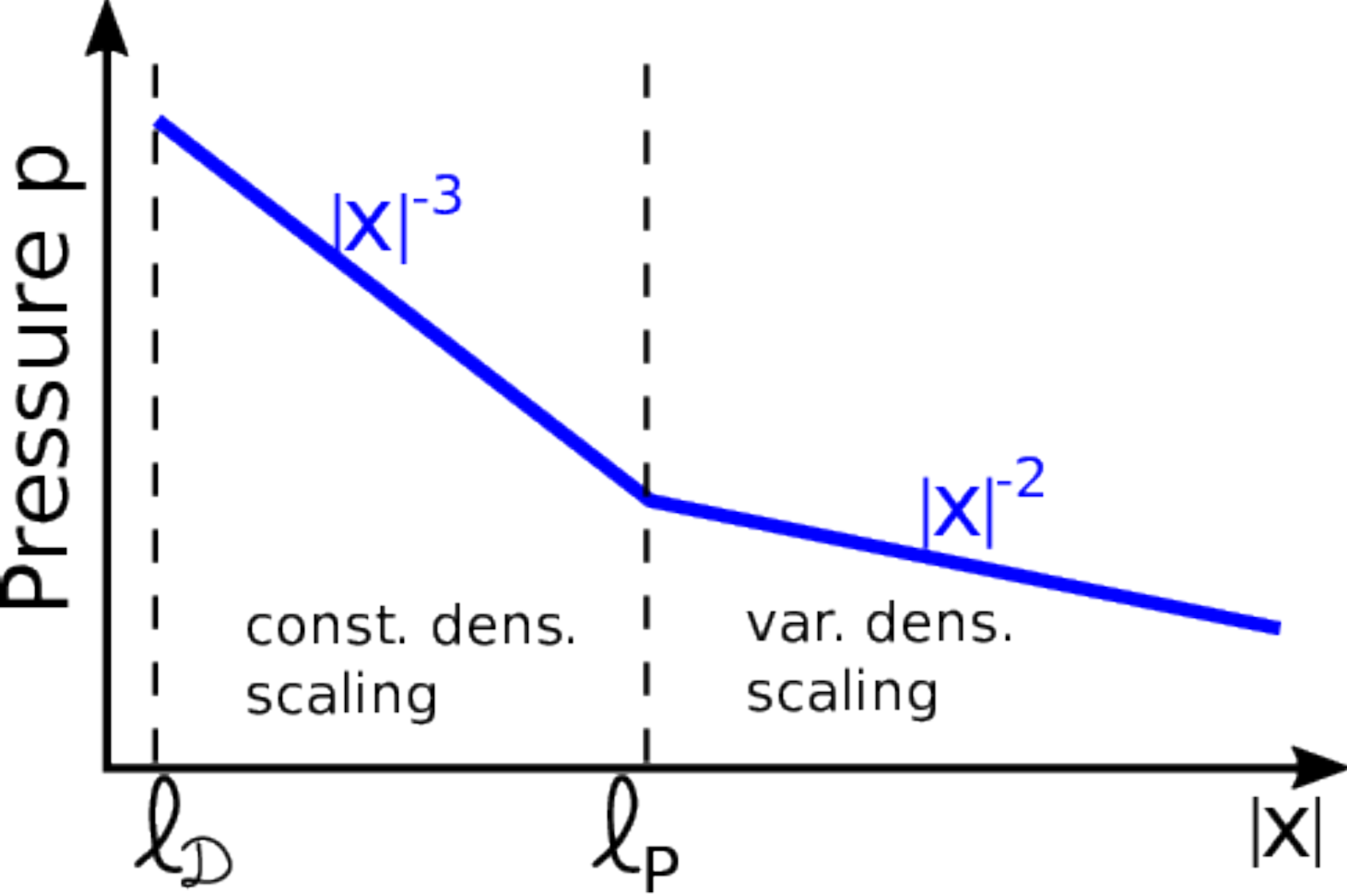}
\caption{\label{fig:blob_pressure}
Schematic representation of the pressure scaling.
}
\end{figure}

With this definition of $\ell_P$, we deduce from Eq. \eqref{eq:blob_press} that $p$ verifies :
\begin{align} \label{eq:p_vd}
\text{for } |\x| \gg \ell_\mathcal{D}, \ell_P  \coma p(\x) \approx |\x|^{-2} \nx_i \partial_t L_i 
\period
\end{align}
If $\ell_P \lesssim \ell_\mathcal{D}$, this relation is  verified as soon as $|\x| \gg \ell_\mathcal{D}$. 
In that case, density variations affect the pressure scaling for all points located far from the eddy core.
However, if $\ell_P \gg \ell_\mathcal{D}$, an intermediate large-scale range exists between  $\ell_\mathcal{D}$ and $\ell_P$.
In this intermediate range, $p$ obeys the following scaling according to Eq. \eqref{eq:blob_press}:
\begin{align}\label{eq:p_cd}
\text{for } \ell_\mathcal{D} \ll  |\x| \ll \ell_P  \coma p(\x) \propto |\x|^{-3}
\period
\end{align}
Thus, in that case, $\ell_P$  separates two different scalings of the pressure field, as schematized in Fig. \ref{fig:blob_pressure}.
For $|\x| \ll \ell_P$, there is a first range of scales where the pressure field obeys its usual constant density scaling.
At larger distances, for $|\x| \gg \ell_P$, a second range of scales exists where the pressure field follows its variable-density scaling \eqref{eq:p_vd}.

The existence of these two ranges has also an impact on the velocity field.
Indeed, if the linear impulse of the eddy is initially null ($\bs{L}=0$), the pressure field can only modify the scaling of the velocity far-field in the second range $|\x| \gg \ell_P$. According to Eq. \eqref{eq:rhou_scaling}, it would transition from a $|\x|^{-4}$ to a $|\x|^{-3}$ scaling.
However, in the first range, $\ell_\mathcal{D} \ll |\x| \ll \ell_P$, the velocity field would keep its initial $|\x|^{-4}$ scaling.
Piecing these remarks together, it appears that the constant density case corresponds to the limit $\ell_P \to \infty$.
Variable-density effects increase when $\ell_P$ decreases and become significant when $\ell_P \sim \ell_\mathcal{D}$.

%=======================================================================
\section{About the velocity spectrum} \label{app:vel_spec}
%=======================================================================

%-------------------------------------------------------------------
\subsection{Evolution of the velocity spectrum at small wavenumbers}
%--------------------------------------------------------------------
The crucial quantity for understanding the permanence of large-eddies in a variable-density flow is the solenoidal momentum $\bs{\qs}$ and its spectrum $\mQ$.
Still, it is interesting to examine the properties of the velocity spectrum at large scales.

To this end, we start by writing the evolution equation of the fluctuating velocity field:
\begin{subequations}  
\begin{align} 
\label{eq:fluc_u}
\partial_t u'_i + \partial_j \big( u'_i  u'_j\big) &= -\partial_j \big( (\tau p)' \delta_{ij} + K'_{ij} \big) 
+ f'_i
\coma
\\
\partial_j u'_j &= 0
\period
\end{align}
\end{subequations}
Then, we apply the Fourier transform to obtain:
\begin{align}  
\label{eq:tf_u}
\partial_t \tf{u}'_i  &= 
-\imath k \mathcal{P}_{ijk}(\bnk) \Big( \tf{u_j' u'_k}
+ \tf{K'_{jk}}
\Big)
+ P_{ij}(\bnk) \tf{f_j}
\coma
\end{align}
with $\bnk$, $\proj_{ij}$ and $\pp_{ijk}$ defined in Sec. \ref{sec:qs_spec}.

The first term on the right-hand side  of Eq. \eqref{eq:tf_u} is the only non-linear one  that would be obtained in a constant density setting. It corresponds to the quadratic product of the velocity field appearing in Eq. \eqref{eq:vd_u} and its redistribution by the pressure field. 
The second term on the right-hand side of Eq. \eqref{eq:tf_u} accounts for viscous and diffusive effects.
The third term appears only when density is variable.
Comparing its expression to the right-hand side of Eq. \eqref{eq:lin_imp}, we see that it involves the same quantity as the one responsible for the variation of the linear impulse of a single eddy and of its far-field scaling.
As will be seen below, this third term is expected to play a similar role for homogeneous turbulence.

Knowing the governing equation of $\tf{\bs{u'}}$, we can deduce the evolution of the velocity spectrum $\mE$.
The latter is defined by:
$$
\mE(k,t) = k^2 \oint E(\K,t) \ud \bnk
\with 
E(\K,t) \delta\left(\K-\K'\right)
= \frac{1}{2} \rey{\tf{u'_i}(\K,t) {\tf{u'_i}^{\cjgt}(\K',t)}}
\period
$$
The evolution equation of $E(\K)$ can be formally written as:
\begin{align}
 \partial_t E(\K,t) =  T_\mE^{(0)}(\K,t) & + \int_0^t \mathcal{T}_\mE(\K,t,s)  \ud s\coma
\\
\nonumber
\with   
T_\mE^{(0)}(\K,t)\delta\left(\K-\K'\right)
= \Re\left( \rey{\tf{u'_i}^{(0)}\hspace{-0.5ex}(\K) \partial_t\tf{u'_i}^\cjgt(\K',t)   } \right)
& \andd 
\mathcal{T}_\mE(\K,t,s) \delta\left(\K-\K'\right)
= \Re\left( \rey{\partial_t\tf{u'_i}(\K,t)\partial_t\tf{u'_i}^\cjgt(\K',s) } \right)
\coma
\end{align}
and that of $\mE$ as:
\begin{align} \label{eq:mt_me}
\partial_t \mE(k,t) = \mT_\mE(k,t) \with \mT_\mE(k,t) = k^2 \oint T_\mE^{(0)}(\K,t) & \ud \bnk + k^2 \int_0^t \oint\mathcal{T}_\mE(\K,t,s) \ud \bnk \ud s
\period 
\end{align}

The notation  $\tf{u'_i}^{(0)}$ refers to the the value of $\tf{u'_i}$ at $t=0$  and $\Re$  to the real part of a given quantity.
The second component of the transfer term can be expressed as the real part of:
\begin{align} \label{eq:Tnl_E}
 \rey{\partial_t\tf{u'_i}(\K,t)\partial_t\tf{u'_i}^\cjgt(\K',s)} =&
k^2 \; \nk_i \nk_k P_{jl}(\bnk) \; \rey{\tf{u'_iu'_j}(\K,t)\tf{u'_ku'_l}^\cjgt(\K',s) } 
\\ \nonumber &
 - \frac{\imath k}{2}\mathcal{P}_{ijk}(\bnk) \!\!\left( \rey{ \tf{u'_ju'_k}(\K,t)  \tf{f_i}^\cjgt(\K',s)} \!-\! \rey{ \tf{u'_ju'_k}^\cjgt(\K',s)  \tf{f_i}(\K,t)}  \right)
+ 
P_{ij}(\bnk) \; \rey{ \tf{f_j}(\K,t)\tf{f_i}^\cjgt(\K',s) }
\\ \nonumber& 
+ \text{terms linked to $\tf{\bs{K'}}$}
\period
\end{align}
In order to model $T_\mE^{(0)}$ and $\mathcal{T}_\mE$, we apply the same set of assumptions as those used to model $T^{(0)}$ and $\mathcal{T}$ in Sec. \ref{sec:model}.
The main outcome is that 
$$
\text{for } k\ll k_e(t) \coma T_\mE^{(0)}(\K,t)=0
\coma
$$ 
and that the non-linear transfer term $\mathcal{T}_\mE$ can be simplified into the following expression:
\begin{align} \label{eq:nonlin_u}
\text{for } k\ll k_e(t) \coma 
\mathcal{T}_\mE(\K,t,s) = k^2 \mathcal{T}_\mE^{(1)}(\bnk,t,s) + k \mathcal{T}_\mE^{(2)}(\bnk,t,s) + \mathcal{T}_\mE^{(3)}(\bnk,t,s)
\coma
\end{align}
\begin{align*}
\with &
\mathcal{T}_\mE^{(1)}(\bnk,t,s) \delta\left(\K-\K'\right)
 =\nk_i \nk_k P_{jl}(\bnk) \; \Re\Big( \left.\rey{\tf{u'_iu'_j}(\K,t)\tf{u'_ku'_l}^\cjgt(\K',s) } \right|_{p,q\gtrsim k_e} \Big)
\\ 
& \mathcal{T}_\mE^{(2)}(\bnk,t,s) \delta\left(\K-\K'\right)
 = 
  \frac{k}{2}\mathcal{P}_{ijk}(\bnk) \; \Im \Big( \left.\rey{ \tf{u'_ju'_k}(\K,t)  \tf{f_i}^\cjgt(\K',s)}\right|_{p,q\gtrsim k_e}   \Big)
\\  
& \mathcal{T}_\mE^{(3)}(\bnk,t,s) \delta\left(\K-\K'\right)
 = 
P_{ij}(\bnk) \; \left. \Re\Big(\rey{ \tf{f_j}(\K,t)\tf{f_i}^\cjgt(\K',s) }\right|_{p,q\gtrsim k_e}\Big)
\period
\end{align*}
The notation $\left. \rey{\;\cdot\;}\right|_{p,q\gtrsim k_e}$ refers to the restriction of the fourth-order correlations to the energetic range. As detailed in Sec. \ref{sec:model}, these restrictions are independent from the wave number $\K$, which explains why  $\mathcal{T}_\mE^{(1)}$, $\mathcal{T}_\mE^{(2)}$ and $\mathcal{T}_\mE^{(3)}$ only depend on $\bnk$.

We eventually obtain the following modeled evolution for $\mE$ at large scales:
\begin{align} \label{eq:dEdt}
\text{for } k\ll k_e(t) \coma 
\partial_t \mE(k,t) =  k^4 {T}_\mE^{(1)}(t) + k^3 {T}_\mE^{(2)}(t) +k^2 {T}_\mE^{(3)}(t)
\end{align}
\begin{align*} 
\with 
{T}_\mE^{(1)}(t) =  \int_0^t \oint \mathcal{T}_\mE^{(1)}(\bnk,t,s) \ud \bnk \, \ud s 
\;,\;
{T}_\mE^{(2)}(t) =  \int_0^t \oint \mathcal{T}_\mE^{(2)}(\bnk,t,s) \ud \bnk \, \ud s
\;,\;
{T}_\mE^{(3)}(t) = \int_0^t \oint  \mathcal{T}_\mE^{(3)}(\bnk,t,s) \ud \bnk \, \ud s
\period
\end{align*}
Thus, following our assumptions, the evolution of $\mE$ is driven by three non-linear terms  with scalings  ranging from $k^4$ to $k^2$.
The first one arises from the the quadratic product between the  components of the velocity field, and their non-local propagation by the pressure field. 
The third one emerges from the non-linear products entering the definition of the variable-density force $\bs{f}$.
The second one is a mix of the two others and is always bounded by them because of Schwartz inequalities.

In a constant density flow, only the first term of Eq. \eqref{eq:dEdt} is present: non-linear interactions have a spectrum scaling as $k^4$.
This is the classical scaling already predicted by several models \cite{proudman54,lesieur08,llor11}. 
When density is variable, additional non-linear terms, stemming from correlations involving the density gradient, arise.
The leading order term  has a  spectrum scaling as $k^2$. 

The difference between the constant and variable-density cases mirrors the various pressure scalings derived in the single eddy case.
As explained in  Sec. \ref{sec:press_field}, these variations on the single eddy pressure field were responsible for modifications in the far-field velocity and its invariance.
Similarly, the differences on the non-linear terms between the constant and variable-density cases  have important consequences concerning the permanence of large eddies.

~\\ \indent
Starting from the modeled equation \eqref{eq:dEdt}, we are now ready to discuss the invariance of $\mE$ at small wavenumbers.
At initial time, we suppose that the velocity spectrum $\mE$  obeys a power law:
\begin{align} \label{eq:mE_init}
\mE(k,t=0) = C k^{\se}
\coma
\end{align}
with $C$ a constant and $\se$ the initial infrared exponent.
Integrating Eq. \eqref{eq:dEdt} yields:
\begin{align} \label{eq:E_ls}
\text{for } k\ll k_e(t) \coma
\mE(k,t) = C k^{\se}+ k^4 \int_0^t{T}_\mE^{(1)}(s) \ud s + k^3 \int_0^t{T}_\mE^{(2)}(s)\ud s +  \int_0^t{T}_\mE^{(3)}(s)\ud s
\period
\end{align}
The spectrum $\mE(k,t)$ is said to be permanent at large scales if it is approximately equal to its initial condition in the limit $k\to0$. In other words, the permanence of large eddies is reached provided the last three non-linear terms on the right-hand side of Eq. \eqref{eq:E_ls} decay faster with $k$ when $k\to0$ than the initial condition $C k^{\se}$.

Comparing the infrared exponents of the different terms of Eq. \eqref{eq:E_ls}, we see that, in the limit $k\to 0$, initial conditions become predominant over non-linear terms only if $\se<2$. Otherwise,  the non-linear component involving $T^{(3)}$ becomes prevalent.
Therefore,  initial spectra with $\se < 2$ are  invariant in the limit $k\to0$ while spectra with $\se>2$ are not. They transition to a spectrum with an infrared exponent  $s=2$. 
Thus, the permanence of the velocity spectrum in variable-density turbulence is only strictly achieved for initial spectra satisfying $\se<2$.

This conclusion is strikingly different from the one obtained for a constant density flow.
Indeed, in that case, ${T}_\mE^{(2)}={T}_\mE^{(3)}=0$ so that only $T_\mE^{(1)}$ remains. Hence, the limit exponent defining whether large eddies are permanent or not is displaced to $\se=4$. Initial spectra with $\se < 4$ are invariant in the limit $k\to0$ while spectra with $\se>4$ are not.
This difference between the constant and variable-density cases echoes what has already been described for the single-eddy configuration concerning the velocity scalings and the invariance of the linear impulse.
It also agrees with what could be deduced from the consideration of homogeneous collections of independent eddies, as described at the beginning of this section.

%--------------------------------------
\subsection{Transition between the constant and variable-density cases}
\begin{figure}[!htb]
\includegraphics[width=0.4\linewidth]{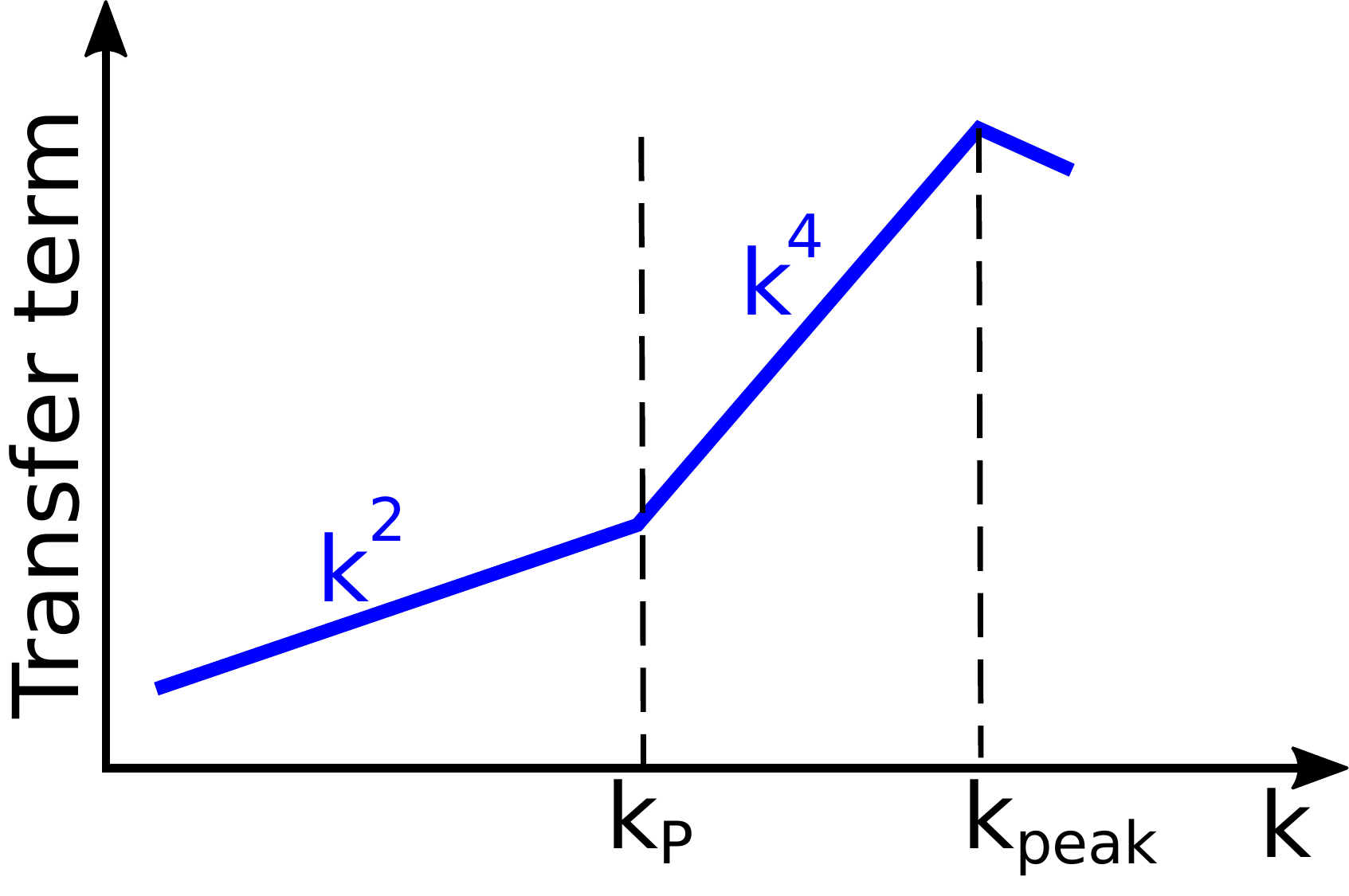}
\caption{\label{fig:transfer}
Schematic representation of the non-linear transfer term.
}
\end{figure}

%--------------------------------------
For a single eddy, a transition between the constant and variable-density cases could be identified thanks to a length scale $\ell_P$ which compared the orders of magnitude of the different contributions of the pressure field.
The same can also be done for homogeneous turbulence, using this time the different contributions to the non-linear transfer terms in Eq. \eqref{eq:dEdt}.
Indeed, the comparison of the orders of magnitude of $k^4 {T}_\mE^{(1)}$ and $k^2 {T}_\mE^{(3)}$ leads to the definition of an additional characteristic wave-number:
\begin{align}\label{eq:def_kp}
k_p(t) = \sqrt{\frac{|{T}^{(3)}_\mE(t)|}{|{T}_\mE^{(1)}(t)| }}
\period
\end{align}
For $k^2 T^{(3)}_\mE$ to be the predominant non-linear term, $k$ must be much smaller than $k_p$:
\begin{align} \label{eq:dEdt_range1}
\text{for } k \ll k_p(t),k_e(t)  \coma
\partial_t \mE(k,t) =  k^2 {T}_\mE^{(3)}(t)
\period
\end{align}
If $k_P \gtrsim k_e$, this relation is  verified as soon as $k \lesssim k_e$. 
In that case, density variations affect the evolution of the kinetic spectrum for the whole large scale range if $\se<2$.
However, if $k_P \ll k_e$, an intermediate large-scale range exists between  $k_P$ and $k_e$.
In this intermediate range, the spectrum obeys the following evolution:
\begin{align}\label{eq:dEdt_range2}
\text{for } k_p \ll  k \ll k_e  \coma
\partial_t \mE(\K,t) =  k^4 {T}_\mE^{(1)}(t)
\period
\end{align}
Thus, in that case, $k_P$  separates two different scalings of the non-linear terms, as schematized in Fig. \ref{fig:transfer}.
For $k_p \ll k \ll k_e$, there is a first range where the spectrum evolution is driven by a $k^4$ backscattering term as in a constant density flow.
For $k \ll k_P$, a second range of scales exists where the spectrum is submitted to a $k^2$ non-linear term.
Note that this schematic description does not account for the possible time variations of $k_p(t)$ and for the possible delays in the establishment of the $k^2$ and $k^4$ backscattering spectra.
Still, it helps convey the important idea that when $k_p\ll k_e$, the range closest to the peak wave number ($k_p\ll k \ll k_e$) behaves as in a constant density flow as far as the permanence of large eddies is concerned.
The differences introduced by density variations only affect the range farthest from the peak wave number ($k\ll k_p$). 

%--------------------------------------
\subsection{Infrared scaling of $\mE$ at late times} \label{app:evol_mE}
%--------------------------------------
Let us assume that the  velocity spectrum $\mE$ obeys the power law \eqref{eq:mE_init} at small wave numbers.
Knowing the conditions under which $\mQ$ is permanent allows to draw some conclusions about the behavior of $\mE$ at small wave numbers.
But first, we must determine the initial condition of $\mQ$ associated with the power law of $\mE$. To this end, we need to relate $\mE$ and $\mQ$. 
This can be done by noting that the solenoidal momentum and the velocity field are related by:
$$
\tf{u'_i}(\K,t) = \rey{\tau} \tf{\qs_i}(\K,t) + \tf{w_i}(\K,t)
\with
\tf{w_i}(\K,t) =  P_{ij}(\bnk) \tf{ \tau' (\rho u_j)'}(\K,t)
\period
$$
Then, if we introduce the spectra $\mF$ of $\tf{w_i}$ and the co-spectrum $\mG$ of $\tf{u'_i}$ and $\tf{w_i}$, we can write that:
\begin{align} \label{eq:Q_E}
\rey{\tau}^2 \mQ(k,t) = \mE(k,t) + \mF(k,t) - 2 \mG(k,t)
\period
\end{align}
The spectrum $E_{ww}$ involves a correlation between two convolution products. In this respect, it is similar to the fourth-order non-linear correlations which were modeled in Sec. \ref{sec:model}.
Thus, the same simplifications can be brought to the expression of $E_{ww}$ as those detailed in Sec. \ref{sec:model}.
This allows to write that:
\begin{align} \label{eq:Eww}
\text{for } k \ll k_e(t) \coma
& 
\mF(k,t) \approx  \frac{k^2}{2} \oint E^\texttt{mod}_{ww}(\bnk,t) \ud \bnk
\\ 
\nonumber
\with & 
E^\texttt{mod}_{ww}(\bnk,t) = P_{ij}(\bnk) 
\Re\Big(
\left.
\rey{
 \tf{\tau' (\rho u_i)'}(\K,t)\tf{\tau' (\rho u_j)'}^\cjgt(\K',t) 
}
\right|_{p,q\gtrsim k_e}
\Big) 
\coma
\end{align}
where $\left. \rey{\;\cdot\;}\right|_{p,q\gtrsim k_e}$ refers to the restriction of the fourth-order correlation to the energetic range detailed in Eq. \eqref{eq:model_Tnl}. For the reasons explained in  Sec. \ref{sec:model}, this restriction is independent of the wavenumber $k$ in the limit $k\ll k_e$.
Therefore, the spectrum $\mF$ has a $k^2$ scaling.

As a result, Eqs. \eqref{eq:Q_E} and \eqref{eq:Eww} show that, at initial time, $\mQ$ is formed by the superposition of two main scalings. The first one is associated with $\mE$ and has an infrared exponent $\se$. The second one is associated with $\mF$ and has an infrared exponent $s=2$.
The component $\mG$ appearing in Eq. \eqref{eq:Q_E} is bound by the Schwartz conditions and remains between $\mE$ and $\mF$. It can contribute to a transition between the two main scalings but does impose one of its own.

\begin{figure}[!htb]
\includegraphics[height=0.4\linewidth,angle=-90]{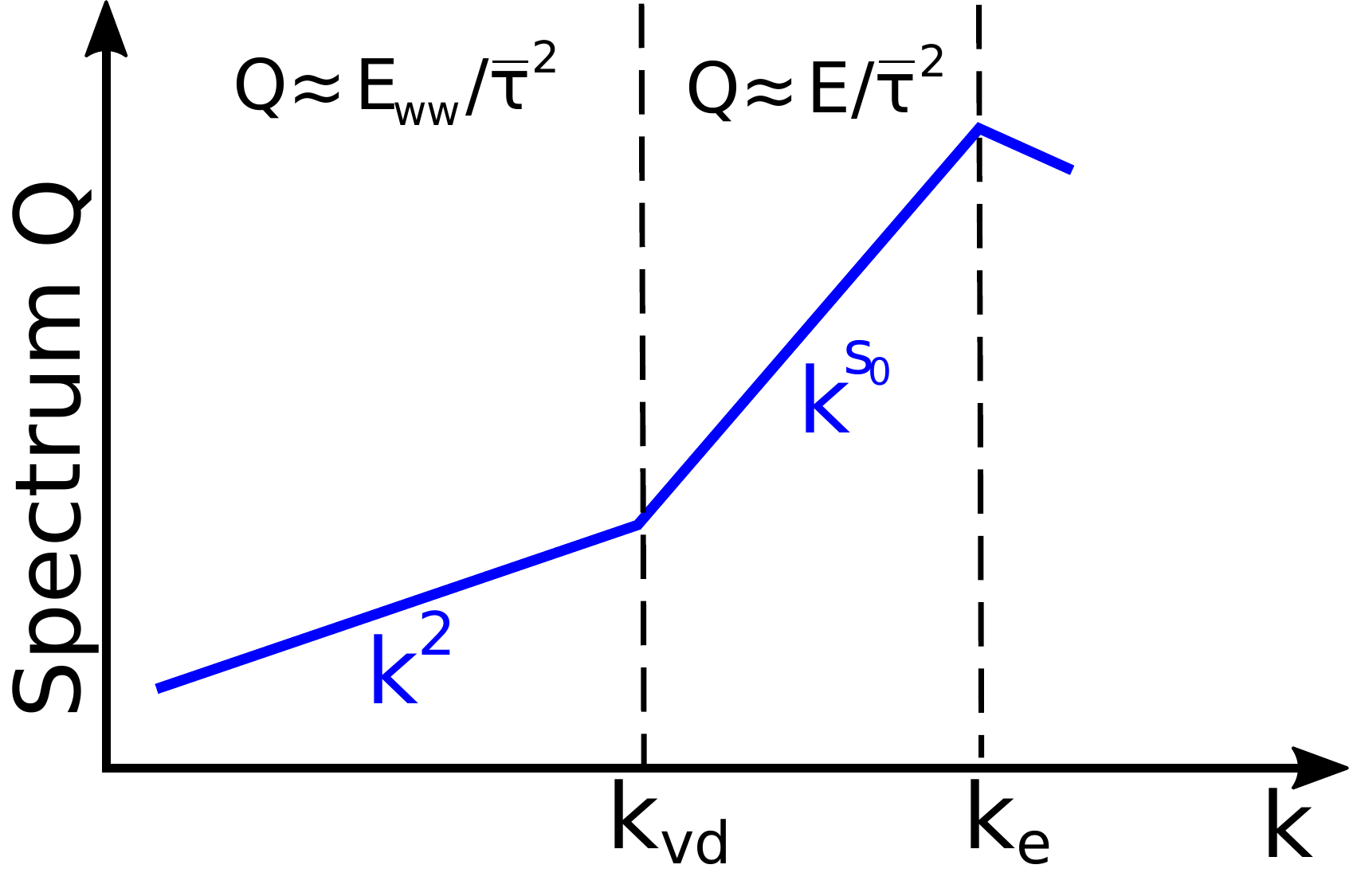}
\caption{\label{fig:Q_init_scaling}
Schematic representation of $\mQ$ when $\mE \propto k^{\se}$ at small wavenumbers.
}
\end{figure}

When $\se > 2$, $\mQ$ displays a transition between a $k^2$ scaling and a steeper $k^{\se}$ scaling.
The order of magnitude of the wave-number $k_\texttt{vd}$ at which this transitions occurs can be estimated as follows.
Starting from the relations $\int_0^{k_e} \mE \ud k \sim \rey{u'_iu'_i}/2$ and $\int_0^{k_e} \mF \ud k \sim \rey{w_iw_i}/2$ and using the scalings of $\mE$ and $\mF$ in the infrared range,  we deduce that:
$$
\mE(k,t=0) \sim (\se+1) \rey{u'_iu'_i} \left(\frac{k}{k_e}\right)^{\se} k_e^{-1}
\andd
\mF(k,t=0) \sim  \rey{w_i w_i} \left(\frac{k}{k_e}\right)^{2} k_e^{-1}
\period
$$
As a crude estimate, we also evaluate the order of $\rey{w_iw_i}$ as:
$$
\rey{w_iw_i} \sim \frac{\rey{\tau'^2}}{\rey{\tau}^2}\rey{u'_iu'_i}
\period
$$
Then, comparing the magnitudes of $\mE$ and $\mF$ in the infrared range  leads to the definition of the following wavenumber:
\begin{align}
\frac{\kvd}{k_e} =\left( \frac{1}{\se+1}   \frac{\rey{\tau'^2}}{\rey{\tau}^2} \right)^{1/(\se-2)}
\period
\end{align}
When the density contrasts is small, then so is the ratio $\kvd/k_e$.
In that case, the infrared range of $\mQ$ is separated into two intervals, as  illustrated in Fig. \ref{fig:Q_init_scaling}:
\begin{subequations} \label{eq:init_mQ}
\begin{align}
\text{When } \kvd \ll k_e \coma
 \mQ \approx \mF/\rey{\tau}^2 \propto k^2 \for k \ll \kvd 
\andd
\mQ \approx \mE/\rey{\tau}^2 \propto k^{\se} \for \kvd \ll k \ll k_e
\period
\end{align}
Note that the coexistence of two different scalings at large scales has been studied in detail in \cite{mons14}.
When the density contrast is high, then the ration $\kvd/k_e$ becomes on the order of one or larger. In that case,  $\mQ$ displays only one infrared scaling:
\begin{align}
\text{When } \kvd \gtrsim k_e \coma
 \mQ \approx \mF/\rey{\tau}^2 \propto k^2 \for  k \ll k_e
\period
\end{align}
\end{subequations}
The last element that needs to be mentioned is that density fluctuations decrease with time so that the difference between $\mE$ and $\mQ$ vanishes:
\begin{align} \label{eq:late_mE2}
\for t \to \infty \coma \mE \to \mQ
\period
\end{align}

Knowing the initial condition of $\mQ$ (Eq. \eqref{eq:init_mQ}), the conditions under which $\mQ$ is permanent (Sec. \ref{sec:perm_mQ}) and the late time behavior of $\mE$ (Eq. \eqref{eq:late_mE2}), we can now conclude on the behavior of $\mE$ at large scales.
\begin{itemize}
\item If the initial density contrast is high enough to ensure the condition $\kvd\gtrsim k_e$ then $\mQ \propto k^2$ at initial time and also at later times since $\mQ$ is then permanent.
Consequently, at late times, $\mE$ will tend to a $k^2$ spectrum independently from the value of $\se$.

\item If the initial density contrast is small enough to lead to $\kvd\ll k_e$ then $\mQ$ has two large-scale ranges at initial time.
\begin{itemize}
\item In the range $k\ll \kvd$, $\mQ$ has a $k^2$ scaling and is permanent. Therefore, at late times, $\mE$ will also tend to a $k^2$ spectrum for $k\ll \kvd$.
\item In the range $\kvd \ll k \ll k_e$, $\mE \approx \rey{\tau}^2 \mQ$ and both have an initial $k^{\se}$ scaling. Therefore, in this range, $\mE$ follows the same evolution as $\mQ$: $\mE$ is permanent if $\se<4$ and transitions to a $k^4$ scaling if $\se>4$.
\end{itemize}
\end{itemize}
It is worth noting that when $\kvd \ll k_e$, the range closest to the peak wave number ($\kvd \ll k \ll k_e$) behaves as in a constant density flow as far as the permanence of large eddies is concerned.
The differences introduced by density variations only affect the range farthest from the peak wave number ($k\ll \kvd$). 
In this respect, the constant density case corresponds to the limit $\kvd \to 0$.
For the spectrum $\mE$, variable-density effects increase when $\kvd$ increases and become significant when $\kvd \sim k_e$. 

Note that $k_P(t)$ and $\kvd$ are two different scales. The former depends on time and marks the influence of non-linear variable-density terms at a given time.
The latter is set by initial conditions and is constant. 
Note also that when instead of $\mE$, it is the infrared slope $\sq$ of $\mQ$ that is imposed at initial time, a similar reasoning can be applied. In particular, $\mE$ will display a $k^2$ scaling for wavenumbers smaller that $\kvd \propto k_e (   {\rey{\tau'^2}}/{\rey{\tau}^2} )^{1/(\sq-2)}$

\end{document}